\documentclass[apj,revtex4]{emulateapj}

\slugcomment{Accepted for publication in ApJ}
\shorttitle{Protocluster search at $z\sim3\mathrm{-}6$}
\shortauthors{Toshikawa et al.}

\begin{document}

\title{A Systematic Survey of Protoclusters at $z\sim3\mathrm{-}6$ in the CFHTLS Deep Fields}
\author{Jun Toshikawa\altaffilmark{1}, Nobunari Kashikawa\altaffilmark{1,2}, Roderik Overzier\altaffilmark{3},
    Matthew A. Malkan\altaffilmark{4}, Hisanori Furusawa\altaffilmark{1}, Shogo Ishikawa\altaffilmark{2},
    Masafusa Onoue\altaffilmark{2}, Kazuaki Ota\altaffilmark{5,6}, Masayuki Tanaka\altaffilmark{1},
    Yuu Niino\altaffilmark{1}, Hisakazu Uchiyama\altaffilmark{2}}
\email{jun.toshikawa@nao.ac.jp}
\altaffiltext{1}{Optical and Infrared Astronomy Division, National Astronomical Observatory, 
    Mitaka, Tokyo 181-8588, Japan.}
\altaffiltext{2}{Department of Astronomy, School of Science, Graduate University for Advanced Studies, 
    Mitaka, Tokyo 181-8588, Japan.}
\altaffiltext{3}{Observat\'{o}rio Nacional, Rua Jos\'{e} Cristino, 77. CEP 20921-400, S\~{a}o Crist\'{o}v\~{a}o,
    Rio de Janeiro-RJ, Brazil}
\altaffiltext{4}{Department of Physics and Astronomy, University of California, Los Angeles, CA 90095-1547.}
\altaffiltext{5}{Kavli Institute for Cosmology, University of Cambridge, Madingley Road, Cambridge CB3 0HA, UK}
\altaffiltext{6}{Cavendish Laboratory, University of Cambridge, 19 J.J. Thomson Avenue, Cambridge CB3 0HE, UK}

\begin{abstract}
We present the discovery of three protoclusters at $z\sim3\mathrm{-}4$ with spectroscopic confirmation in the
Canada-France-Hawaii Telescope (CFHT) Legacy Survey Deep Fields.
In these fields, we investigate the large-scale projected sky distribution of $z\sim3\mathrm{-}6$ Lyman break
galaxies and identify 21 protocluster candidates from regions that are overdense at more than $4\sigma$
overdensity significance.
Based on cosmological simulations, it is expected that more than $76\%$ of these candidates will evolve into a
galaxy cluster of at least a halo mass of $10^{14}\,\mathrm{M_\odot}$ at $z=0$.
We perform follow-up spectroscopy for eight of the candidates using Subaru/FOCAS, KeckII/DEIMOS, and Gemini-N/GMOS.
In total we target 462 dropout candidates and obtain 138 spectroscopic redshifts.
We confirm three real protoclusters at $z=3\mathrm{-}4$ with more than five members spectroscopically identified,
and find one to be an incidental overdense region by mere chance alignment.
The other four candidate regions at $z\sim5\mathrm{-}6$ require more spectroscopic follow-up in order to be
conclusive.
A $z=3.67$ protocluster, which has eleven spectroscopically confirmed members, shows a remarkable core-like
structure composed of a central small region ($<0.5\,\mathrm{physical\>Mpc}$) and an outskirts region
($\sim1.0\,\mathrm{physical\>Mpc}$).
The Ly$\alpha$ equivalent widths of members of the protocluster are significantly smaller than those of field
galaxies at the same redshift while there is no difference in the UV luminosity distributions.
These results imply that some environmental effects start operating as early as at $z\sim4$ along with the
growth of the protocluster structure.
This study provides an important benchmark for our analysis of protoclusters in the upcoming Subaru/HSC imaging
survey and its spectroscopic follow-up with the Subaru/PFS that will detect thousands of protoclusters up to
$z\sim6$.
\end{abstract}


\section{INTRODUCTION}
Galaxy clusters in the early universe provide key clues to the relation between structure formation and galaxy
evolution.
In the local universe, galaxy clusters are located in the densest peaks of the dark matter distribution at the
intersections of filaments \citep[e.g.,][]{lapparent86}.
Their galaxies have significantly different properties from field galaxies as evidenced by the morphology-density
relation and the cluster red sequence \citep[e.g.,][]{dressler80,visvanathan77}.
In this way, cluster formation is closely linked to the large-scale structure and environmental effects on
galaxy properties.
The direct observation of protoclusters, which are overdense regions of galaxies in the high-redshift universe,
will provide us clues to how these relations are formed.

In the local universe, clusters contain a large number of passive galaxies, but the fraction of star-forming
galaxies in clusters is gradually increasing with redshift \citep[e.g.,][]{butcher84}.
 Around $z\sim1$, some studies reported that galaxies in high-density environments form stars more actively
than those in low-density environments \citep[e.g.,][]{tran10,popesso11}.
Beyond $z\sim2$, young and star-forming galaxies appear to be a dominant galaxy population even in the most
overdense regions.
These protoclusters are often identified by using Lyman break galaxies (LBGs) or Ly$\alpha$ emitters (LAEs), which
enable us to trace large-scale structures at high redshift
\citep[e.g.,][]{steidel98,venemans07,kuiper10,galametz13}.
 The highest-redshift protocluster discovered to date is at $z=6.01$ \citep{toshikawa12}, and is composed of
at least ten galaxies \citep{toshikawa14}.
Some candidates without spectroscopic confirmation beyond $z=6$ have also been found \citep{trenti12,ishigaki15}.
Although the majority of protocluster members are young and star-forming galaxies, a red sequence, composed of
bright and red galaxies, is found to appear in protoclusters around $z\sim2\mathrm{-}3$
\citep[e.g.,][]{kurk04,kodama07,zirm08,kubo13,lemaux14}.
These color differences between protocluster and field galaxies is the result of different galaxy properties,
such as age, dust, or metallicity.
For example, stellar mass is a basic and readily observable property that can be used to determine details of the
star-formation history; protocluster galaxies appear to have higher stellar masses than their field counterparts
at $z\sim2\mathrm{-}3$ \citep{steidel05,kuiper10}.
However, \citet{hatch11} reported that star formation rate (SFR) is similar between protocluster and field
galaxies at $z\sim2$.
These results suggest that the differences in stellar mass at $z\sim2\mathrm{-}3$ between protocluster and field
galaxies may be attributed to differences in star-formation duration or the formation epoch.
Besides ordinary galaxies, very rare objects, such as Ly$\alpha$ blobs, submillimeter galaxies (SMGs), and active
galactic nuclei (AGN), are also frequently discovered in high-density environments
\citep{lehmer09,digby10,tamura10,matsuda11}.
Some contradictory results in the mass-metallicity relation have been revealed among protoclusters at the
same redshift \citep{kulas13,valentino15,shimakawa15,kacprzak15}.
Furthermore, \citet{cucciati14} have found a large amount of cold gas surrounding a $z=2.9$ protocluster, which
may serve as the reservoir for significant future star formation.

As described above, some distinguishing features have been identified in each protocluster by comparing with
field galaxies, but the still relatively small number of protoclusters known (see the overview Table in
\citet{chiang13}) makes it difficult to determine which are common features.
In addition to the sample size, it should be noted that many previous studies used radio galaxies (RGs) or quasars
(QSOs) as probes of protoclusters \citep[e.g.,][]{miley04,venemans07,wylezalek13,adams15}, because the
host galaxies of these objects are thought to be embedded in massive dark matter halos.
\citet{hatch14} showed that RGs tend to reside in high-density environments; on the other hand, protoclusters
are also found in the regions without RGs or QSOs \citep[e.g.,][]{steidel98,ouchi05}.
Despite the efficiency of searching for protoclusters around RG or QSO fields, biased protoclusters might be
selected because strong radiation from RGs and QSOs may provide negative feedback and suppress nearby galaxy
formation, especially for low-mass galaxies \citep[e.g.,][]{barkana99,kashikawa07}.
 Furthermore, the mechanisms triggering AGN activity may not be present in all overdense environments and,
given that AGN activity is a transient phenomenon with, in some cases, relatively short timescales
\citep[e.g.,][]{hopkins12}, it is likely that many protoclusters are missed when using such objects as beacons.
In addition to RGs and QSOs, SMGs have been used as similar probes of overdensities at high redshift
\citep[e.g.,][]{capak11,walter12,casey15,smolcic16}.
However, only cursory systematic studies of such objects have been performed \citep[e.g.,][]{aravena10} finding
that SMGs have a complicated relationship with environment, and, indeed, simulations appear to underscore the
complexity of this relationship \citep{miller15}.
Therefore it is preferable, for any systematic study of such environments, to use a population that minimizes such
biases and complexities.
This would allow us to address cluster formation and environmental effects on galaxy evolution based on
systematic and less biased samples of protoclusters.

Here, we present a systematic survey of protoclusters at $z\gtrsim3$ based on wide-field imaging and follow-up
spectroscopy.
This is a complementary approach to protocluster research compared with previous surveys targeting RG/QSO fields
using a relatively small field of view (FoV).
This survey was performed using the wide ($\sim4\,\mathrm{deg^2}$) Canada-France-Hawaii Telescope Legacy Survey
(CFHTLS) Deep Fields, whose depth and area are more adequate to measuring the overdensity of high-redshift
galaxies to identify distant protoclusters than the CFHTLS Wide Fields.
 Although our protocluster candidates were detected by using a less biased method, our method still relies on
the presence of Ly$\alpha$ emission, which will bring another potential bias, in confirming the protocluster
members by spectroscopy.
Section \ref{sec:phot} describes the imaging data and the $z\sim3\mathrm{-}6$ dropout galaxy sample used in this
study.
In Section \ref{sec:over}, we quantify overdensity based on the sky distribution of $z\sim3\mathrm{-}6$ galaxies,
and select the best protocluster candidates by comparing the most overdense regions with expectations from a
cosmological simulation.
The configuration and results of our follow-up spectroscopy are shown in Section \ref{sec:spec}.
In Section \ref{sec:discuss}, we discuss the structure and properties of confirmed protoclusters and compare with
those of field galaxies.
The conclusions are given in Section \ref{sec:conc}.
We assume the following cosmological parameters: $\Omega_\mathrm{M}=0.3, \Omega_\Lambda=0.7,
\mathrm{H}_0=70 \mathrm{\,km\,s^{-1}\,Mpc^{-1}}$, and magnitudes are given in the AB system.

\section{SAMPLE SELECTION} \label{sec:phot}
\subsection{Photometric Data}
We made use of publicly available data from the CFHTLS \citep[T0007:][]{gwyn12}, which was obtained with MegaCam
mounted at the prime focus of the CFHT.
The Deep Fields of the CFHTLS were used in this study, which consist of four independent fields of about
$1\,\mathrm{deg^2}$ area each ($\sim4\,\mathrm{deg^2}$ area in total) observed in the $u^*$-, $g'$-, $r'$-, $i'$-,
and $z'$-bands.
The field center and limiting magnitudes of each field are summarized in Table \ref{tab:photo}.
The seeing size (FWHM) and pixel scale of all the images are $\sim0\farcs7$ and $0\farcs186$, respectively.
Although data at other wavelengths, such as near- or mid-infrared imaging, are available in a part of the
CFHTLS Deep Fields, the depth and coverage are significantly different from field to field.
Therefore, our protocluster search is conducted only based on the optical data to make an uniform survey.
\begin{deluxetable*}{ccccccccccccc}
\tabletypesize{\scriptsize}
\tablecaption{Photometric data and the number of dropout galaxies \label{tab:photo}}
\tablewidth{0pt}
\tablehead{Field & R.A. & Decl. & area & $u^*$\tablenotemark{a} & $g'$\tablenotemark{a} &
    $r'$\tablenotemark{a} & $i'$\tablenotemark{a} & $z'$\tablenotemark{a} & $N_u$\tablenotemark{b} &
    $N_g$\tablenotemark{b} & $N_r$\tablenotemark{b} & $N_i$\tablenotemark{b} \\
    & (J2000) & (J2000) & ($\mathrm{arcmin^2}$) & (mag) & (mag) & (mag) & (mag) & (mag) & & & &}
\startdata
D1 & 02:25:59 & $-$04:29:40 & 3063 & 28.12 & 28.32 & 27.77 & 27.30 & 26.39 & 17110 & 10416 & 2433 & 148 \\
D2 & 10:00:28 & $+$02:12:30 & 2902 & 28.07 & 28.19 & 27.70 & 27.30 & 26.45 & 14515 & 11160 & 2539 & 231 \\
D3 & 14:19:27 & $+$52:40:56 & 3161 & 28.14 & 28.38 & 27.91 & 27.48 & 26.43 & 21454 & 14896 & 2579 & 232 \\
D4 & 22:15:31 & $-$17:43:56 & 3035 & 27.96 & 28.19 & 27.67 & 27.17 & 26.26 & 10484 & 11288 & 1926 & 188
\enddata
\tablenotetext{a}{$3\sigma$ limiting magnitude in a $1\farcs4$ aperture.}
\tablenotetext{b}{The number of $u$-, $g$-, $r$-, or $i$-dropout galaxies.}
\end{deluxetable*}

We created two multi-color catalogs with SExtractor \citep[version 2.8.6;][]{bertin96}, in which $i'$- and
$z'$-band images were used as detection images respectively.
The detection images were first smoothed with Gaussian function, then objects were detected by requiring a
minimum of three adjacent pixels each above $1\sigma$ of the sky background rms noise.
Then, the magnitudes and several other photometric parameters were measured in the other band images at exactly
the same positions and with the same aperture of $1.4\,\mathrm{arcsec}$ as in the detection-band images using
the ``double image mode.''
The Galactic extinction was removed for each field based on the measurement of \citet{schlafly11}. 
The individual catalogues were masked to remove the regions where the detection and photometric measurements
of objects may have been seriously affected.
These regions are around bright stars, diffraction and bleed spikes from bright stars.
The regions near the frame edges, whose depth is systematically shallow, were also excluded.
As noted in \citet{gwyn12}, bright stars with $\lesssim 9\,\mathrm{mag}$ produce a large halo, whose radius is
$\sim3.5\,\mathrm{arcmin}$.
The total masked regions were $\sim15\mathrm{-}20\%$ of the FoV, and the effective areas of our analysis are
shown in Table \ref{tab:photo}.
Finally, $\sim$330,00--420,000 and 230,000--270,000 objects were detected down to the $3\sigma$ limiting
magnitudes,  defined as the magnitude corresponding to three times the standard deviation in the sky flux
measured in empty $1.4\,\mathrm{arcsec}$ apertures, of the $i'$- and $z'$-bands in each field, respectively.
To estimate the detection completeness of each $i'$- and $z'$-band image, we used the IRAF task {\sf mkobjects}
to create artificial objects on the original images.
Artificial objects, which were given a Gaussian profile with a FWHM the same as the seeing size, were randomly
distributed on the real image outside of twice the FWHM of the real objects to avoid blending artificial objects
with real objects.
We generated 50,000 artificial objects in the $20\mathrm{-}30\,\mathrm{mag}$ range, and extracted them using
SExtractor with the same parameter set.
This procedure was repeated ten times, and the detection completeness was 70--50\% at the $3\sigma$ limiting
magnitudes of $i'$- and $z'$-bands in each field.
 At fainter than $3\sigma$ limiting magnitude, the detection completeness drops sharply to $\sim10\%$.
We confirmed that the results of our completeness tests are consistent with the values described in the CFHTLS
data release (we find $\sim84\%$ completeness at the same magnitude at which $\sim80\%$ is expected).
It should be noted that detection completeness at bright magnitudes depends on blending with neighbor objects
and we carefully masked out bright objects.

\subsection{Selection of Dropout Galaxies at $z\sim3-6$ \label{sec:select}}
We selected $z\sim3-6$ galaxy candidates using the Lyman break technique ($u$-, $g$-, $r$-, and $i$-dropout
galaxies).
The $i'$-band detection catalog was used for the selection of $u$-, $g$-, and $r$-dropout galaxies, and
$i$-dropout galaxies were selected from the $z'$-band detection catalog, based on the following color selection
criteria \citep{burg10,toshikawa12}:
\begin{eqnarray*}
u\mathrm{-dropouts} &:& 1.0<(u^*-g') \wedge -1.0<(g'-r')<1.2 \\
 & & \hspace{5mm} \wedge 1.5(g'-r')<(u^*-g')-0.75, \\
g\mathrm{-dropouts} &:& 1.0<(g'-r') \wedge -1.0<(r'-i')<1.0 \\
 & & \hspace{5mm} \wedge 1.5(r'-i')<(g'-r')-0.80 \\
 & & \hspace{5mm} \wedge u^*>m_\mathrm{lim,2\sigma}, \\
r\mathrm{-dropouts} &:& 1.2<(r'-i') \wedge -1.0<(i'-z')<0.7 \\
 & & \hspace{5mm} \wedge 1.5(i'-z')<(r'-i')-1.00 \\
 & & \hspace{5mm} \wedge u^*,g'>m_\mathrm{lim,2\sigma}, \\
i\mathrm{-dropouts} &:& (i'-z')>1.5 \wedge u^*,g',r'>m_\mathrm{lim,2\sigma},
\end{eqnarray*}
where $m_\mathrm{lim,2\sigma}$ is a $2\sigma$ limiting magnitude.
In redder bands than Lyman break (e.g. $g'$- and $r'$-bands for $u$-dropout galaxies), we only used objects
detected with more than $2\sigma$ significance in order to accurately estimate their color; on the other hand,
$2\sigma$ limiting magnitude was used as color limit if objects were detected less than $2\sigma$ significance
in bluer bands (e.g. $u^*$-band for $u$-dropout galaxies).
 The results do not significantly change even if we use 1 or $3\sigma$ significance as the limiting magnitude.
We estimated the redshift distribution resulting from these criteria by using the population synthesis model code
GALAXEV \citep{BC03} and the absorption of intergalactic medium (IGM) \citep{madau95}.
In GALAXEV, we simulated a large variety of galaxy spectral energy distributions (SEDs) using the Padova 1994
simple stellar population model.
We assumed a \citet{salpeter55} initial mass function with lower and upper mass cutoffs $m_L=0.1\,\mathrm{M_\sun}$
and $m_U=100\,\mathrm{M_\sun}$, two metallicities (0.2 and 0.02 $\mathrm{Z_\sun}$), and  two star formation
histories (SFHs) of constant and instantaneous star formation.
We extracted model spectra with ages between $5\,\mathrm{Myr}$ and the age of the universe at that redshift
and applied the reddening law of \citet{calzetti00} with $E(B-V)$ between 0.00 and 1.50.
The model magnitudes were estimated by convolving these simulated SEDs with the filter transmission curves.
\begin{figure}
\epsscale{1.0}
\plotone{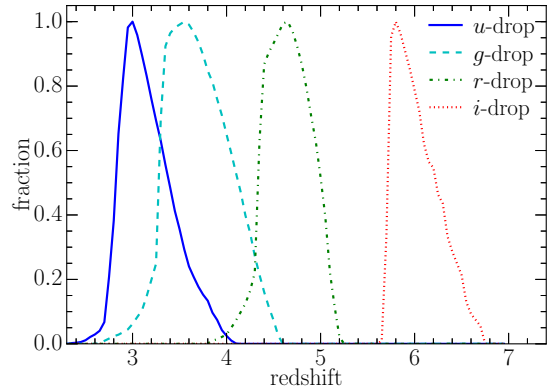}
\caption{Expected redshift distribution of $u$-, $g$-, $r$-, and $i$-dropout galaxies.}
\label{fig:select}
\end{figure}
We then added the photometric noise, which is typically $\Delta m=0.04\,\mathrm{mag}$ and $0.13\,\mathrm{mag}$ at
the $10\sigma$ and $3\sigma$ limiting magnitudes, respectively.
In this process, we assumed that the magnitude distribution of the simulated galaxies and the observed dropout
galaxies were the same.
Then, the redshift distributions of the $u$-, $g$-, $r$-, and $i$-dropout galaxies are estimated by applying the
same color selection criteria of dropout galaxies to these simulated SEDs (Figure \ref{fig:select}).
The expected redshift ranges of the dropout selection are $z\sim2.8\mathrm{-}3.7$, $3.3\mathrm{-}4.3$,
$4.3\mathrm{-}5.1$, and $5.7\mathrm{-}6.5$ for $u$-, $g$-, $r$-, and $i$-dropout galaxies, respectively.
It should be pointed out that these estimates rely on some assumptions (e.g., the model of IGM absorption,
SFH).
Although we used the IGM model of \citet{madau95}, other models have been proposed \citep{meiksin06,inoue14}.
Furthermore, \citet{thomas14} found that IGM absorption  can vary significantly among different lines of
sight, but there is a degeneracy between IGM absorption and dust extinction.
However, as long as the properties of the galaxies in overdense regions and in the field are not too different,
both populations will be affected in the same manner, implying that we can still search for relative overdensities
as a tracer of protoclusters.
The numbers of selected dropout galaxies detected in each field are shown in Table \ref{tab:photo}.
Note that the limiting magnitude of this study corresponds to about $M^*_\mathrm{UV}+2.6$,
$M^*_\mathrm{UV}+2.4$, $M^*_\mathrm{UV}+1.7$, and $M^*_\mathrm{UV}$  (where $M^*_\mathrm{UV}$ is the
characteristic magnitude of the Schechter functions fitted the dropout galaxies at each redshift) for $u$-, $g$-,
$r$-, and $i$-dropout galaxies, respectively
\citep{bouwens07,burg10}.
Thus, the survey depth reaches at least the typical brightness of dropout galaxies even for $i$-dropout galaxies.

\begin{figure}
\epsscale{1.2}
\plotone{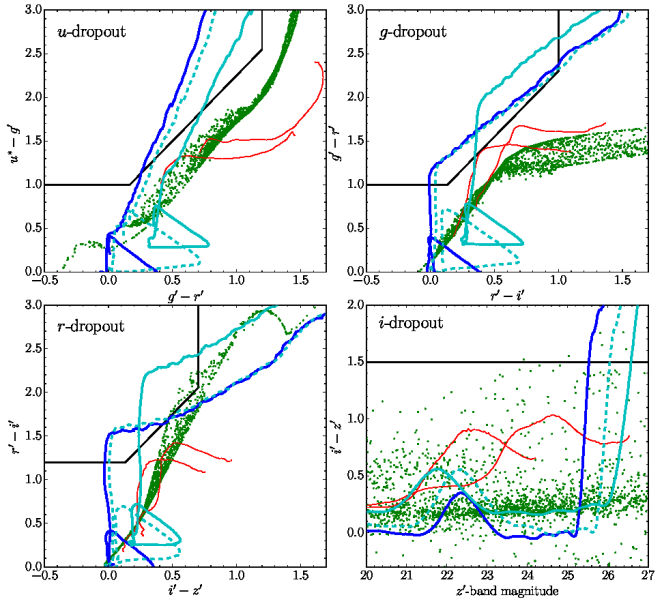}
\caption{Demonstration of dropout galaxy selection on two color and color-magnitude diagrams.
    Thick black lines show the borders of our dropout galaxy selection (see Section \ref{sec:select}).
    Blue lines indicate redshift evolution tracks of young star-forming galaxies
    ($\mathrm{age}=100\,\mathrm{Myr}$, $\mathrm{E(B-V)}=0.1$), and cyan lines indicate the same as blue lines but
    for $\mathrm{E(B-V)}=0.4$ (solid), $\mathrm{age}=600\,\mathrm{Myr}$ (dashed).
    Two red lines are redshift evolution tracks of elliptical galaxies at $z=0\mathrm{-}1.5$ with ages of 1 and
    $7\,\mathrm{Gyr}$, and green dots are dwarf stars estimated by the TRILEGAL galactic model \citep{girardi05}.
    Note that redshift evolution tracks in the $i$-dropout panel can shift horizontally depending assumption of
    stellar mass since the $x$-axis is magnitude not color.}
\label{fig:demo}
\end{figure}

We evaluated the contamination rate for these color-selection criteria by comparing the dropout selection regions
with the positions of contamination on two-color diagram (Figure \ref{fig:demo}).
The major sources of the contamination are dwarf stars and old elliptical galaxies, the latter being possible
to satisfy the color criteria due to the 4000{\AA}/Balmer break.
To estimate the contamination rate of dwarf stars, we use the TRILEGAL code \citep{girardi05}, which can simulate
number count and broad-band photometry of stars in any Galaxy field.
Since this model enables us to set up various structural parameters of thin disc, thick disc, halo, and bulge,
we used an exponential disk model with default values of scale length and height, and a Chabrier IMF was applied.
The galactic latitudes were set to be the same as those of the observations ($|b|=40-60^\circ$).
Then, photometry of the simulated dwarf stars was calculated for the CFHT/Megacam's filter set.
Next, we simulated old galaxy SEDs using the GALAXEV, assuming two relatively high metallicities
($\mathrm{Z_\sun}$ and $2.5\mathrm{Z_\sun}$), and extracted model spectra with ages of
$1.0\mathrm{-}10.0\,\mathrm{Gyr}$.
The redshift tracks of old galaxies are away from all dropout selection regions.
And, although a few dwarf stars are located only within the $r$- and $i$-dropout selection regions, the main
locus of dwarf stars lie far from these regions.
Actually, the contamination rate of dwarf stars in the $r$-dropout samples is expected to be $2.2-7.8\%$
depending on the galactic latitude, and the contamination rate in $i$-dropout samples is $3.4-6.4\%$ in the
CFHTLS Deep Fields.
Based on these simulations of dwarf stars and old galaxies, the dropout selection criteria used in this study
are confirmed to be able to separate high-redshift galaxies from contaminations.

\section{PROTOCLUSTER CANDIDATES} \label{sec:over}
\subsection{Sky Distribution and Selection of Significant Overdensities of Dropout Galaxies} \label{sec:sky}
We have estimated the local surface number density by counting the number of dropout galaxies within a fixed
aperture in order to determine the overdensity significance quantitatively.
\citet{chiang13} presented a useful definition of the characteristic radius of protoclusters based on cosmological
simulations, which encloses 65\% of the mass, based on a combination of $N$-body dark matter simulations and
semi-analytical galaxy formation models.
Although their result was based on the three-dimensional distribution of protocluster galaxies, it is still useful
guide for constructing a map of the projected local surface number density when searching for protoclusters.
According to the characteristic radius of protoclusters having a descendant halo mass of
$1\mathrm{-}3\times10^{14}\,\mathrm{M_\sun}$ at $z=0$, the radius of $0.75\,\mathrm{physical\>Mpc}$ was
used for $u$-, $g$-, and $r$-dropout galaxies, which corresponds to $1.6$, $1.8$, and $1.9\,\mathrm{arcmin}$,
respectively.
Although the characteristic radius is expected to be larger for protoclusters having more massive descendant
masses, these still show a clear overdensity even on $0.75\,\mathrm{physical\>Mpc}$ scales; thus, the aperture
size of $0.75\,\mathrm{physical\>Mpc}$ should be effective to find protoclusters with a descendant halo mass of
$>1\times10^{14}\,\mathrm{M_\sun}$.
However, we still have to consider the projection effects resulting from the large redshift uncertainty of our
dropout selection.
We discuss the effectiveness of our protocluster search more quantitatively by using a theoretical model in the
following subsection.
It should be noted that protoclusters have a nearly constant physical size at $z\gtrsim3$
\citep{chiang13,muldrew15}.
\begin{figure}
\epsscale{1.2}
\plotone{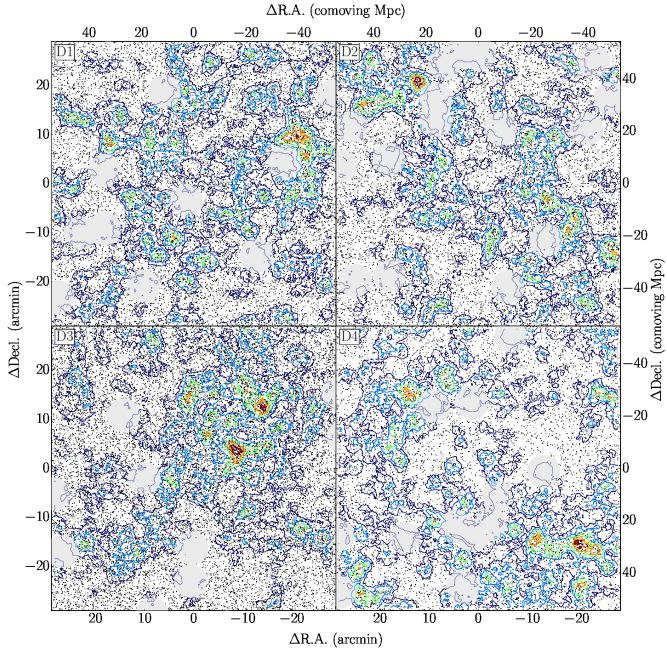}
\caption{Sky distribution of $u$-dropout galaxies (dots) with surface number density contours (lines) in the D1
    (upper left), D2 (upper right), D3 (lower left), and D4 (lower right) field.
    The lines correspond to contours of surface overdense significance from $4\sigma$ to $0\sigma$ (mean) with
    a step of $1\sigma$.
    North is up, and east is to left.
    The comoving scale projected to $z=3.1$ is shown along the axes, and masked regions are also shown by gray
    region.}
\label{fig:cntr_udrop}
\end{figure}
\begin{figure}
\epsscale{1.2}
\plotone{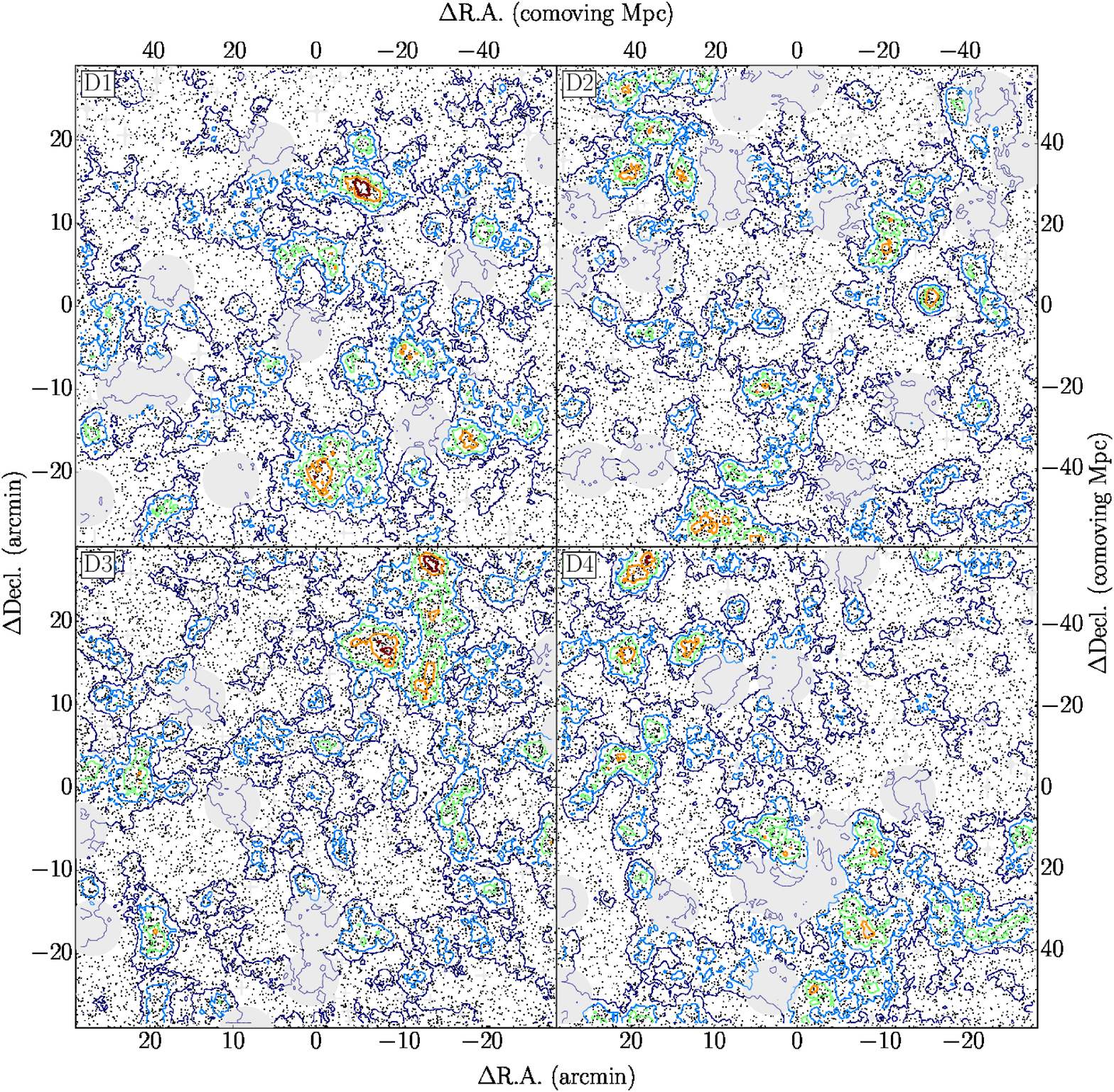}
\caption{Same as Figure \ref{fig:cntr_udrop}, but for the $g$-dropout galaxies.
    The comoving scale projected to $z=3.8$ is shown along the axes.}
\label{fig:cntr_gdrop}
\end{figure}
\begin{figure}
\epsscale{1.2}
\plotone{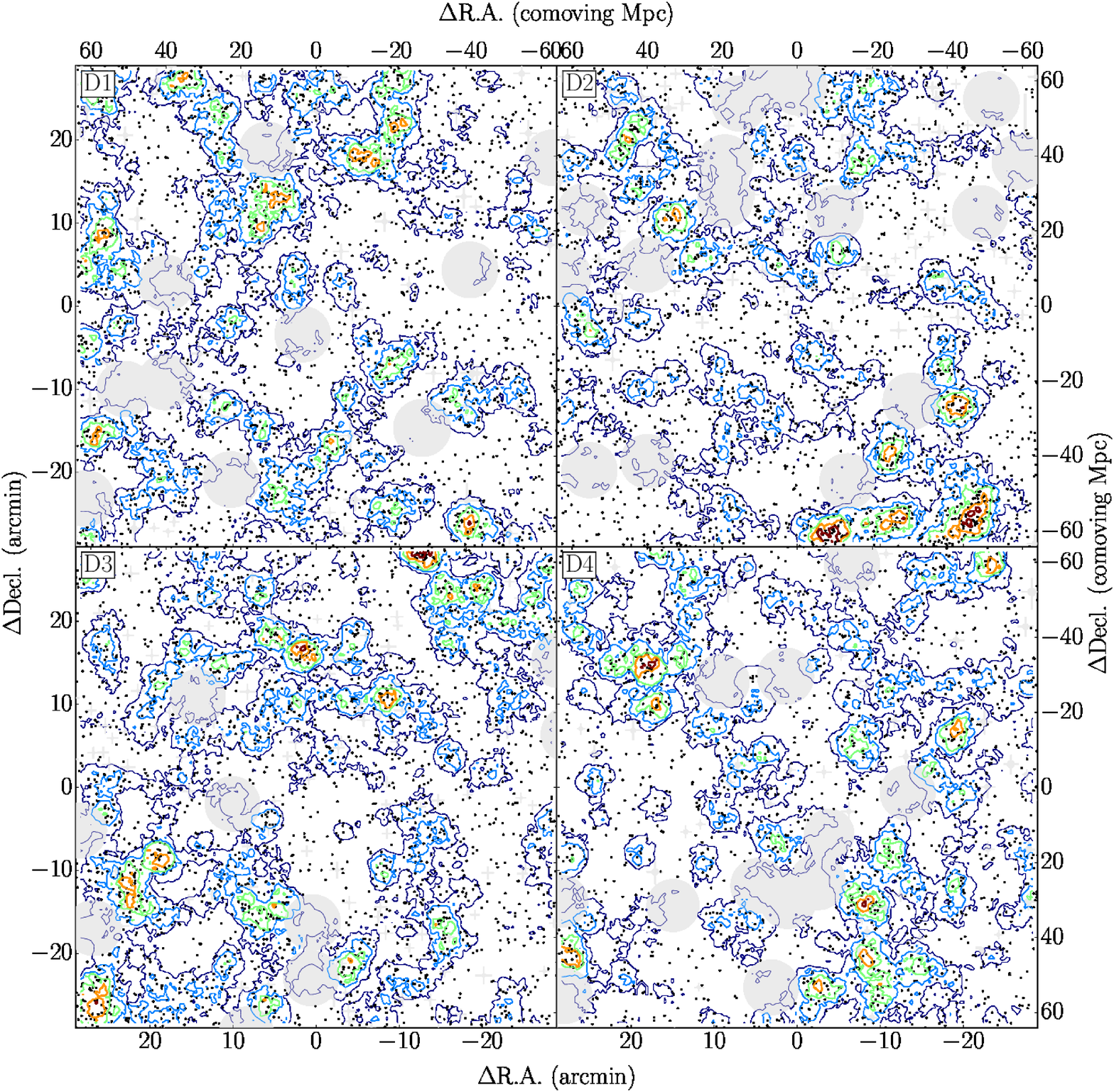}
\caption{Same as Figure \ref{fig:cntr_udrop}, but for the $r$-dropout galaxies.
    The comoving scale projected to $z=4.7$ is shown along the axes.}
\label{fig:cntr_rdrop}
\end{figure}
\begin{figure}
\epsscale{1.2}
\plotone{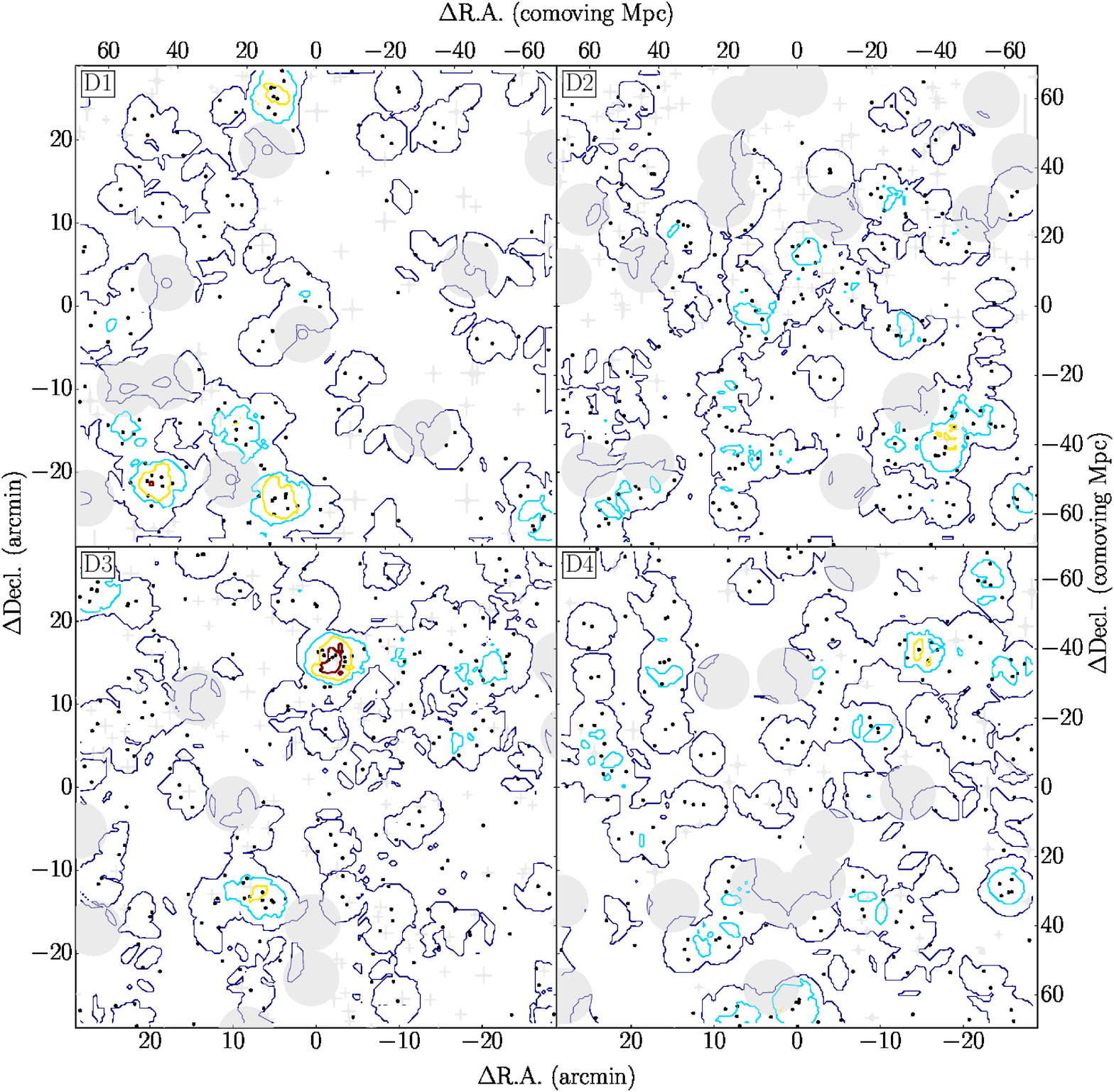}
\caption{Same as Figure \ref{fig:cntr_udrop}, but for the $i$-dropout galaxies, and the lines show overdensity
    significance from $6\sigma$ to $0\sigma$ (mean) with a step of $2\sigma$.
    The comoving scale projected to $z=5.9$ is shown along the axes.}
\label{fig:cntr_idrop}
\end{figure}
We used a slightly larger radius of $1.0\,\mathrm{physical\>Mpc}$ ($2.9\,\mathrm{arcmin}$) for $i$-dropout
galaxies in order to reduce the large Poisson error resulting from a too small aperture.
The apertures were distributed over the CFHTLS Deep fields in a grid pattern at intervals of $\sim20\arcsec$.
We measured the mean and the dispersion, $\sigma$, of the number of galaxies in an aperture over the field.
The surface number density in masked regions was assumed to be the same as the mean surface number density.
Apertures in which more than $5\%$ area is masked are not used in the following analysis.
Using the mean and $\sigma$ of the number of dropout galaxies in an aperture, surface number density contours
of $u$-, $g$-, $r$-, and $i$-dropout galaxies were calculated and are plotted in Figures \ref{fig:cntr_udrop},
\ref{fig:cntr_gdrop}, \ref{fig:cntr_rdrop}, and \ref{fig:cntr_idrop}, respectively.
We note that this is the same procedure that was applied to the $i$-dropout galaxies in the Subaru Deep Field
(SDF) to draw their surface number density contour, which led to the discovery of the protocluster at $z=6.01$
from \citet{toshikawa12,toshikawa14}.
In order to study the effects of incompleteness, we performed the same overdensity measurement but for $u$-,
$g$-, and $r$-dropout galaxies brighter than $26.0\,\mathrm{mag}$ in $i'$-band (for $i$-dropout galaxies, it is
hard to select a brighter subset due to the small sample size).
From this analysis, we confirmed that the change of limiting magnitude does not have a significant effect on our
protocluster selection.

\subsection{Comparison with Cosmological Simulations}
Although we can clearly see some overdense regions in the sky distribution maps presented in Figures
\ref{fig:cntr_udrop}, \ref{fig:cntr_gdrop}, \ref{fig:cntr_rdrop}, and \ref{fig:cntr_idrop}, it is not
straightforward to find plausible protocluster candidates since the large redshift uncertainty of dropout
technique, which is $\Delta z\sim1$ ($\sim 230-60\,\mathrm{physical\>Mpc}$ at $z\sim3.1-6.0$), hampers the
identification of clustering structure in three-dimensional space. 
On one hand, overdensities associated with real protoclusters could be weakened by fore/background galaxies; on
the other hand, chance alignments of the large-scale structure or superpositions of filaments could erroneously
enhance the surface overdensity.
Therefore, we will make use of predictions from simulations to understand the relation between surface overdensity
significance and the probability of finding real protoclusters.

To this end, we used a set of light-cone models constructed by \citet{henriques12}.
A brief outline of the light-cone models is presented below.
First, the assembly history of the dark matter halos was traced using an $N$-body simulation \citep{springel05},
in which the length of the simulation box was $500\,h^{-1}\,\mathrm{Mpc}$ and the particle mass was
$8.6\times10^8\,h^{-1}\,\mathrm{M_\sun}$.
The distributions of dark matter halos were stored at discrete epochs.
Next, the processes of baryonic physics were added to dark matter halos at each epoch using a semi-analytic galaxy
formation model \citep{guo11}.
Based on the intrinsic parameters of galaxies predicted by the semi-analytic model, such as stellar mass, star
formation history, metallicity, and dust content, the photometric properties of simulated galaxies were estimated 
from the stellar population synthesis models developed by \citet{BC03}.
Then, these simulated galaxies in boxes at different epochs were projected along the line-of-sight, and
intergalactic medium (IGM) absorption was applied \citep{madau95} in order to mimic a pencil-beam survey as
described in \citet{overzier13}.
Finally, 24 light-cone models with $1.4\times1.4\,\mathrm{deg^2}$ FoV were extracted using these procedures.
\begin{deluxetable*}{cccccccc}
\tablecaption{Overview of the protocluster candidates \label{tab:target}}
\tablewidth{0pt}
\tablehead{Name & R.A. (J2000) & Decl. (J2000) & Field & Population & Overdensity\tablenotemark{a} &
    $N_\mathrm{galaxy}$\tablenotemark{b} & Spec.\tablenotemark{c}}
\startdata
D1UD01 & 02:24:35.4 & $-$04:19:58.9 & D1 & $u$-dropout & $4.2\sigma$ & 244 & Yes \\
D2UD01 & 10:01:18.6 & $+$02:33:20.3 & D2 & $u$-dropout & $4.6\sigma$ & 182 & No \\
D3UD01 & 14:18:29.1 & $+$52:44:05.3 & D3 & $u$-dropout & $4.8\sigma$ & 300 & No \\
D3UD02 & 14:17:52.0 & $+$52:53:03.2 & D3 & $u$-dropout & $4.4\sigma$ & 268 & No \\
D4UD01 & 22:14:03.4 & $-$17:58:43.4 & D4 & $u$-dropout & $4.4\sigma$ & 157 & Yes \\
D1GD01 & 02:25:36.3 & $-$04:15:57.4 & D1 & $g$-dropout & $5.5\sigma$ & 162 & Yes \\
D1GD02 & 02:25:56.2 & $-$04:48:30.4 & D1 & $g$-dropout & $4.2\sigma$ & 153 & No \\
D3GD01 & 14:18:28.9 & $+$52:57:06.5 & D3 & $g$-dropout & $4.7\sigma$ & 214 & No \\
D3GD02 & 14:17:55.6 & $+$53:07:37.6 & D3 & $g$-dropout & $4.5\sigma$ & 201 & No \\
D4GD01 & 22:16:47.3 & $-$17:16:52.7 & D4 & $g$-dropout & $4.3\sigma$ & 153 & Yes \\
D1RD01 & 02:24:45.3 & $-$04:55:56.5 & D1 & $r$-dropout & $4.4\sigma$ & 40 & Yes \\
D2RD01 & 10:00:14.1 & $+$01:44:03.0 & D2 & $r$-dropout & $4.9\sigma$ & 48 & No \\
D2RD02 & 09:59:04.6 & $+$01:47:27.5 & D2 & $r$-dropout & $4.5\sigma$ & 64 & No \\
D3RD01 & 14:19:36.8 & $+$52:57:44.6 & D3 & $r$-dropout & $4.5\sigma$ & 39 & No \\
D4RD01 & 22:14:58.1 & $-$17:58:07.2 & D4 & $r$-dropout & $4.2\sigma$ & 31 & No \\
D4RD02 & 22:16:46.1 & $-$17:29:16.7 & D4 & $r$-dropout & $4.1\sigma$ & 31 & Yes \\
D1ID01 & 02:27:18.4 & $-$04:50:58.9 & D1 & $i$-dropout & $6.1\sigma$ & 10 & Yes \\
D1ID02 & 02:26:19.9 & $-$04:51:55.0 & D1 & $i$-dropout & $5.5\sigma$ & 9 & No \\
D3ID01 & 14:19:14.2 & $+$52:55:15.7 & D3 & $i$-dropout & $7.6\sigma$ & 16 & Yes \\
D3ID02 & 14:20:09.3 & $+$52:28:17.1 & D3 & $i$-dropout & $4.7\sigma$ & 10 & No \\
D4ID01 & 22:14:29.6 & $-$17:27:25.4 & D4 & $i$-dropout & $4.1\sigma$ & 7 & No
\enddata
\tablenotetext{a}{Overdensity at the peak.}
\tablenotetext{b}{The number of dropout galaxies within $3\,\mathrm{arcmin}$ radius from its overdensity peak.}
\tablenotetext{c}{The protocluster candidates observed by follow-up spectroscopy are marked as ``Yes''.}
\end{deluxetable*}

\begin{figure}
\epsscale{1.2}
\plotone{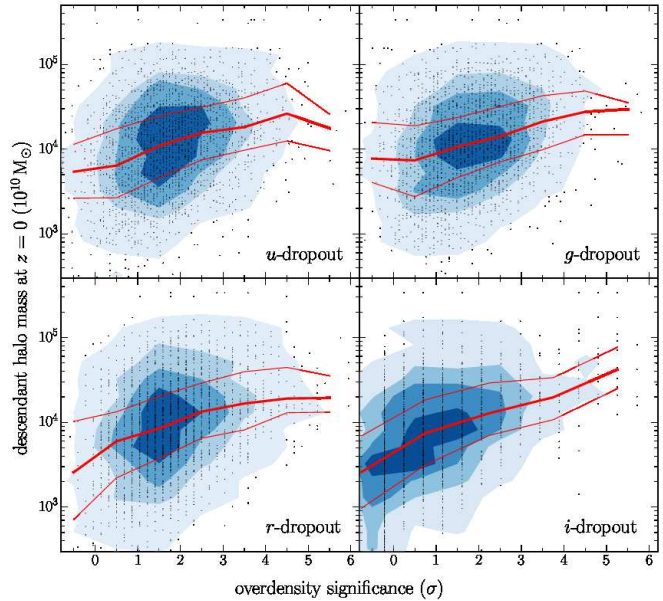}
\caption{Relation between surface overdensity of $u$-, $g$-, $r$-, and $i$-dropout galaxies and descendant halo
    mass at $z=0$.
    The thick and thin red lines are the median, upper, and lower quartiles.
    The background contours show the 25, 50, 75, and 95\% region from dark to light.}
\label{fig:prob}
\end{figure}

The simulated $u$-, $g$-, $r$-, and $i$-dropout galaxy catalogs were made by matching the expected redshift
distribution of each dropout galaxy sample (Figure \ref{fig:select})
\footnote{Ideally, we should select the dropout galaxies from the simulations by applying the same color selection
criteria to the simulated catalogs.
However, because there exist some systematic differences between the simulated and real galaxies in color-color
space, this method will be explored in the future as the quality of the simulated catalogs improves.}.
We also applied the same limiting magnitude cut only for the detection band used for the observations to
the simulated catalogs ($i$-band for $u$-, $g$-, $r$-dropout galaxies, and $z$-band for $i$-dropout
galaxies, respectively).
The average stellar mass of these simulated dropout galaxies is $\sim2\times10^9\,\mathrm{M_\odot}$, and the 90\%
quantile range is $\sim2\times10^8\mathrm{-}1\times10^{10}\,\mathrm{M_\odot}$.
This is consistent with that observed \citep[e.g.,][]{stark09}, implying that simulated dropout galaxies trace
similar structures.
Based on these simulated catalogs, we calculated local number density maps as in Section \ref{sec:sky} and
selected overdense regions in the same way as in the CFHTLS Deep Fields.
For each overdense region of dropout galaxies, we selected the strongest spike in the redshift distribution, and
the dominant dark matter structure was defined by the most massive halo in that redshift spike.
The descendant halos at $z=0$ for each overdense region were then easily identified by tracing the halo merger
tree of those halos.
As shown in Figure \ref{fig:prob}, although there is a large scatter, the significance of the overdensity at high
redshift is clearly correlated with the descendant halo mass at $z=0$, whose probability of no correlation is
$<0.01$ based on the Spearman rank correlation test. 
Thus, it is possible to select reliable protocluster candidates through the surface overdensity.
In this study, we set the criterion of protocluster candidates at $>4\sigma$ overdensity significance in
order to obtain a high purity of real protoclusters.
Based on this criterion, $76\,(90)\%$ of these candidates of $u$-dropout ($i$-dropout) galaxies are expected to
be in real protoclusters.
However, the completeness is very small ($\sim5\%$ and 10\% for $u$- and $i$-dropout galaxies) mainly due to the
projection effect of the dropout technique.
It should be noted that the average descendant mass of protoclusters with $>4\sigma$ overdensity significance is
$\sim5\times10^{14}\,\mathrm{M_\odot}$ (Figure \ref{fig:prob}).
A total of 21 candidates were identified from $z\sim3$ to $z\sim6$ (five, five, six, and five candidates for the
maps of $u$-, $g$-, $r$-, and $i$-dropout galaxies, respectively).
The coordinates and overdensity of the protocluster candidates are listed in Table \ref{tab:target}.
Since these numbers of protocluster candidates are consistent with the model prediction, in which
$\sim2.9\mathrm{-}6.4$ candidates per observed area ($\sim4\,\mathrm{deg^2}$) are found in each redshift
bin from $z\sim3$ to $z\sim6$, most of the candidates are expected to be real protoclusters.
To summarize, from the wide field imaging of the CFHTLS Deep Fields, we have made a large protocluster
sample at $z\sim3\mathrm{-}6$ without the aid of any special probes, such as  QSOs, RGs, or SMGs; thus, this
sample is not only large but also complementary with previous studies targeting  QSO, RG, or SMGs fields.

\section{FOLLOW-UP SPECTROSCOPY} \label{sec:spec}
Despite our calibration of the selection of the protocluster candidates using light-cone projections described in
the previous section, the overdense regions discovered could still be attributed to mere chance of alignments
along the line-of-sight, given that the dropout technique samples a broad range of redshifts.
Another possibility is that the overdense significance could be affected by the presence of highly clustered
contaminating populations.
 This possibility is negligible for the $u$-, $g$-, and $r$-dropout samples because of their high number
density.
However, for the $i$-dropout galaxies, the average number density per aperture is only 1.6 due to the shallow
$z'$-band depth.
Since the contamination rate of $i$-dropout galaxies is $\sim5\%$ (Section \ref{sec:select}), the number of
contaminants in an aperture is $0.0^{+1.8}_{-0.0}$ on average, which could result in an overdensity $\sim2\sigma$
higher at worst.
Therefore, further confirmation of clustering in redshift space is required to see whether our candidates are
real or not.
In addition to protocluster confirmation, spectroscopic observations enables us to inquire into the internal
structure or line properties of protoclusters.
It is necessary for revealing cluster formation to take various viewpoints.
\begin{deluxetable*}{ccccccccc}
\tabletypesize{\scriptsize}
\tablecaption{Overview of our spectroscopic observations \label{tab:obs_spec}}
\tablewidth{0pt}
\tablehead{Date & Instrument & Target & Grism & resolution (\AA) & coverage (\AA) &
    $t_\mathrm{exp}$ (min.) & $N_\mathrm{mask}$ & seeing}
\startdata
2012 May 13 \& 14 & GMOS & D3ID01 & R600 & 4.5 & 7500-10000 & 330 & 1 & $0\farcs5$ \\
2012 Oct. 21 & FOCAS & D1ID01 & VPH900 & 5.7 & 7500-10100 & 220 & 1 & $0\farcs9$ \\
2014 Aug. 24 & DEIMOS & D1GD01 & 600ZD & 3.5 & 5000-9300 & 120 & 1 & $0\farcs7$ \\
 & & D4GD01 & 600ZD & 3.5 & 5000-9300 & 120 & 1 & $0\farcs7$ \\
2014 Oct. 20 \& 21 & FOCAS & D1RD01 & VPH650 & 5.5 & 6000-8300 & 280 & 1 & $0\farcs7$ \\
 & & D1GD01 & VPH520 & 2.5 & 4900-6500 & 100 & 1 & $0\farcs9$ \\
 & & D1UD01 & VPH520 & 2.5 & 4300-5900 & 60 & 4 & $0\farcs6$ \\
 & & D4GD01 & VPH520 & 2.5 & 4900-6500 & 120 & 2 & $0\farcs7$ \\
 & & D4UD01 & VPH520 & 2.5 & 4300-5900 & 60 & 3 & $0\farcs8$ \\
2014 Oct. 24 \& 25 & FOCAS & D4RD02 & VPH650 & 5.5 & 6000-8300 & 120 & 1 & $0\farcs8$
\enddata
\end{deluxetable*}

\begin{figure}
\epsscale{1.2}
\plotone{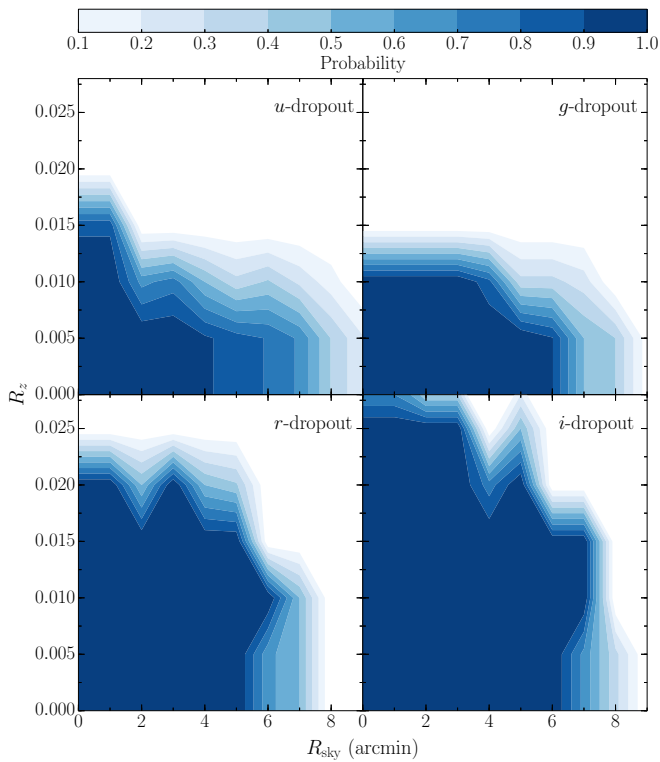}
\caption{probability of protocluster member as a function of distance from the center of a protocluster.
    The horizontal and vertical axes indicate spatial and redshift directions, and color contours show the
    probability.
    At $z=3.0\,(5.9)$, $2\,\mathrm{physical\>Mpc}$ corresponds to $\Delta z=0.009\,(0.032)$ and
    $4.3\,(5.8)\,\mathrm{arcmin}$.}
\label{fig:3D_model}
\end{figure}

Before performing the follow-up spectroscopic observations, we first investigated how far protocluster members
are typically spread from the center, again using the light-cone model.
In the model, protocluster members are defined as galaxies whose descendants at $z=0$ reside in
$>10^{14}\,\mathrm{M_\sun}$ halos \citep{overzier09a,chiang13}.
The center of a protocluster in three-dimensional space was estimated by using the median R.A., Decl., and
redshift of all protocluster members.
The positional difference between the protocluster center defined as above and the peak of the surface
overdensity observed, is typically less than $0.5\,\mathrm{arcmin}$ and less than $2\,\mathrm{arcmin}$ at worst.
Then, we investigated the three-dimensional distribution of protocluster members in the overdense regions.
Although each protocluster has a different structural morphology, such as filamentary or sheet-like, we simply
estimated the probability of protocluster membership as a function of the distance to the center by calculating
the ratio between protocluster members and non-members at a certain distance from the center.
We finally derive a probability map by taking the median stack of the probability maps of all the protocluster
regions computed for each redshift.
Figure \ref{fig:3D_model} shows the probability map of protocluster members of $u$-, $g$-, $r$-, and $i$-dropout
galaxies.
We found that galaxies lying within the volume of $R_\mathrm{sky}<4\,(6)\,\mathrm{arcmin}$ and
$R_z<0.010\,(0.025)$ at $z\sim3\,(6)$, will be protocluster members with a probability of $>80\%$.
Based on this estimate, we defined the protocluster region as a sphere of $2\,\mathrm{physical\>Mpc}$
radius.
It should be noted that we will evaluate protocluster existence based on significance of excess from homogeneous
distribution rather than the absolute number of confirmed galaxies, because actual observations are incomplete
and rely on Ly$\alpha$ emission to identify the redshifts of dropout galaxies.
The fraction of Ly$\alpha$ emitting galaxies among dropout galaxies has been investigated
\citep[e.g.,][]{stark11,curtis12}; however, it would yet to be explored for the fraction in overdense regions
because the previous studies are mainly based on field galaxies.
\citet{overzier08} found that a protocluster around a radio galaxy at $z=4.1$ exhibits a high overdensity of
both LAEs and LBGs, while \citet{kashikawa07} reported that there is no correlation between the distributions
of LAEs and LBGs around a QSO at $z=4.9$.
Hence, it is not yet fully understood what fraction of LBGs emits Ly$\alpha$, especially in overdense regions.
Although this may potentially bias protocluster identification, most protoclusters at $z\sim2\mathrm{-}3$ tend to
have high overdensities of both LBGs and LAEs \citep[e.g.,][]{kuiper10}.

\subsection{Observations}
We carried out spectroscopic observations using Subaru/FOCAS \citep{kashikawa02}, KeckII/DEIMOS \citep{faber03},
and Gemini-N/GMOS \citep{hook04}.
In these observations, eight protocluster candidates from $z\sim3$ to $z\sim6$ were observed in total (two at
each redshift).
The target protocluster candidates and the configuration of spectroscopic observations are described in Table
\ref{tab:target} and \ref{tab:obs_spec}, respectively.
All these observations were conducted with Multi-Object Spectroscopy (MOS) mode.
The slits typically had a length of $6\mathrm{-}8\,\mathrm{arcsec}$ and a width of
$0.8\mathrm{-}1.0\,\mathrm{arcsec}$.
The used grisms were selected in order to have the highest efficiency at the wavelength of the redshifted
Ly$\alpha$ line of targeted dropout galaxies and the spectral resolution of
$<2.8(1+z_\mathrm{[OII]})\,\mathrm{\AA}$, where $2.8\,\mathrm{\AA}$ is the wavelength separation of the
[O{\sc ii}] doublet ($\lambda=3726.0,\,3728.8\,\mathrm{\AA}$) in the rest-frame.
Therefore, our spectroscopic observations were set up to have a resolution that is sufficient to resolve
the [O{\sc ii}] emission into the doublet to check for contamination by foreground interlopers.
The wavelength coverage is also wide enough to cover the expected redshift range of the dropout galaxies.
In the FOCAS observations, the telescope was dithered along the slit to enable more accurate sky subtraction
between exposures, and we used Nod-and-Shuffle mode, which allows increase in the accuracy of sky subtraction
by real-time flipping to the sky position in the GMOS observation.
Although higher priority was given to brighter galaxies, we designed slit masks so as to allocate as many objects
as possible.
Furthermore, a slit of each mask was allocated for a bright star ($\sim20\,\mathrm{mag}$) to monitor the
time variations of seeing size or atmospheric transmission between exposures.
Although the sky condition was good and stable during all observing nights, we removed only a few poor-quality
frames by checking the bright star.
Long slit exposures of one of the following spectroscopic standard stars HZ44, BD+28d4211, and G191-B2B were
taken each night with all configurations used in the night, and we corrected the difference of airmass between
science targets and standard stars in the flux calibration.
The data taken by FOCAS and GMOS were reduced in a standard manner with IRAF, and the pipeline spec2d\footnote{The
data reduction pipeline was developed at the University of California, Berkeley, with support from National
Science Foundation grant AST 00-71048.} was used for the reduction of the data taken by DEIMOS.
The $3\sigma$ detection limits of emission line are typically $5.0\times10^{-18}$, $4.0\times10^{-18}$,
$1.0\times10^{-18}$, and $1.3\times10^{-18}, \,\mathrm{erg\,s^{-1}\,cm^{-2}}$ for $u$-, $g$-, $r$-, and
$i$-dropout galaxies, respectively, assuming the line width of $\mathrm{FWHM}=5.0\,\mathrm{\AA}$.

\subsection{Line Contaminations \label{sec:contami}}
All emission lines we detected are single emission lines, which are not likely to be H$\beta$ or [\ion{O}{3}]
emission lines because the wavelength coverage of our observation is wide enough to detect all these multiple
lines simultaneously.
Only [\ion{O}{3}]$\lambda5007$ emission, which is generally the strongest emission among them, might resemble 
a single emission line if the other lines are too faint to be detected.
However, according to the typical line flux ratio of [\ion{O}{3}]$\lambda5007$ and [\ion{O}{3}]$\lambda4959$
($f_{\lambda5007}\sim3\times f_{\lambda4959}$), even
[\ion{O}{3}]$\lambda4959$ should be detected in our spectroscopic observations since the signal-to-noise ratio
(S/N) of detected emission lines is $\sim10$ on average (and always $>3$).
Therefore, we investigated the possibility that H$\alpha$ and [\ion{O}{2}] emission lines contaminate dropout
galaxy samples based on both imaging and spectroscopic data.

Only $r$- and $i$-dropout galaxies can be contaminated by H$\alpha$ emitting galaxies according to its wavelength
($\lambda_\mathrm{H\alpha}=6562.8\,\mathrm{\AA}$).
Since higher-redshift dropout galaxies are selected using a redder color criterion to detect the strong Lyman
break, it is almost impossible to mimic this color by a Balmer break object at lower redshift based on the
expected color of passive galaxies as described in Section \ref{sec:select}.
Even if Balmer breaks of passive and old galaxies were strong enough to satisfy the dropout color criteria,
these stellar population are unlikely to have H$\alpha$ emission, as a diagnostic of the star-formation activity.
On the other hand, dusty starburst galaxies with strong Balmer breaks could be contaminated because they would
show an H$\alpha$ emission line as well as a very red color, resulting from the combination of Balmer break and
dust reddening.
However, their H$\alpha$ emissions can be discriminated from Ly$\alpha$ emissions, which appear right at the
Lyman break.
Therefore, we consider the possibility of finding foreground H$\alpha$ emission negligible.

Regarding [\ion{O}{2}] doublet emission lines, it is possible to distinguish between Ly$\alpha$ and [\ion{O}{2}]
emission lines based on the line profile.
The spectral resolution of most of our spectroscopic observation was set high enough to resolve [\ion{O}{2}]
emission lines as doublets ($\Delta\lambda=3.8-6.3\,\mathrm{\AA}$ at $z\sim0.3-1.3$), although it would be
practically difficult to resolve these in most cases due to low S/N.
In this case, the [\ion{O}{2}] emission line is typically skewed blueward, while the Ly$\alpha$ emission line
from high redshift galaxies is skewed redward.
Therefore, the skewness of the line profile allows us to distinguish between Ly$\alpha$ and [\ion{O}{2}] emission
lines; however, it should be noted that [OII] emission is sometimes skewed redward when assuming an exotic physical
properties of \ion{H}{2} region (e.g., election density).
To improve the way of distinguishing, \citet{kashikawa06} introduced ``weighted'' skewness, which makes use of
line width as well as line profile.
Ly$\alpha$ emission usually has a larger line width than [\ion{O}{2}] emission because Ly$\alpha$ emission
typically emerges in an outflow or galactic wind.
Therefore, we calculated the weighted skewness of all spectroscopically detected galaxies.
The asymmetric emission lines with $S_w>3$ are clear evidence of Ly$\alpha$ emission from high redshift galaxies,
though it would be more difficult to distinguish them from nearby emission line galaxies at $z\sim3$, where the
IGM attenuation is weaker than at higher redshifts.
As shown in Table \ref{tab:spec_all}, most of the emission lines of this study have large $S_w$.
However, 17\% of all identified emission lines have $S_w<3$; although this would be caused by strong sky
line residuals and low S/N data, we could not rule out the possibility of [\ion{O}{2}] emission lines.
In order to make the line profile measurement more accurate by reducing the effect of sky noise and low S/N data,
we made a composite spectrum of all 24 emission lines with $S_w<3$ by taking a median in the rest-frame, assuming
they were Ly$\alpha$, and normalized by the peak flux.
The weighted skewness of the composite spectrum was found to be $S_w=8.1\pm1.2$, indicating that most of the
emission lines even with $S_w<3$ in individual spectra are real Ly$\alpha$ emission lines from high-redshift
galaxies.
In addition to the line profile, the possibility of [\ion{O}{2}] emission can further be reduced by  taking
account of photometric data.
Although [\ion{O}{2}] emission is closer to Balmer break than H$\alpha$ emission, it will still be difficult
to find a sharp break near the emission line except for peculiar galaxies such as dusty starburst galaxies.

From these considerations, it is unlikely that H$\alpha$ or [\ion{O}{2}] emission lines contaminate our dropout
samples, and Ly$\alpha$ is the most plausible interpretation to explain both photometric and spectral features.
Since the major contamination in the photometric selection is completely different from that in the spectroscopic
observation, the combination of photometric and spectroscopic observations enables us to select a clean sample of
high-redshift galaxies.
We can regard all single emission lines detected from our dropout sample as Ly$\alpha$ emission lines.

\subsection{Results \label{sec:spec_result}}
 Our protocluster confirmation completely depends on the detection of Ly$\alpha$ emission; therefore, we might
only select a part of the galaxy populations in these protoclusters.
We might miss protoclusters, if they were mainly composed of passive or dusty galaxies without Ly$\alpha$ emission.
This may lead to a possible selection bias in our protocluster search.
However, protoclusters at $z\sim2\mathrm{-}3$ have been found to mainly contain star-forming galaxies.
It is worth noting that known protoclusters that include a large number of older or dustier galaxies, like the
SSA22 ($z=3.1$) and the spider-web ($z=2.2$) protoclusters, also show a significant overdensity in LAEs as well
\citep{steidel00,kuiper10,kubo13}.
These results suggest that the possible bias introduced by tracing protoclusters only by LAEs is not probably
significant.

The number of dropout galaxies located in each protocluster candidate region is listed in Table
\ref{tab:target}, and the number of spectroscopically observed galaxies is shown in Table \ref{tab:cluster}.
From these numbers,  about half, at least 35\%, of dropout galaxies in protocluster candidate regions were
observed by our follow-up spectroscopic observations.
We carefully discriminated real emission lines from sky lines or noise by examining both two-dimensional and
one-dimensional spectra, and all emission lines identified in this study are shown in Figure \ref{fig:spec_all}.
We estimate the observed properties of the spectroscopically confirmed galaxies, such as UV absolute magnitude
at 1300{\AA} in the rest-frame ($M_\mathrm{UV}$), Ly$\alpha$ luminosity ($L_\mathrm{Ly\alpha}$), and
rest-frame Ly$\alpha$ equivalent width ($EW_0$), shown in Table \ref{tab:spec_all}.
The redshifts were derived by the peak wavelength of the Ly$\alpha$ emission line, assuming the rest wavelength of
Ly$\alpha$ to be 1215.6{\AA}.
These measurements could be overestimated if there was a galactic outflow.
When emission lines are located near strong sky lines, the position of the peak could be shifted.
These effects of sky lines and the wavelength resolution are taken into account when estimating the error.
Observed line flux, $f_\mathrm{Ly\alpha}$, corresponds to the total amount of the flux within the line profile.
The slit loss was corrected based on the ratio of slit 
\begin{turnpage}
\begin{deluxetable}{cccccccccc}
\tabletypesize{\scriptsize}
\tablecaption{Observed properties of all spectroscopically confirmed dropout galaxies. \label{tab:spec_all}}
\tablewidth{0pt}
\tablehead{
    \colhead{ID} & \colhead{R.A.} & \colhead{Decl.} & \colhead{$m$\tablenotemark{a}} & \colhead{redshift} &
        \colhead{$M_\mathrm{UV}$} & \colhead{$f_\mathrm{Ly\alpha}$} & \colhead{$L_\mathrm{Ly\alpha}$} &
        \colhead{$EW_0$} & \colhead{$S_w$} \\
    \colhead{} & \colhead{(J2000)} & \colhead{(J2000)} & \colhead{(mag)} & \colhead{} & \colhead{(mag)} &
        \colhead{($10^{-18}\,\mathrm{erg\,s^{-1}\,cm^{-2}}$)} & \colhead{($10^{42}\,\mathrm{erg\,s^{-1}}$)} &
        \colhead{(\AA)} & \colhead{(\AA)}
}
\startdata
\multicolumn{10}{c}{D1ID01 (three galaxies)} \\
1 & 02:27:18.8 & -04:50:08.3 & $25.45\pm0.06$ & $5.966^{+0.002}_{-0.004}$ & $-21.57\pm0.06$ & $2.66\pm0.41$ & $1.05\pm0.16$ & $2.41\pm0.46$ & $2.25\pm1.21$ \\
2 & 02:27:21.0 & -04:50:49.3 & $25.97\pm0.10$ & $6.044^{+0.002}_{-0.002}$ & $-20.85\pm0.13$ & $19.55\pm0.72$ & $7.92\pm0.29$ & $40.05\pm6.13$ & $5.21\pm0.72$ \\
3 & 02:27:19.0 & -04:53:48.0 & $26.30\pm0.13$ & $6.325^{+0.002}_{-0.003}$ & $-20.76\pm0.24$ & $31.73\pm0.46$ & $14.29\pm0.21$ & $81.79\pm24.95$ & $7.85\pm0.59$ \\
\hline\\
\multicolumn{10}{c}{D3ID01 (two galaxies)} \\
1 & 14:19:22.5 & +52:57:22.5 & $25.21\pm0.05$ & $5.749^{+0.002}_{-0.002}$ & $-21.49\pm0.05$ & $5.80\pm0.88$ & $2.09\pm0.32$ & $3.79\pm0.84$ & $5.41\pm1.16$ \\
2 & 14:19:17.2 & +52:56:14.4 & $25.74\pm0.08$ & $5.756^{+0.002}_{-0.002}$ & $-20.94\pm0.08$ & $8.16\pm1.10$ & $2.95\pm0.40$ & $10.84\pm2.14$ & $5.38\pm3.85$ \\
\hline\\
\multicolumn{10}{c}{D1RD01 (six galaxies)} \\
1 & 02:24:45.467 & -04:58:52.83 & $26.37\pm0.06$ & $4.431^{+0.002}_{-0.002}$ & $-19.85\pm0.14$ & $1.60\pm0.39$ & $0.31\pm0.08$ & $4.09\pm1.14$ & $3.08\pm5.41$ \\ 
2 & 02:24:45.957 & -04:56:57.69 & $26.14\pm0.05$ & $4.602^{+0.002}_{-0.002}$ & $-20.14\pm0.12$ & $2.75\pm0.46$ & $0.59\pm0.10$ & $5.84\pm1.17$ & $0.72\pm4.11$ \\ 
3 & 02:24:42.586 & -04:58:36.00 & $26.81\pm0.09$ & $4.742^{+0.002}_{-0.003}$ & $-19.49\pm0.21$ & $5.22\pm0.46$ & $1.20\pm0.10$ & $21.69\pm5.06$ & $6.73\pm2.10$ \\ 
4 & 02:24:38.212 & -04:57:15.05 & $26.16\pm0.05$ & $4.840^{+0.002}_{-0.002}$ & $-20.21\pm0.13$ & $5.67\pm0.73$ & $1.37\pm0.18$ & $12.79\pm2.28$ & $15.74\pm5.69$ \\ 
5 & 02:24:45.964 & -04:54:34.80 & $26.12\pm0.05$ & $4.890^{+0.002}_{-0.002}$ & $-20.35\pm0.12$ & $1.73\pm0.31$ & $0.43\pm0.08$ & $3.52\pm0.74$ & $3.11\pm2.06$ \\ 
6 & 02:24:43.757 & -04:54:31.19 & $26.32\pm0.05$ & $4.894^{+0.002}_{-0.002}$ & $-20.19\pm0.14$ & $1.56\pm0.29$ & $0.39\pm0.07$ & $3.69\pm0.85$ & $0.58\pm6.76$ \\ 
\hline\\
\multicolumn{10}{c}{D4RD02 (three galaxies)} \\
1 & 22:16:46.722 & -17:28:02.00 & $26.00\pm0.05$ & $4.630^{+0.002}_{-0.002}$ & $-20.27\pm0.12$ & $1.27\pm0.17$ & $0.28\pm0.04$ & $2.44\pm0.43$ & $3.47\pm21.52$ \\ 
2 & 22:16:39.959 & -17:31:34.58 & $25.94\pm0.04$ & $4.865^{+0.002}_{-0.002}$ & $-20.41\pm0.12$ & $10.31\pm0.45$ & $2.52\pm0.11$ & $19.51\pm2.48$ & $14.12\pm2.55$ \\ 
3 & 22:16:45.765 & -17:29:19.89 & $26.07\pm0.05$ & $4.952^{+0.004}_{-0.002}$ & $-20.36\pm0.14$ & $9.98\pm0.35$ & $2.54\pm0.09$ & $20.67\pm2.93$ & $15.00\pm2.88$ \\ 
\hline\\
\multicolumn{10}{c}{D1GD01 (36 galaxies)} \\
1 & 02:25:28.536 & -04:17:14.12 & $26.93\pm0.07$ & $3.435^{+0.001}_{-0.001}$ & $-18.85\pm0.15$ & $12.50\pm1.49$ & $1.34\pm0.16$ & $43.82\pm8.47$ & $3.32\pm8.74$ \\ 
2 & 02:25:30.408 & -04:15:56.70 & $26.92\pm0.07$ & $3.623^{+0.001}_{-0.001}$ & $-18.92\pm0.16$ & $6.94\pm0.92$ & $0.84\pm0.11$ & $25.90\pm5.31$ & $2.74\pm2.46$ \\ 
3 & 02:25:32.014 & -04:17:03.56 & $27.17\pm0.09$ & $3.705^{+0.001}_{-0.001}$ & $-18.66\pm0.21$ & $6.85\pm1.01$ & $0.88\pm0.13$ & $34.18\pm8.92$ & $8.91\pm4.93$ \\ 
4 & 02:25:25.565 & -04:17:12.58 & $27.15\pm0.09$ & $3.717^{+0.001}_{-0.001}$ & $-18.73\pm0.20$ & $4.51\pm0.90$ & $0.58\pm0.12$ & $21.26\pm6.04$ & $1.08\pm1.34$ \\ 
5 & 02:26:12.550 & -04:18:41.23 & $26.69\pm0.06$ & $3.733^{+0.001}_{-0.001}$ & $-18.94\pm0.17$ & $21.41\pm1.71$ & $2.79\pm0.22$ & $83.76\pm15.69$ & $-0.32\pm3.13$ \\ 
6 & 02:25:17.290 & -04:14:02.23 & $25.40\pm0.02$ & $3.738^{+0.001}_{-0.001}$ & $-20.49\pm0.04$ & $21.57\pm1.23$ & $2.82\pm0.16$ & $20.45\pm1.43$ & $6.11\pm0.86$ \\ 
7 & 02:25:51.739 & -04:14:37.26 & $26.07\pm0.03$ & $3.744^{+0.001}_{-0.001}$ & $-19.91\pm0.07$ & $3.07\pm0.79$ & $0.40\pm0.10$ & $4.97\pm1.32$ & $0.23\pm2.27$ \\ 
8 & 02:25:39.708 & -04:14:20.73 & $25.22\pm0.01$ & $3.754^{+0.001}_{-0.001}$ & $-20.76\pm0.04$ & $9.32\pm1.58$ & $1.23\pm0.21$ & $6.97\pm1.20$ & $9.41\pm9.42$ \\ 
9 & 02:26:11.563 & -04:19:21.65 & $25.88\pm0.03$ & $3.755^{+0.001}_{-0.001}$ & $-19.82\pm0.08$ & $39.82\pm2.21$ & $5.27\pm0.29$ & $70.40\pm6.78$ & $8.13\pm1.98$ \\ 
10 & 02:26:10.246 & -04:18:18.50 & $26.81\pm0.06$ & $3.759^{+0.001}_{-0.001}$ & $-18.69\pm0.21$ & $25.61\pm1.08$ & $3.40\pm0.14$ & $128.41\pm28.60$ & $5.91\pm0.54$ \\ 
11 & 02:25:33.011 & -04:14:45.24 & $25.28\pm0.02$ & $3.766^{+0.001}_{-0.001}$ & $-20.53\pm0.04$ & $47.32\pm2.23$ & $6.30\pm0.30$ & $43.78\pm2.74$ & $12.15\pm1.04$ \\ 
12 & 02:26:07.202 & -04:17:12.22 & $26.73\pm0.06$ & $3.793^{+0.001}_{-0.001}$ & $-19.17\pm0.15$ & $9.89\pm1.63$ & $1.34\pm0.22$ & $32.77\pm7.27$ & $3.60\pm4.24$ \\ 
13 & 02:25:59.907 & -04:15:45.42 & $26.23\pm0.04$ & $3.797^{+0.001}_{-0.001}$ & $-19.74\pm0.09$ & $10.03\pm2.29$ & $1.36\pm0.31$ & $19.73\pm4.83$ & $7.09\pm4.96$ \\ 
14 & 02:25:57.460 & -04:18:10.27 & $26.01\pm0.03$ & $3.799^{+0.001}_{-0.001}$ & $-19.99\pm0.07$ & $7.82\pm1.38$ & $1.06\pm0.19$ & $12.24\pm2.32$ & $8.82\pm4.36$ \\ 
15 & 02:25:42.923 & -04:15:38.74 & $26.41\pm0.04$ & $3.800^{+0.001}_{-0.001}$ & $-19.44\pm0.12$ & $17.61\pm1.24$ & $2.40\pm0.17$ & $45.41\pm6.08$ & $9.85\pm1.19$ \\ 
16 & 02:25:44.405 & -04:14:11.81 & $25.32\pm0.02$ & $3.803^{+0.001}_{-0.001}$ & $-20.64\pm0.04$ & $24.22\pm2.34$ & $3.30\pm0.32$ & $20.78\pm2.16$ & $10.09\pm2.61$ \\ 
17 & 02:25:34.147 & -04:14:21.23 & $26.40\pm0.04$ & $3.809^{+0.001}_{-0.001}$ & $-19.59\pm0.11$ & $8.27\pm1.39$ & $1.13\pm0.19$ & $18.65\pm3.66$ & $6.15\pm4.06$ \\ 
\enddata
\end{deluxetable}
\clearpage
\setcounter{table}{3}
\begin{deluxetable}{cccccccccc}
\tabletypesize{\scriptsize}
\tablecaption{(continued)}
\tablewidth{0pt}
\tablehead{
    \colhead{ID} & \colhead{R.A.} & \colhead{Decl.} & \colhead{$m$\tablenotemark{a}} & \colhead{redshift} &
        \colhead{$M_\mathrm{UV}$} & \colhead{$f_\mathrm{Ly\alpha}$} & \colhead{$L_\mathrm{Ly\alpha}$} &
        \colhead{$EW_0$} & \colhead{$S_w$} \\
    \colhead{} & \colhead{(J2000)} & \colhead{(J2000)} & \colhead{(mag)} & \colhead{} & \colhead{(mag)} &
        \colhead{($10^{-18}\,\mathrm{erg\,s^{-1}\,cm^{-2}}$)} & \colhead{($10^{42}\,\mathrm{erg\,s^{-1}}$)} &
        \colhead{(\AA)} & \colhead{(\AA)}
}
\startdata
18 & 02:25:56.529 & -04:17:27.85 & $26.77\pm0.06$ & $3.818^{+0.001}_{-0.001}$ & $-19.15\pm0.16$ & $10.76\pm1.85$ & $1.48\pm0.25$ & $36.77\pm8.52$ & $3.17\pm8.69$ \\ 
19 & 02:25:30.087 & -04:15:15.84 & $25.85\pm0.03$ & $3.827^{+0.001}_{-0.001}$ & $-20.20\pm0.06$ & $6.02\pm0.94$ & $0.83\pm0.13$ & $7.83\pm1.30$ & $8.59\pm1.02$ \\ 
20 & 02:25:49.845 & -04:14:53.42 & $26.57\pm0.05$ & $3.829^{+0.001}_{-0.001}$ & $-19.35\pm0.13$ & $13.03\pm1.37$ & $1.81\pm0.19$ & $37.33\pm6.22$ & $13.66\pm4.39$ \\ 
21 & 02:25:41.772 & -04:16:06.53 & $25.70\pm0.02$ & $3.843^{+0.002}_{-0.001}$ & $-20.30\pm0.06$ & $21.31\pm1.34$ & $2.98\pm0.19$ & $25.65\pm2.14$ & $-0.62\pm0.96$ \\ 
22 & 02:25:56.593 & -04:15:15.20 & $26.89\pm0.07$ & $3.859^{+0.001}_{-0.001}$ & $-18.85\pm0.21$ & $19.31\pm1.22$ & $2.73\pm0.17$ & $88.80\pm19.46$ & $5.15\pm0.78$ \\ 
23 & 02:25:39.324 & -04:14:40.82 & $26.79\pm0.06$ & $3.866^{+0.001}_{-0.001}$ & $-19.12\pm0.17$ & $13.16\pm1.11$ & $1.87\pm0.16$ & $47.68\pm8.84$ & $5.37\pm0.92$ \\ 
24 & 02:26:11.555 & -04:17:39.51 & $26.38\pm0.04$ & $3.886^{+0.001}_{-0.001}$ & $-19.35\pm0.14$ & $33.92\pm2.51$ & $4.87\pm0.36$ & $100.23\pm15.93$ & $1.63\pm1.82$ \\ 
25 & 02:25:27.362 & -04:16:43.59 & $27.28\pm0.10$ & $3.891^{+0.001}_{-0.001}$ & $-18.37\pm0.32$ & $17.16\pm1.63$ & $2.47\pm0.24$ & $125.83\pm44.70$ & $2.87\pm0.77$ \\ 
26 & 02:25:25.816 & -04:16:38.94 & $27.61\pm0.13$ & $3.927^{+0.001}_{-0.001}$ & $-18.04\pm0.43$ & $13.14\pm1.83$ & $1.93\pm0.27$ & $133.94\pm67.59$ & $-1.16\pm6.97$ \\ 
27 & 02:26:01.853 & -04:14:41.24 & $27.73\pm0.15$ & $4.000^{+0.001}_{-0.001}$ & $-17.65\pm0.61$ & $15.71\pm2.24$ & $2.41\pm0.34$ & $238.11\pm183.28$ & $5.40\pm1.71$ \\ 
28 & 02:25:31.239 & -04:15:49.81 & $26.37\pm0.04$ & $4.054^{+0.001}_{-0.001}$ & $-19.76\pm0.12$ & $18.39\pm1.82$ & $2.92\pm0.29$ & $41.22\pm6.28$ & $5.63\pm2.24$ \\ 
29 & 02:25:35.470 & -04:14:15.68 & $26.25\pm0.04$ & $4.119^{+0.001}_{-0.001}$ & $-19.94\pm0.11$ & $23.18\pm1.90$ & $3.82\pm0.31$ & $45.59\pm6.16$ & $3.35\pm4.05$ \\ 
30 & 02:25:34.433 & -04:15:05.46 & $26.37\pm0.04$ & $4.185^{+0.001}_{-0.001}$ & $-19.83\pm0.13$ & $25.45\pm1.46$ & $4.35\pm0.25$ & $57.91\pm8.23$ & $8.31\pm0.57$ \\ 
31 & 02:25:39.830 & -04:14:53.33 & $26.20\pm0.04$ & $4.236^{+0.001}_{-0.002}$ & $-20.19\pm0.11$ & $23.32\pm2.01$ & $4.11\pm0.35$ & $39.12\pm5.27$ & $6.86\pm1.12$ \\ 
32 & 02:25:11.525 & -04:16:20.17 & $25.91\pm0.03$ & $4.276^{+0.001}_{-0.001}$ & $-19.10\pm0.29$ & $107.86\pm2.43$ & $19.42\pm0.44$ & $505.39\pm153.02$ & $2.40\pm0.33$ \\ 
33 & 02:25:57.659 & -04:14:24.90 & $26.54\pm0.05$ & $4.385^{+0.001}_{-0.001}$ & $-18.83\pm0.40$ & $61.69\pm2.19$ & $11.79\pm0.42$ & $393.36\pm177.53$ & $5.26\pm0.81$ \\ 
34 & 02:25:21.473 & -04:16:01.50 & $27.04\pm0.08$ & $4.391^{+0.001}_{-0.002}$ & $-19.20\pm0.30$ & $23.76\pm2.33$ & $4.55\pm0.45$ & $107.90\pm36.20$ & $5.27\pm1.87$ \\ 
35 & 02:25:08.716 & -04:15:24.74 & $27.10\pm0.08$ & $4.395^{+0.001}_{-0.001}$ & $-19.34\pm0.27$ & $52.76\pm2.69$ & $10.13\pm0.52$ & $210.28\pm60.76$ & $7.74\pm0.99$ \\ 
36 & 02:25:28.097 & -04:14:54.46 & $27.68\pm0.14$ & $4.442^{+0.001}_{-0.001}$ & $-18.01\pm0.78$ & $22.59\pm2.06$ & $4.45\pm0.41$ & $316.27\pm334.36$ & $4.13\pm1.99$ \\ 
\hline\\
\multicolumn{10}{c}{D4GD01 (42 galaxies)} \\
1 & 22:16:55.191 & -17:25:51.91 & $24.53\pm0.01$ & $3.568^{+0.001}_{-0.001}$ & $-21.32\pm0.02$ & $11.47\pm1.23$ & $1.34\pm0.14$ & $4.54\pm0.50$ & $7.39\pm2.56$ \\ 
2 & 22:17:00.190 & -17:25:06.37 & $26.39\pm0.05$ & $3.569^{+0.001}_{-0.001}$ & $-19.45\pm0.11$ & $9.05\pm1.27$ & $1.06\pm0.15$ & $20.01\pm3.48$ & $4.83\pm1.81$ \\ 
3 & 22:16:55.670 & -17:20:49.98 & $27.16\pm0.10$ & $3.581^{+0.001}_{-0.001}$ & $-18.67\pm0.21$ & $6.39\pm0.65$ & $0.76\pm0.08$ & $29.05\pm6.79$ & $1.78\pm1.80$ \\ 
4 & 22:16:49.872 & -17:21:53.02 & $25.42\pm0.02$ & $3.622^{+0.001}_{-0.001}$ & $-20.44\pm0.05$ & $14.01\pm1.24$ & $1.70\pm0.15$ & $12.85\pm1.26$ & $5.97\pm4.29$ \\ 
5 & 22:17:01.326 & -17:20:52.55 & $26.81\pm0.07$ & $3.624^{+0.001}_{-0.001}$ & $-19.04\pm0.15$ & $5.40\pm1.15$ & $0.66\pm0.14$ & $17.95\pm4.70$ & $-0.73\pm2.60$ \\ 
6 & 22:16:54.811 & -17:28:37.91 & $26.93\pm0.08$ & $3.626^{+0.001}_{-0.001}$ & $-18.89\pm0.18$ & $10.59\pm1.15$ & $1.29\pm0.14$ & $40.75\pm8.47$ & $3.86\pm1.26$ \\ 
7 & 22:16:58.872 & -17:28:33.27 & $26.33\pm0.04$ & $3.628^{+0.001}_{-0.001}$ & $-19.53\pm0.10$ & $6.92\pm1.32$ & $0.84\pm0.16$ & $14.68\pm3.15$ & $5.65\pm3.60$ \\ 
8 & 22:17:07.296 & -17:28:45.15 & $26.22\pm0.04$ & $3.654^{+0.001}_{-0.001}$ & $-19.61\pm0.10$ & $16.61\pm1.61$ & $2.06\pm0.20$ & $33.43\pm4.49$ & $4.80\pm3.23$ \\ 
9 & 22:16:51.756 & -17:24:57.97 & $26.26\pm0.04$ & $3.666^{+0.001}_{-0.001}$ & $-19.59\pm0.10$ & $11.90\pm0.89$ & $1.49\pm0.11$ & $24.67\pm3.00$ & $1.03\pm2.79$ \\ 
10 & 22:16:42.993 & -17:15:53.36 & $26.96\pm0.08$ & $3.669^{+0.001}_{-0.001}$ & $-18.90\pm0.18$ & $4.98\pm0.89$ & $0.62\pm0.11$ & $19.56\pm4.98$ & $7.49\pm2.83$ \\ 
11 & 22:16:50.981 & -17:18:49.87 & $26.71\pm0.06$ & $3.670^{+0.001}_{-0.001}$ & $-19.10\pm0.15$ & $11.16\pm2.01$ & $1.40\pm0.25$ & $36.23\pm8.51$ & $5.17\pm4.24$ \\ 
12 & 22:16:53.509 & -17:19:06.60 & $25.74\pm0.03$ & $3.671^{+0.001}_{-0.001}$ & $-20.08\pm0.06$ & $23.63\pm1.76$ & $2.96\pm0.22$ & $31.08\pm2.99$ & $5.67\pm2.85$ \\ 
13 & 22:16:49.716 & -17:17:00.96 & $26.45\pm0.05$ & $3.671^{+0.001}_{-0.001}$ & $-19.41\pm0.12$ & $8.69\pm1.12$ & $1.09\pm0.14$ & $21.37\pm3.67$ & $7.45\pm1.80$ \\ 
14 & 22:16:53.576 & -17:19:07.20 & $24.96\pm0.01$ & $3.671^{+0.001}_{-0.001}$ & $-20.92\pm0.03$ & $17.97\pm1.17$ & $2.25\pm0.15$ & $10.92\pm0.77$ & $14.77\pm4.05$ \\ 
15 & 22:16:51.410 & -17:17:50.44 & $26.02\pm0.03$ & $3.672^{+0.001}_{-0.001}$ & $-19.83\pm0.08$ & $14.83\pm1.31$ & $1.86\pm0.16$ & $24.75\pm2.90$ & $4.22\pm1.98$ \\ 
16 & 22:16:54.326 & -17:18:34.98 & $25.95\pm0.03$ & $3.675^{+0.001}_{-0.001}$ & $-19.95\pm0.07$ & $6.62\pm0.81$ & $0.83\pm0.10$ & $9.85\pm1.39$ & $5.83\pm2.25$ \\ 
17 & 22:16:57.890 & -17:21:51.88 & $26.42\pm0.05$ & $3.675^{+0.001}_{-0.001}$ & $-19.37\pm0.12$ & $18.62\pm1.19$ & $2.34\pm0.15$ & $47.33\pm6.39$ & $4.28\pm0.83$ \\ 
18 & 22:16:51.591 & -17:18:12.00 & $26.30\pm0.04$ & $3.681^{+0.001}_{-0.001}$ & $-19.61\pm0.10$ & $4.78\pm0.95$ & $0.60\pm0.12$ & $9.83\pm2.17$ & $4.43\pm2.62$ \\ 
\enddata
\end{deluxetable}
\clearpage
\setcounter{table}{3}
\begin{deluxetable}{cccccccccc}
\tabletypesize{\scriptsize}
\tablecaption{(continued)}
\tablewidth{0pt}
\tablehead{
    \colhead{ID} & \colhead{R.A.} & \colhead{Decl.} & \colhead{$m$\tablenotemark{a}} & \colhead{redshift} &
        \colhead{$M_\mathrm{UV}$} & \colhead{$f_\mathrm{Ly\alpha}$} & \colhead{$L_\mathrm{Ly\alpha}$} &
        \colhead{$EW_0$} & \colhead{$S_w$} \\
    \colhead{} & \colhead{(J2000)} & \colhead{(J2000)} & \colhead{(mag)} & \colhead{} & \colhead{(mag)} &
        \colhead{($10^{-18}\,\mathrm{erg\,s^{-1}\,cm^{-2}}$)} & \colhead{($10^{42}\,\mathrm{erg\,s^{-1}}$)} &
        \colhead{(\AA)} & \colhead{(\AA)}
}
\startdata
19 & 22:16:55.554 & -17:20:14.08 & $26.66\pm0.06$ & $3.681^{+0.001}_{-0.001}$ & $-19.14\pm0.15$ & $12.33\pm1.31$ & $1.55\pm0.17$ & $38.80\pm7.05$ & $5.73\pm2.17$ \\ 
20 & 22:16:48.909 & -17:15:31.09 & $26.51\pm0.05$ & $3.685^{+0.001}_{-0.001}$ & $-19.35\pm0.12$ & $8.96\pm1.38$ & $1.13\pm0.17$ & $23.36\pm4.59$ & $-3.55\pm6.95$ \\ 
21 & 22:16:55.005 & -17:21:00.75 & $25.78\pm0.03$ & $3.717^{+0.001}_{-0.001}$ & $-20.06\pm0.07$ & $22.37\pm1.51$ & $2.89\pm0.19$ & $31.09\pm2.93$ & $9.26\pm1.83$ \\ 
22 & 22:16:46.962 & -17:21:06.42 & $25.93\pm0.03$ & $3.719^{+0.001}_{-0.001}$ & $-19.74\pm0.09$ & $41.34\pm1.66$ & $5.35\pm0.21$ & $77.34\pm7.55$ & $5.17\pm0.88$ \\ 
23 & 22:16:46.961 & -17:17:10.24 & $27.19\pm0.10$ & $3.720^{+0.001}_{-0.001}$ & $-18.64\pm0.24$ & $6.30\pm0.94$ & $0.81\pm0.12$ & $32.32\pm9.19$ & $3.09\pm2.67$ \\ 
24 & 22:16:42.903 & -17:17:35.09 & $25.47\pm0.02$ & $3.721^{+0.001}_{-0.001}$ & $-20.44\pm0.05$ & $14.86\pm1.46$ & $1.92\pm0.19$ & $14.53\pm1.58$ & $2.22\pm2.25$ \\ 
25 & 22:16:50.522 & -17:18:22.62 & $26.00\pm0.03$ & $3.723^{+0.001}_{-0.001}$ & $-19.87\pm0.08$ & $15.39\pm1.44$ & $2.00\pm0.19$ & $25.62\pm3.13$ & $11.86\pm2.51$ \\ 
26 & 22:16:49.533 & -17:16:44.13 & $26.38\pm0.05$ & $3.728^{+0.001}_{-0.001}$ & $-19.50\pm0.11$ & $8.94\pm1.05$ & $1.16\pm0.14$ & $20.82\pm3.35$ & $10.33\pm3.45$ \\ 
27 & 22:17:09.126 & -17:28:52.31 & $26.73\pm0.06$ & $3.730^{+0.001}_{-0.001}$ & $-19.07\pm0.17$ & $11.89\pm1.54$ & $1.55\pm0.20$ & $41.26\pm8.65$ & $3.25\pm1.57$ \\ 
28 & 22:16:56.467 & -17:17:20.11 & $26.68\pm0.06$ & $3.831^{+0.001}_{-0.001}$ & $-19.12\pm0.17$ & $19.18\pm0.72$ & $2.66\pm0.10$ & $67.69\pm11.99$ & $4.13\pm0.77$ \\ 
29 & 22:17:01.475 & -17:23:58.97 & $27.01\pm0.08$ & $3.837^{+0.001}_{-0.001}$ & $-18.82\pm0.23$ & $13.81\pm1.17$ & $1.92\pm0.16$ & $64.73\pm16.19$ & $5.90\pm1.67$ \\ 
30 & 22:17:04.114 & -17:29:22.86 & $26.57\pm0.06$ & $3.839^{+0.001}_{-0.001}$ & $-19.29\pm0.16$ & $19.06\pm1.23$ & $2.66\pm0.17$ & $58.23\pm9.70$ & $3.59\pm1.20$ \\ 
31 & 22:16:59.785 & -17:26:15.22 & $25.50\pm0.02$ & $3.852^{+0.001}_{-0.003}$ & $-20.57\pm0.05$ & $12.90\pm1.35$ & $1.81\pm0.19$ & $12.16\pm1.40$ & $4.41\pm1.16$ \\ 
32 & 22:17:00.167 & -17:27:32.72 & $26.70\pm0.06$ & $3.854^{+0.001}_{-0.001}$ & $-19.03\pm0.19$ & $23.89\pm1.47$ & $3.36\pm0.21$ & $93.40\pm19.22$ & $3.68\pm2.03$ \\ 
33 & 22:16:44.680 & -17:17:48.48 & $25.93\pm0.03$ & $3.856^{+0.001}_{-0.001}$ & $-19.49\pm0.13$ & $75.34\pm1.29$ & $10.62\pm0.18$ & $192.60\pm24.90$ & $1.53\pm0.42$ \\ 
34 & 22:16:49.846 & -17:17:16.49 & $26.41\pm0.05$ & $4.026^{+0.001}_{-0.001}$ & $-19.11\pm0.22$ & $48.03\pm2.06$ & $7.50\pm0.32$ & $193.21\pm43.99$ & $8.03\pm0.93$ \\ 
35 & 22:16:51.997 & -17:26:10.95 & $25.85\pm0.03$ & $4.076^{+0.001}_{-0.001}$ & $-20.35\pm0.08$ & $27.36\pm1.87$ & $4.40\pm0.30$ & $36.05\pm3.71$ & $5.22\pm1.24$ \\ 
36 & 22:16:53.458 & -17:20:03.45 & $27.06\pm0.09$ & $4.093^{+0.001}_{-0.001}$ & $-19.19\pm0.22$ & $7.49\pm0.66$ & $1.22\pm0.11$ & $29.18\pm6.98$ & $7.24\pm1.49$ \\ 
37 & 22:16:52.593 & -17:29:00.63 & $26.89\pm0.08$ & $4.109^{+0.001}_{-0.001}$ & $-19.17\pm0.23$ & $17.48\pm1.22$ & $2.86\pm0.20$ & $70.02\pm17.27$ & $5.76\pm1.73$ \\ 
38 & 22:16:59.778 & -17:22:16.93 & $25.75\pm0.03$ & $4.126^{+0.001}_{-0.001}$ & $-20.12\pm0.11$ & $68.15\pm2.28$ & $11.28\pm0.38$ & $115.08\pm12.30$ & $4.28\pm0.67$ \\ 
39 & 22:17:03.102 & -17:25:52.33 & $25.23\pm0.02$ & $4.170^{+0.001}_{-0.001}$ & $-20.63\pm0.07$ & $120.44\pm3.00$ & $20.43\pm0.51$ & $130.16\pm9.74$ & $11.81\pm0.64$ \\ 
40 & 22:16:48.708 & -17:15:41.17 & $26.41\pm0.05$ & $4.182^{+0.002}_{-0.001}$ & $-19.79\pm0.15$ & $25.01\pm1.35$ & $4.27\pm0.23$ & $59.09\pm9.28$ & $1.45\pm0.88$ \\ 
41 & 22:16:49.635 & -17:15:26.63 & $27.50\pm0.13$ & $4.220^{+0.001}_{-0.001}$ & $-18.24\pm0.54$ & $17.35\pm1.48$ & $3.03\pm0.26$ & $173.91\pm113.30$ & $6.64\pm1.53$ \\ 
42 & 22:16:56.050 & -17:24:57.12 & $26.65\pm0.06$ & $4.258^{+0.001}_{-0.001}$ & $-19.59\pm0.19$ & $22.63\pm1.57$ & $4.03\pm0.28$ & $66.48\pm13.79$ & $6.99\pm1.76$ \\ 
\hline\\
\multicolumn{10}{c}{D1UD01 (30 galaxies)} \\
1 & 02:24:33.775 & -04:22:05.64 & $27.48\pm0.08$ & $2.730^{+0.001}_{-0.001}$ & $-17.14\pm0.33$ & $36.73\pm4.16$ & $2.26\pm0.26$ & $357.30\pm132.78$ & $6.12\pm1.92$ \\ 
2 & 02:24:24.047 & -04:19:30.14 & $26.75\pm0.04$ & $2.936^{+0.001}_{-0.001}$ & $-18.82\pm0.09$ & $11.49\pm1.98$ & $0.84\pm0.15$ & $28.27\pm5.46$ & $5.96\pm3.75$ \\ 
3 & 02:24:38.501 & -04:19:31.91 & $25.96\pm0.02$ & $2.954^{+0.001}_{-0.001}$ & $-19.63\pm0.04$ & $24.14\pm1.82$ & $1.80\pm0.14$ & $28.80\pm2.48$ & $5.79\pm1.33$ \\ 
4 & 02:24:32.251 & -04:20:05.64 & $26.36\pm0.03$ & $2.961^{+0.001}_{-0.001}$ & $-19.18\pm0.07$ & $24.91\pm1.67$ & $1.87\pm0.13$ & $45.24\pm4.18$ & $10.26\pm1.40$ \\ 
5 & 02:24:30.247 & -04:20:25.53 & $24.45\pm0.01$ & $2.977^{+0.001}_{-0.001}$ & $-21.21\pm0.01$ & $52.64\pm3.69$ & $3.99\pm0.28$ & $14.85\pm1.05$ & $11.87\pm1.47$ \\ 
6 & 02:24:35.414 & -04:20:32.25 & $26.00\pm0.02$ & $3.124^{+0.001}_{-0.001}$ & $-19.61\pm0.05$ & $58.84\pm2.09$ & $5.01\pm0.18$ & $81.48\pm5.02$ & $8.38\pm1.17$ \\ 
7 & 02:24:32.181 & -04:18:52.41 & $27.00\pm0.05$ & $3.127^{+0.001}_{-0.001}$ & $-18.75\pm0.12$ & $11.77\pm2.07$ & $1.01\pm0.18$ & $36.17\pm7.54$ & $12.53\pm5.35$ \\ 
8 & 02:24:26.931 & -04:18:09.40 & $25.10\pm0.01$ & $3.130^{+0.001}_{-0.001}$ & $-20.77\pm0.02$ & $16.28\pm1.97$ & $1.39\pm0.17$ & $7.81\pm0.95$ & $1.18\pm1.96$ \\ 
9 & 02:24:32.111 & -04:19:01.04 & $26.73\pm0.04$ & $3.131^{+0.001}_{-0.001}$ & $-19.09\pm0.09$ & $9.38\pm1.71$ & $0.80\pm0.15$ & $21.03\pm4.21$ & $7.64\pm2.41$ \\ 
10 & 02:24:32.361 & -04:18:33.93 & $27.32\pm0.07$ & $3.132^{+0.001}_{-0.001}$ & $-18.45\pm0.15$ & $8.49\pm1.34$ & $0.73\pm0.12$ & $34.32\pm7.41$ & $0.35\pm3.73$ \\ 
11 & 02:24:38.052 & -04:17:50.69 & $25.76\pm0.02$ & $3.150^{+0.001}_{-0.001}$ & $-20.08\pm0.04$ & $18.81\pm2.10$ & $1.64\pm0.18$ & $17.21\pm2.00$ & $6.70\pm2.73$ \\ 
12 & 02:24:36.424 & -04:20:40.01 & $27.41\pm0.08$ & $3.193^{+0.001}_{-0.001}$ & $-18.27\pm0.18$ & $17.26\pm1.44$ & $1.55\pm0.13$ & $86.42\pm17.51$ & $-1.63\pm4.53$ \\ 
13 & 02:24:39.007 & -04:17:25.43 & $26.14\pm0.02$ & $3.200^{+0.001}_{-0.001}$ & $-19.77\pm0.05$ & $14.54\pm1.65$ & $1.31\pm0.15$ & $18.37\pm2.25$ & $6.27\pm1.62$ \\ 
14 & 02:24:35.609 & -04:19:31.99 & $27.32\pm0.07$ & $3.220^{+0.001}_{-0.001}$ & $-18.46\pm0.16$ & $14.64\pm1.81$ & $1.34\pm0.17$ & $62.97\pm12.76$ & $3.07\pm2.45$ \\ 
\enddata
\end{deluxetable}
\clearpage
\setcounter{table}{3}
\begin{deluxetable}{cccccccccc}
\tabletypesize{\scriptsize}
\tablecaption{(continued)}
\tablewidth{0pt}
\tablehead{
    \colhead{ID} & \colhead{R.A.} & \colhead{Decl.} & \colhead{$m$\tablenotemark{a}} & \colhead{redshift} &
        \colhead{$M_\mathrm{UV}$} & \colhead{$f_\mathrm{Ly\alpha}$} & \colhead{$L_\mathrm{Ly\alpha}$} &
        \colhead{$EW_0$} & \colhead{$S_w$} \\
    \colhead{} & \colhead{(J2000)} & \colhead{(J2000)} & \colhead{(mag)} & \colhead{} & \colhead{(mag)} &
        \colhead{($10^{-18}\,\mathrm{erg\,s^{-1}\,cm^{-2}}$)} & \colhead{($10^{42}\,\mathrm{erg\,s^{-1}}$)} &
        \colhead{(\AA)} & \colhead{(\AA)}
}
\startdata
15 & 02:24:36.250 & -04:19:11.89 & $25.81\pm0.02$ & $3.258^{+0.001}_{-0.001}$ & $-20.04\pm0.04$ & $56.13\pm2.21$ & $5.29\pm0.21$ & $57.99\pm3.23$ & $12.49\pm1.98$ \\ 
16 & 02:24:36.988 & -04:18:09.47 & $25.44\pm0.01$ & $3.274^{+0.002}_{-0.001}$ & $-20.53\pm0.03$ & $42.95\pm2.36$ & $4.10\pm0.23$ & $28.53\pm1.73$ & $14.83\pm2.07$ \\ 
17 & 02:24:27.725 & -04:17:48.50 & $27.03\pm0.06$ & $3.284^{+0.001}_{-0.001}$ & $-18.83\pm0.13$ & $20.07\pm1.68$ & $1.93\pm0.16$ & $64.53\pm9.60$ & $5.80\pm1.54$ \\ 
18 & 02:24:35.157 & -04:17:00.64 & $26.61\pm0.04$ & $3.344^{+0.001}_{-0.001}$ & $-19.48\pm0.08$ & $8.56\pm1.33$ & $0.86\pm0.13$ & $15.65\pm2.68$ & $7.98\pm2.43$ \\ 
19 & 02:24:28.399 & -04:20:01.41 & $26.27\pm0.03$ & $3.351^{+0.001}_{-0.001}$ & $-19.77\pm0.06$ & $23.60\pm1.80$ & $2.38\pm0.18$ & $33.30\pm3.16$ & $10.49\pm1.71$ \\ 
20 & 02:24:38.367 & -04:17:16.11 & $27.34\pm0.07$ & $3.357^{+0.001}_{-0.001}$ & $-18.71\pm0.15$ & $7.92\pm1.12$ & $0.80\pm0.11$ & $29.73\pm6.09$ & $3.94\pm1.62$ \\ 
21 & 02:24:41.996 & -04:18:59.15 & $27.01\pm0.05$ & $3.400^{+0.001}_{-0.001}$ & $-19.11\pm0.11$ & $12.78\pm1.57$ & $1.33\pm0.16$ & $34.43\pm5.69$ & $10.15\pm2.55$ \\ 
22 & 02:24:37.488 & -04:19:20.22 & $26.12\pm0.02$ & $3.426^{+0.001}_{-0.001}$ & $-20.02\pm0.05$ & $34.92\pm1.95$ & $3.71\pm0.21$ & $41.50\pm3.12$ & $9.26\pm0.97$ \\ 
23 & 02:24:28.416 & -04:21:30.12 & $26.78\pm0.04$ & $3.435^{+0.001}_{-0.001}$ & $-19.32\pm0.10$ & $23.86\pm1.70$ & $2.55\pm0.18$ & $54.13\pm6.44$ & $3.95\pm0.73$ \\ 
24 & 02:24:36.548 & -04:18:26.31 & $26.56\pm0.04$ & $3.454^{+0.001}_{-0.001}$ & $-19.61\pm0.08$ & $21.02\pm1.16$ & $2.28\pm0.13$ & $37.07\pm3.42$ & $7.23\pm1.34$ \\ 
25 & 02:24:35.602 & -04:16:54.03 & $27.31\pm0.07$ & $3.455^{+0.001}_{-0.001}$ & $-18.60\pm0.19$ & $26.89\pm1.59$ & $2.92\pm0.17$ & $119.80\pm23.53$ & $3.58\pm1.71$ \\ 
26 & 02:24:44.626 & -04:19:35.65 & $26.24\pm0.03$ & $3.463^{+0.001}_{-0.001}$ & $-19.95\pm0.06$ & $30.63\pm1.60$ & $3.34\pm0.17$ & $39.74\pm3.03$ & $7.64\pm2.19$ \\ 
27 & 02:24:38.037 & -04:22:12.46 & $27.49\pm0.08$ & $3.529^{+0.001}_{-0.001}$ & $-18.67\pm0.19$ & $20.88\pm1.47$ & $2.38\pm0.17$ & $91.63\pm19.14$ & $9.51\pm2.56$ \\ 
28 & 02:24:29.531 & -04:21:43.03 & $26.83\pm0.05$ & $3.550^{+0.001}_{-0.001}$ & $-19.45\pm0.10$ & $29.51\pm2.19$ & $3.41\pm0.25$ & $64.39\pm8.00$ & $3.44\pm2.08$ \\ 
29 & 02:24:35.608 & -04:21:10.87 & $27.10\pm0.06$ & $3.551^{+0.001}_{-0.001}$ & $-19.25\pm0.12$ & $14.71\pm1.57$ & $1.70\pm0.18$ & $38.72\pm6.21$ & $11.86\pm3.78$ \\ 
30 & 02:24:24.653 & -04:19:31.71 & $26.99\pm0.05$ & $3.555^{+0.001}_{-0.001}$ & $-19.40\pm0.11$ & $11.07\pm1.24$ & $1.29\pm0.14$ & $25.39\pm3.88$ & $3.50\pm3.09$ \\ 
\hline\\
\multicolumn{10}{c}{D4UD01 (16 galaxies)} \\
1 & 22:14:03.642 & -18:00:09.90 & $26.61\pm0.04$ & $2.973^{+0.001}_{-0.001}$ & $-19.05\pm0.08$ & $6.08\pm1.27$ & $0.46\pm0.10$ & $12.47\pm2.79$ & $3.65\pm6.22$ \\ 
2 & 22:13:58.570 & -17:59:30.91 & $27.05\pm0.06$ & $3.008^{+0.001}_{-0.001}$ & $-18.60\pm0.13$ & $7.66\pm1.28$ & $0.60\pm0.10$ & $24.50\pm5.14$ & $0.09\pm1.53$ \\ 
3 & 22:14:11.265 & -17:59:56.51 & $25.84\pm0.02$ & $3.037^{+0.001}_{-0.001}$ & $-19.70\pm0.05$ & $57.83\pm2.77$ & $4.61\pm0.22$ & $68.72\pm4.69$ & $7.67\pm0.97$ \\ 
4 & 22:13:53.488 & -17:56:54.74 & $26.56\pm0.04$ & $3.046^{+0.001}_{-0.001}$ & $-18.95\pm0.10$ & $33.35\pm2.43$ & $2.67\pm0.19$ & $80.04\pm9.72$ & $5.06\pm1.72$ \\ 
5 & 22:13:51.171 & -17:57:18.84 & $26.66\pm0.04$ & $3.138^{+0.001}_{-0.001}$ & $-19.15\pm0.09$ & $11.14\pm1.56$ & $0.96\pm0.13$ & $23.74\pm3.92$ & $4.81\pm1.76$ \\ 
6 & 22:13:53.773 & -17:57:40.48 & $27.09\pm0.06$ & $3.210^{+0.001}_{-0.001}$ & $-18.79\pm0.13$ & $9.29\pm1.75$ & $0.84\pm0.16$ & $29.27\pm6.72$ & $4.17\pm1.95$ \\ 
7 & 22:13:54.597 & -17:59:06.04 & $26.79\pm0.05$ & $3.241^{+0.001}_{-0.001}$ & $-19.12\pm0.10$ & $13.84\pm1.33$ & $1.29\pm0.12$ & $32.89\pm4.58$ & $2.97\pm4.01$ \\ 
8 & 22:13:55.114 & -17:59:55.62 & $26.26\pm0.03$ & $3.242^{+0.001}_{-0.001}$ & $-19.68\pm0.06$ & $16.00\pm1.44$ & $1.49\pm0.13$ & $22.77\pm2.46$ & $3.70\pm2.06$ \\ 
9 & 22:14:04.835 & -17:57:44.20 & $26.93\pm0.06$ & $3.243^{+0.001}_{-0.001}$ & $-18.81\pm0.14$ & $27.86\pm1.96$ & $2.60\pm0.18$ & $88.14\pm13.39$ & $9.82\pm1.41$ \\ 
10 & 22:14:04.154 & -18:00:05.58 & $26.72\pm0.05$ & $3.243^{+0.001}_{-0.001}$ & $-19.11\pm0.11$ & $23.82\pm1.90$ & $2.22\pm0.18$ & $56.94\pm7.36$ & $7.18\pm1.45$ \\ 
11 & 22:14:03.430 & -17:59:22.71 & $26.03\pm0.02$ & $3.249^{+0.001}_{-0.001}$ & $-19.90\pm0.05$ & $24.84\pm1.88$ & $2.32\pm0.18$ & $28.98\pm2.62$ & $5.89\pm1.17$ \\ 
12 & 22:14:09.396 & -17:57:58.56 & $26.98\pm0.06$ & $3.336^{+0.001}_{-0.001}$ & $-19.12\pm0.11$ & $3.32\pm0.73$ & $0.33\pm0.07$ & $8.44\pm2.09$ & $4.28\pm10.94$ \\ 
13 & 22:14:16.330 & -17:57:22.22 & $27.54\pm0.10$ & $3.341^{+0.001}_{-0.001}$ & $-18.46\pm0.21$ & $9.19\pm1.56$ & $0.92\pm0.16$ & $43.24\pm11.65$ & $6.27\pm2.88$ \\ 
14 & 22:14:09.371 & -17:58:15.61 & $26.99\pm0.06$ & $3.341^{+0.001}_{-0.001}$ & $-19.02\pm0.13$ & $13.93\pm1.74$ & $1.39\pm0.17$ & $39.00\pm6.87$ & $4.65\pm2.26$ \\ 
15 & 22:13:58.117 & -17:59:46.75 & $26.92\pm0.06$ & $3.560^{+0.001}_{-0.001}$ & $-19.48\pm0.11$ & $12.41\pm1.54$ & $1.44\pm0.18$ & $26.50\pm4.40$ & $5.63\pm1.58$ \\ 
16 & 22:14:07.173 & -18:00:24.05 & $26.48\pm0.04$ & $3.635^{+0.001}_{-0.001}$ & $-20.03\pm0.08$ & $37.94\pm2.08$ & $4.65\pm0.26$ & $51.22\pm4.73$ & $8.40\pm1.04$ \\ 
\enddata
\tablenotetext{a}{The apparent aperture magnitude of detection-band: $i'$-band for $u$-, $g$-, and $r$-dropout,
    and $z'$-band for $i$-dropout galaxies.}
\end{deluxetable}
\end{turnpage}
\clearpage
\begin{deluxetable*}{ccccccc}
\tablecaption{Results of the protocluster confirmation \label{tab:cluster}}
\tablewidth{0pt}
\tablehead{Name & $N_\mathrm{obs}$\tablenotemark{a} & $N_\mathrm{det}$\tablenotemark{b} & Protocluster? &
    $N_\mathrm{member}$\tablenotemark{c} & redshift & $\sigma_v$ ($\mathrm{km\,s^{-1}}$)}
\startdata
D1ID01 & 8 & 3 & unclear & --- & --- & --- \\
D3ID01 & 8 & 2 & possible & 2 & 5.75 & --- \\
D1RD01 & 15 & 6 & possible & 2 & 4.89 & --- \\
D4RD02 & 12 & 3 & unclear & --- & --- & --- \\
D1GD01 & 123 & 36 & No & --- & --- & --- \\
D4GD01 & 144 & 42 & Yes & 11 & 3.67 & $352\pm140$ \\
D1UD01 & 95 & 30 & Yes & 5 & 3.13 & $235\pm75$ \\
D4UD01 & 57 & 16 & Yes & 5 & 3.24 & $61\pm105$
\enddata
\tablenotetext{a}{The number of observed galaxies.}
\tablenotetext{b}{The number of spectroscopically detected galaxies.}
\tablenotetext{c}{The number of protocluster members.}
\end{deluxetable*}
\begin{figure*}
\epsscale{1.0}
\plotone{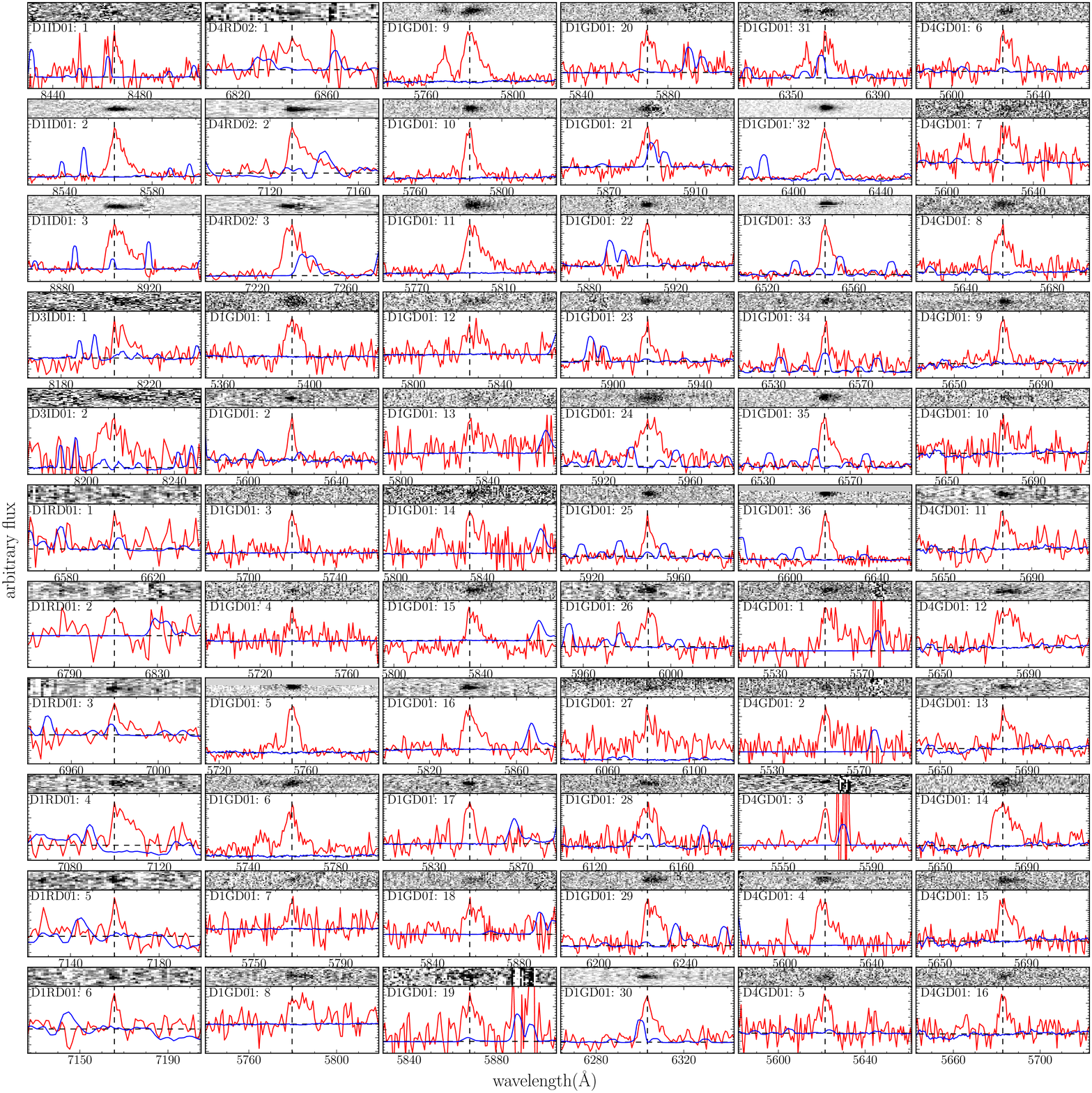}
\caption{Spectra of all dropout galaxies having Ly$\alpha$ emission line.
    The field and object IDs are indicated at the upper left corner (Column 1 of Table \ref{tab:spec_all}).
    The vertical and horizontal dashed lines show the wavelength of Ly$\alpha$ emission and the zero level of
    flux, respectively.}
\label{fig:spec_all}
\end{figure*}
\setcounter{figure}{8}
\begin{figure*}
\epsscale{1.0}
\plotone{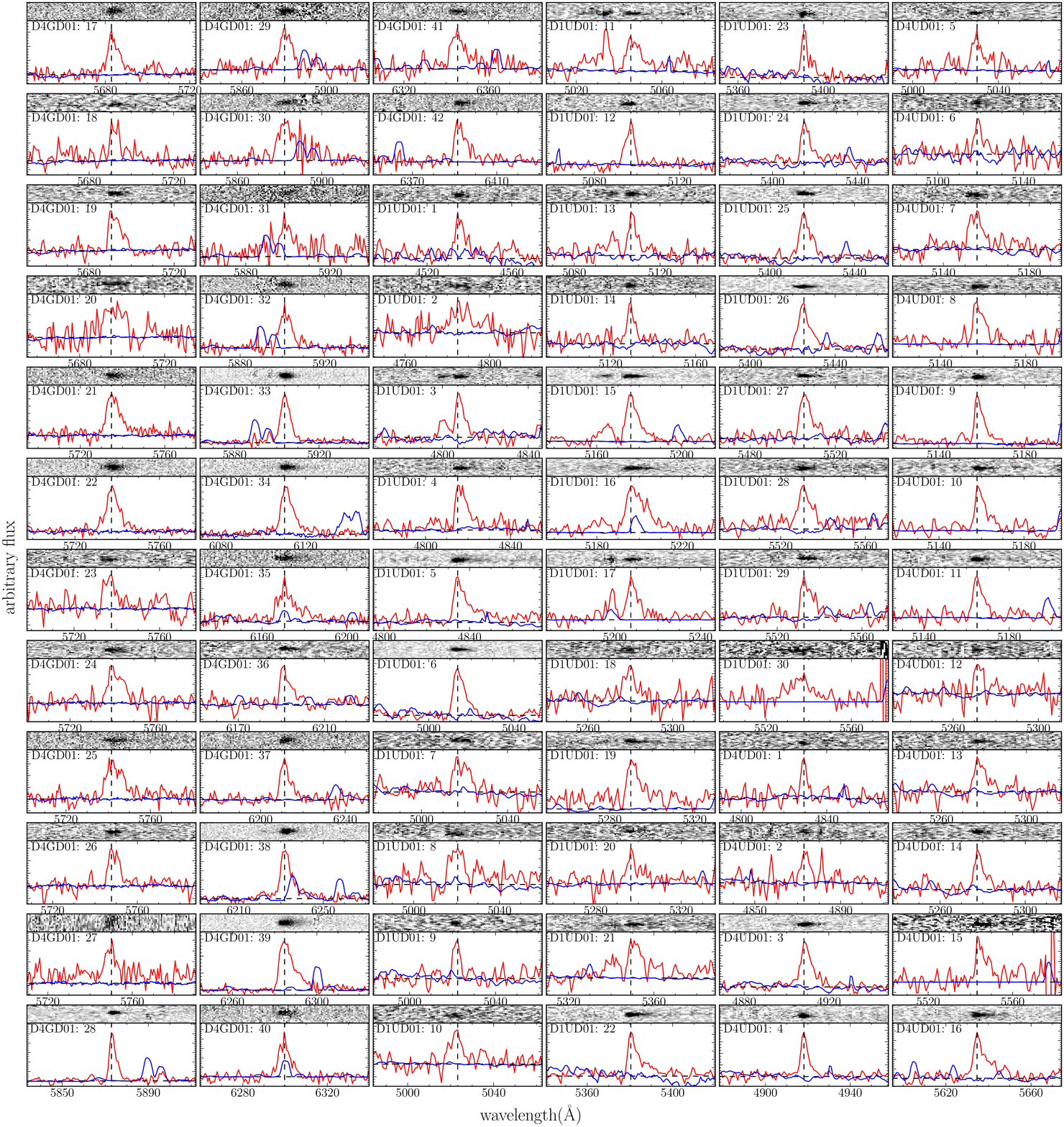}
\caption{(continued)}
\end{figure*}

\noindent
width and seeing size for each observation, and its
flux error was estimated from  the combination of the line width and the noise level per {1\AA} at wavelengths
blueward of Ly$\alpha$.
Since continuum flux was too faint to be detected in the observed spectra, $M_\mathrm{UV}$ was estimated from
the broad-band magnitude ($g'$-, $r'$-, $i'$-, and $z'$-bands for $u$-, $g$-, $r$-, and $i$-dropout
galaxies).
The $M_\mathrm{UV}$ is defined as the absolute magnitude at 1300A in the rest-frame.
It is derived from the broad-band photometry after correcting the contribution of IGM absorption and the
Ly$\alpha$ emission to the broad-band photometry based on the spectroscopic.
In this calculation, we have assumed a UV slope, $\beta$ ($f_\lambda\propto\lambda^\beta$) to be -2, which is
consistent with the previous observations (e.g., Bouwens et al. 2012, ApJ, 754, 83).
The broad-band magnitude has a negligible dependence on UV slope, because the broad-band width is only
$\sim1000\mathrm{-}1500\mathrm{\AA}$.
We have confirmed that $M_\mathrm{UV}$ chages only a few percent at maximum, when UV slope was varied from -3.0
to -1.0.
In addition, $EW_0$ was estimated by combining $f_\mathrm{Ly\alpha}$ and $M_\mathrm{UV}$.
The results found for each region are described in the following subsections.

\subsubsection{The $i$-dropout protocluster candidate in the D1 field}
\begin{figure}
\epsscale{1.0}
\plotone{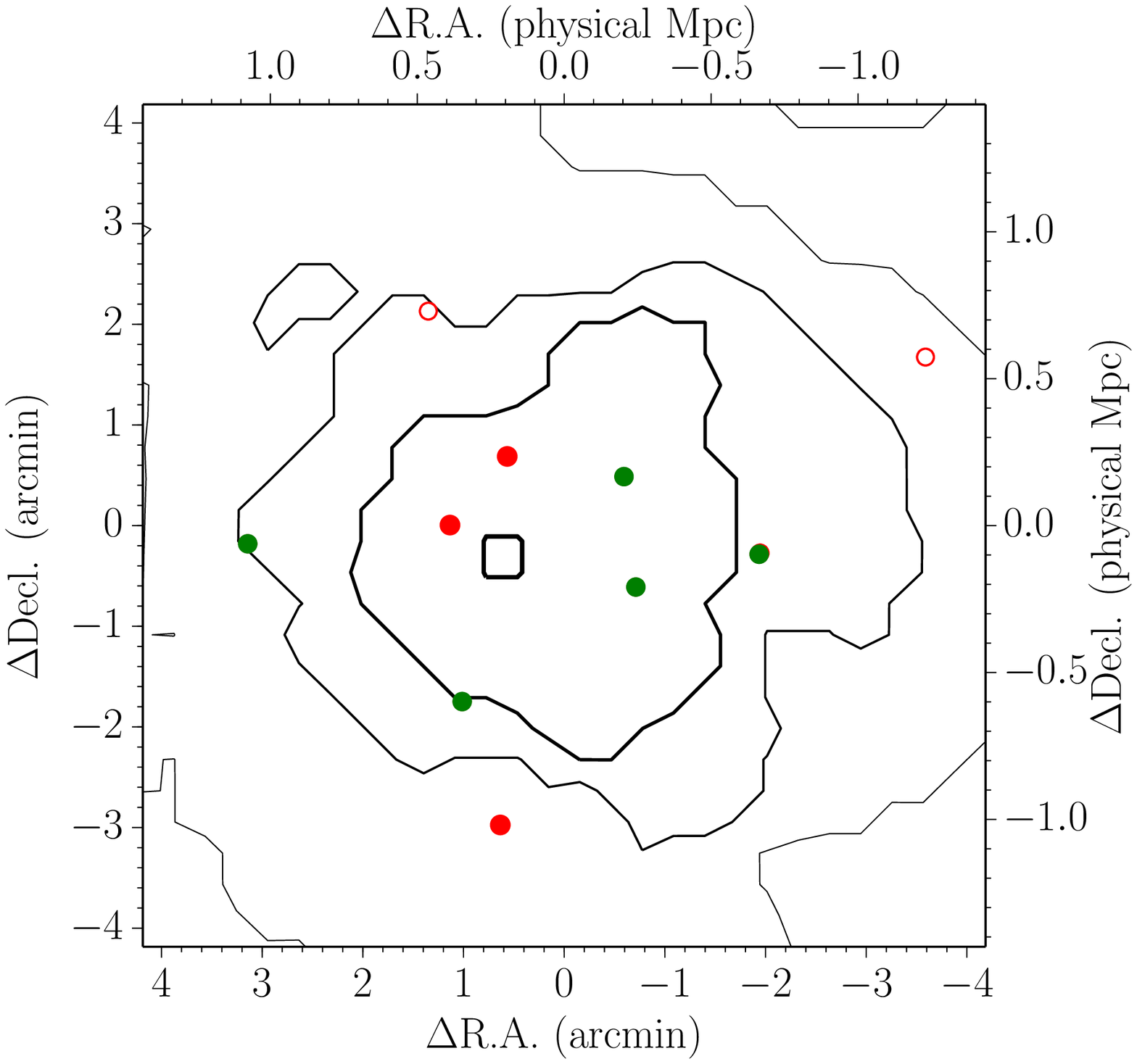}
\caption{Sky distribution of $i$-dropout galaxies and number density contours in the D1ID01 region.
    Spectroscopically observed galaxies are marked by filled circles (red: Ly$\alpha$ detected galaxies, green:
    Ly$\alpha$ undetected galaxies), and spectroscopically unobserved galaxies are shown by open circles.
    The origin (0,0) is $(\mathrm{R.A.}, \mathrm{Decl.})=(02:27:16.5,-04:50:49.6)$, which is defined as the
    center of the figure.
    The lines show the number density contours of $i$-dropout galaxies from $6\sigma$ to $0\sigma$ with a step
    of $2\sigma$.}
\label{fig:spec_obs_D1i}
\end{figure}

We have observed eight $i$-dropout galaxies in the D1ID01 region out of ten candidates.
Almost all $i$-dropout galaxies in the D1ID01 region were spectroscopically observed, as shown in Figure
\ref{fig:spec_obs_D1i}.
Three galaxies clearly have single emission lines, which can be identified as Ly$\alpha$ emission lines of
$z\sim6$ galaxies.
Their photometric and spectroscopic properties are summarized in Table \ref{tab:spec_all}.
Two of three galaxies (ID=1 and 2) have close redshifts with difference of $\Delta z=0.08$, corresponding to
the radial distance of $4.7\,\mathrm{Mpc}$ in physical scale.
From our selection criteria, we can expect $\sim0.2\mathrm{-}0.4$ galaxy in a $\Delta z=0.1$ bin if they were
homogeneously distributed in redshift space.
The possibility to have two galaxies within $\Delta z<0.1$ is 16\%.
Although their distribution is more concentrated than homogeneous, these two galaxies are unlikely to merge into
a single halo by $z=0$ based on our analysis of the possible separations of protocluster galaxies at $z\sim6$
(Figure \ref{fig:3D_model}).
For now, we cannot conclude that there is a real protocluster in the D1ID01 region due to the small number of
statistics.

\subsubsection{The $i$-dropout protocluster candidate in the D3 field}
\begin{figure}
\epsscale{1.0}
\plotone{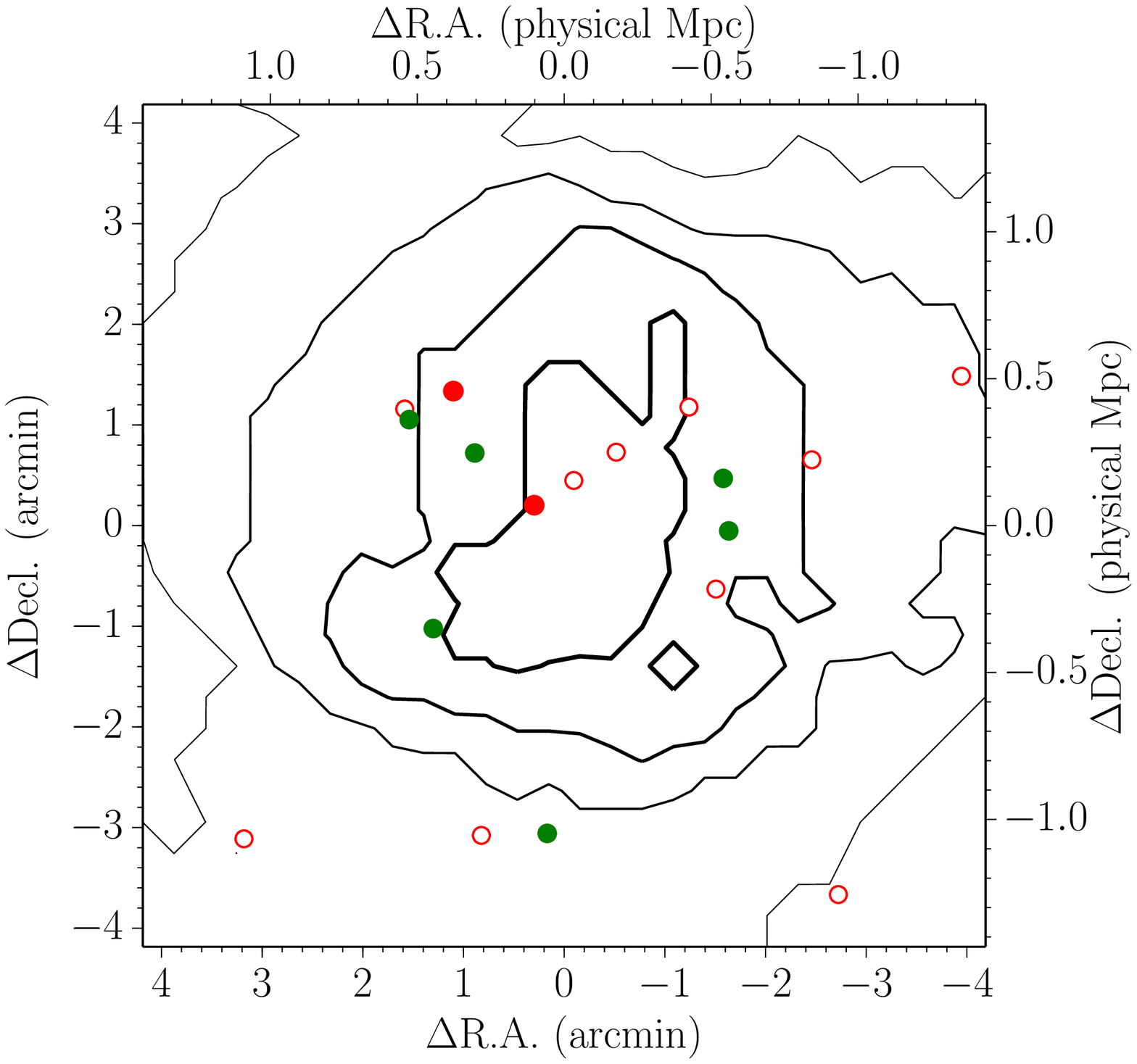}
\caption{Sky distribution of $i$-dropout galaxies and number density contours in the D3ID01 region.
    Spectroscopically observed galaxies are marked by filled circles (red: Ly$\alpha$ detected galaxies, green:
    Ly$\alpha$ undetected galaxies), and spectroscopically unobserved galaxies are shown by open circles.
    The origin (0,0) is $(\mathrm{R.A.}, \mathrm{Decl.})=(14:19:15.2,+52:56:02.2)$, which is defined as the
    center of the figure.
    The lines show the number density contours of $i$-dropout galaxies from $6\sigma$ to $0\sigma$ with a step
    of $2\sigma$.}
\label{fig:spec_obs_D3i}
\end{figure}

As for the D3ID01 region, eight $i$-dropout galaxies were observed out of sixteen candidates.
The completeness of our spectroscopic observation is smaller ($\sim50\%$) than for the D1ID01 region, which
has less protocluster member candidates.
Many faint $i$-dropout galaxies are still to be observed because we assigned the brighter $i$-dropout galaxies 
higher priorities.
Ly$\alpha$ emission lines were detected from two of the eight spectroscopic targets.
The sky distribution of the targets is shown in Figure \ref{fig:spec_obs_D3i}.
Table \ref{tab:spec_all} describes the properties of spectroscopically confirmed galaxies.
These two galaxies have almost the same redshift with the difference of $\Delta z < 0.01$ ($<0.5\,\mathrm{Mpc}$
in physical scale).
The possibility that two galaxies have this small redshift separation is only 1.2\%, and these two galaxies can
certainly be expected to be in the same halo at $z=0$ based on this small separation.
While we could not make a clear conclusion due to the small number of confirmed galaxies, the discovery of a close
galaxy-pair at $z\sim6$ could imply the existence of a protocluster.
Interestingly, \citet{toshikawa14} have found that galaxies tend to be in close pairs in another protocluster at
$z=6.01$.

The relatively small number of confirmed candidate can be attributed to the observational limit since our 
spectroscopic samples are biased to brighter galaxies ($\mathrm{M_{UV}}<-20.75$).
We compared the D3ID01 protocluster candidate with the $z=6.01$ protocluster in the SDF \citep{toshikawa14},
which was also identified by the combination of dropout selection and follow-up spectroscopy targeting Ly$\alpha$
emission like this study.
Applying the same magnitude limit of $\mathrm{M_{UV}}<-20.75$ to the SDF protocluster, the number of remaining
protocluster members was only two our of a total of ten.
Furthermore, \citet{ouchi05} reported the discovery of two protoclusters at $z\sim5.7$.
These were discovered from a narrow-band survey, and six and four LAEs are included in each protocluster.
Although LAE selection is different from our dropout selection, it is useful to check the distribution of the UV
continuum and the Ly$\alpha$ luminosity of protocluster galaxies.
Based on our observational limits of UV continuum and Ly$\alpha$ luminosity, only $\sim2$ LAEs would be
identified for these protoclusters.
Therefore, it is not unreasonable to expect only two confirmed member galaxies in this study even if there is a
real protocluster.

\subsubsection{The $r$-dropout protocluster candidate in the D1 field}
\begin{figure}
\epsscale{1.0}
\plotone{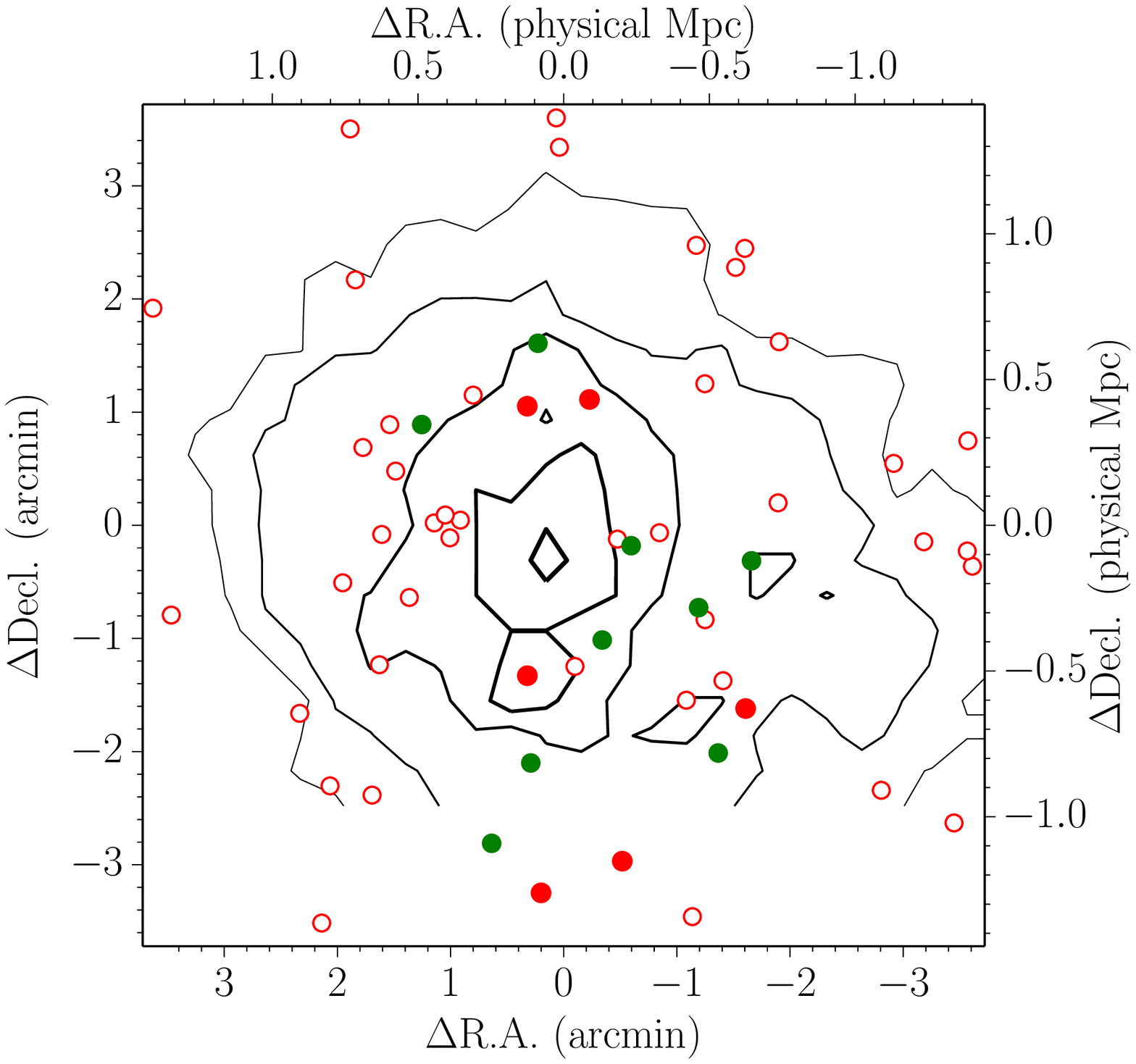}
\caption{Sky distribution of $r$-dropout galaxies and number density contours in the D1RD01 region.
    Spectroscopically observed galaxies are marked by filled circles (red: Ly$\alpha$ detected galaxies, green:
    Ly$\alpha$ undetected galaxies), and spectroscopically unobserved galaxies are shown by open circles.
    The origin (0,0) is $(\mathrm{R.A.}, \mathrm{Decl.})=(02:24:44.7,-04:55:37.9)$, which is defined as the
    center of the figure.
    The lines show the number density contours of $i$-dropout galaxies from $4\sigma$ to $0\sigma$ with a step
    of $1\sigma$.}
\label{fig:spec_obs_D1r}
\end{figure}

We have spectroscopically observed fifteen $r$-dropout galaxies in the D1RD01 region, and detected single
emission lines from six galaxies.
The sky distribution of the observed galaxies is shown in Figure \ref{fig:spec_obs_D1r}.
In the $>1\sigma$ overdense region, there are $\sim40$ galaxies; thus, only $\sim38\%$ $r$-dropout galaxies were
observed by the follow-up spectroscopy.
Two galaxies (ID=5 and 6) out of six are clustering both in spatial ($\Delta \mathrm{sky}=33\,\mathrm{arcsec}$)
and redshift space ($\Delta z=0.004$) at $z=4.89$, whose three-dimensional separation is $0.7\,\mathrm{Mpc}$ in
physical scale.
Considering the observed volume ($3\,\mathrm{arcmin}$ radius and $\Delta z\sim0.8$), it is unlikely ($<1\%$)
that the close pair is reproduced by uniform random distribution of six galaxies in three-dimensional space.
However, it is unclear whether this galaxy pair will grow into a cluster at $z=0$ due to the small number of
confirmed galaxies; at least, these two galaxies are expected to merge into a single halo.
Since there are many spectroscopically unobserved galaxies, further follow-up observation will enable to clarify
whether there is a protocluster or not.

\subsubsection{The $r$-dropout protocluster candidate in the D4 field}
\begin{figure}
\epsscale{1.0}
\plotone{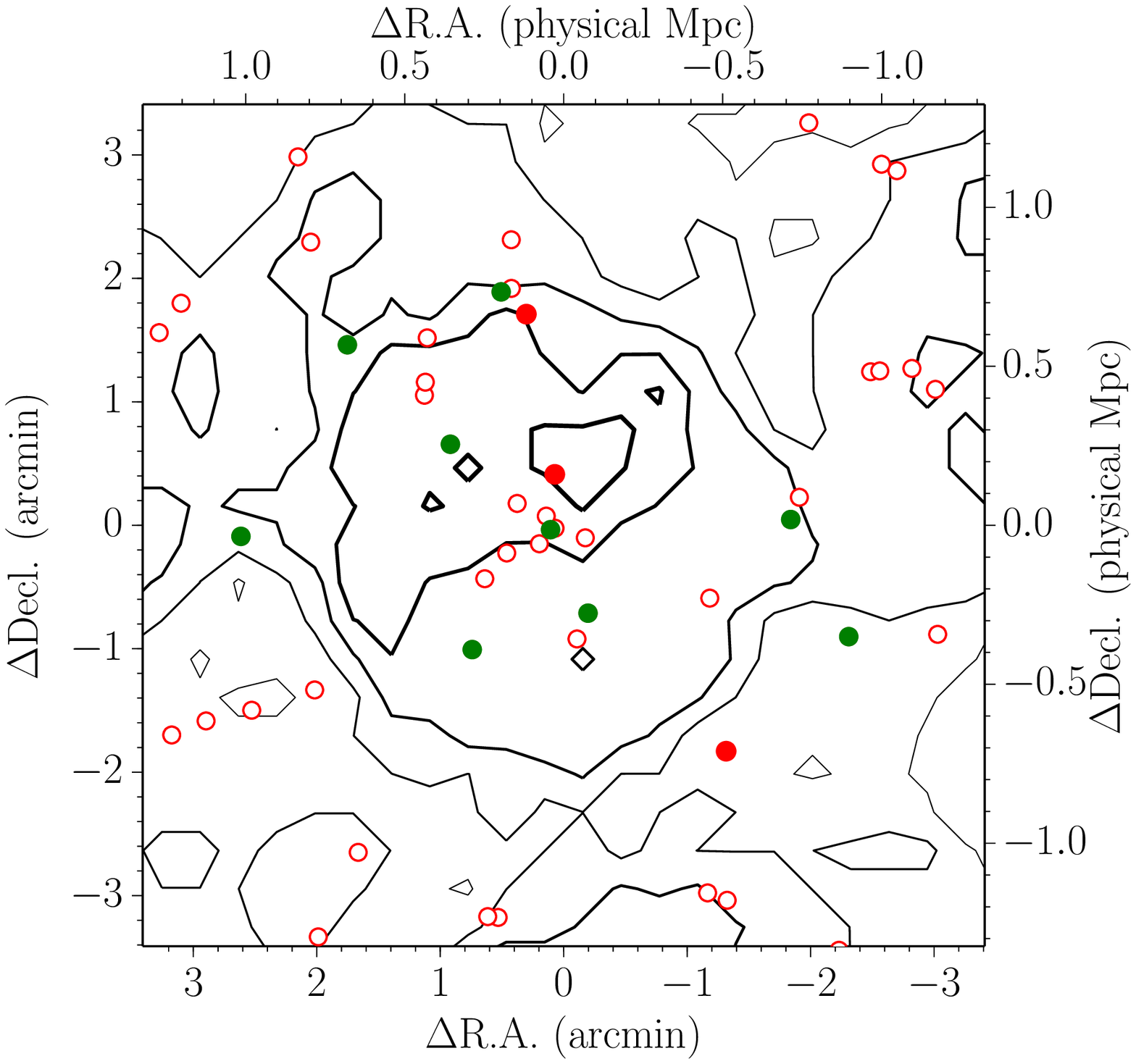}
\caption{Sky distribution of $r$-dropout galaxies and number density contours in the D4RD02 region.
    Spectroscopically observed galaxies are marked by filled circles (red: Ly$\alpha$ detected galaxies, green:
    Ly$\alpha$ undetected galaxies), and spectroscopically unobserved galaxies are shown by open circles.
    The origin (0,0) is $(\mathrm{R.A.}, \mathrm{Decl.})=(22:16:45.5,-17:29:44.7)$, which is defined as the
    center of the figure.
    The lines show the number density contours of $i$-dropout galaxies from $4\sigma$ to $0\sigma$ with a step
    of $1\sigma$.}
\label{fig:spec_obs_D4r}
\end{figure}

In the D4RD02 region, the total integration time of follow-up spectroscopic observation was only two hours,
which was half of that in the D1RD01 region.
Thus, although twelve $r$-dropout galaxies were observed, Ly$\alpha$ emission lines were detected from only three
galaxies.
The sky distribution of the observed galaxies is shown in Figure \ref{fig:spec_obs_D4r}.
These three galaxies are largely separated in redshift space.
Since about 20 $r$-dropout galaxies remain to be spectroscopically observed, further follow-up observation will
be necessary to make a conclusion.

\subsubsection{The $g$-dropout protocluster candidate in the D1 field}
\begin{figure*}
\epsscale{1.0}
\plotone{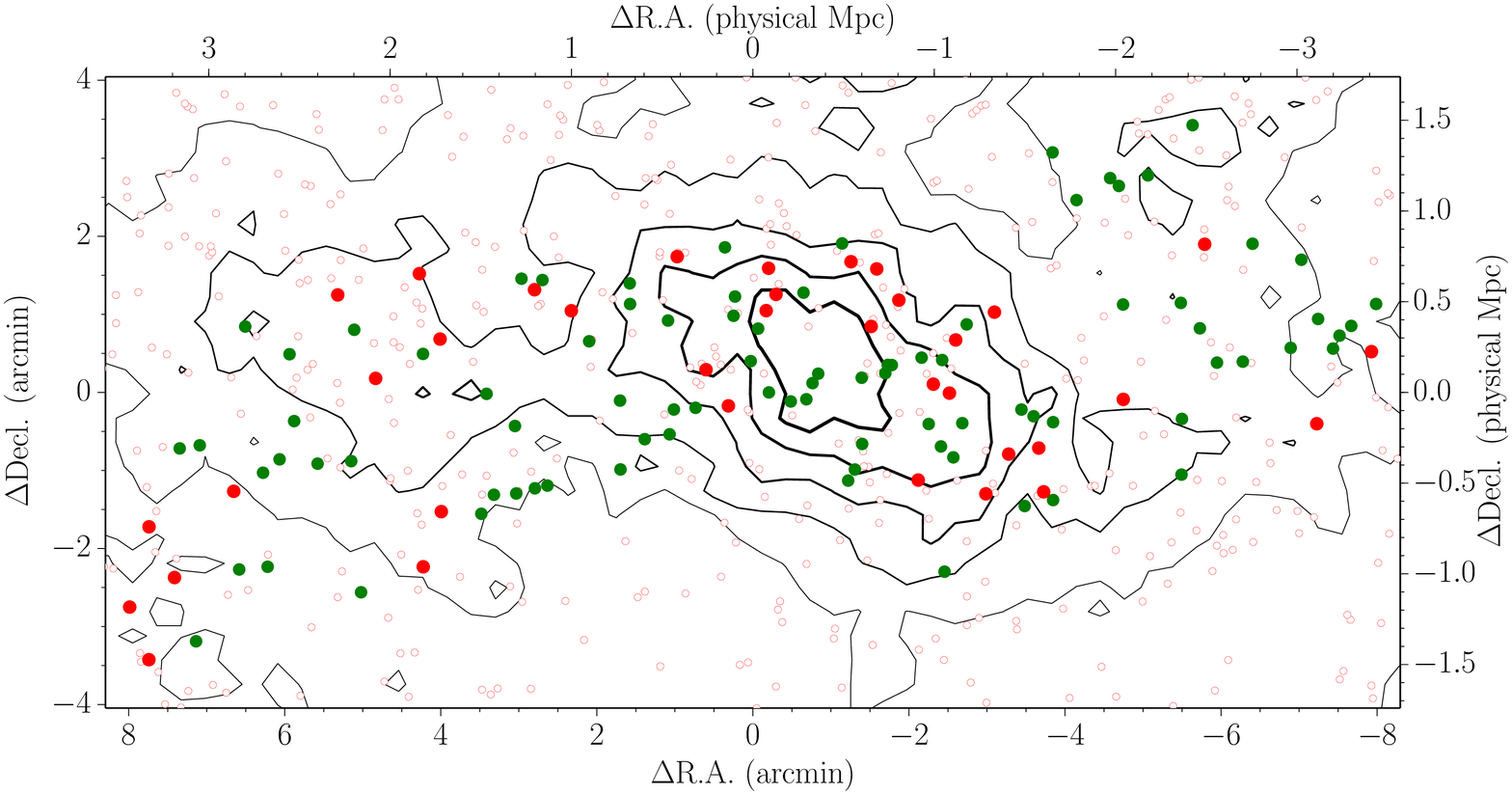}
\caption{Sky distribution of $g$-dropout galaxies and number density contours in/around the D1GD01 region.
    Spectroscopically observed galaxies are marked by filled circles (red: Ly$\alpha$ detected galaxies, green:
    Ly$\alpha$ undetected galaxies), and spectroscopically unobserved galaxies are shown by open circles.
    The origin (0,0) is $(\mathrm{R.A.}, \mathrm{Decl.})=(02:25:40.5,-04:15:56.3)$, which is defined as the
    center of the figure.
    The lines show the number density contours of $i$-dropout galaxies from $4\sigma$ to $0\sigma$ with a step
    of $1\sigma$.}
\label{fig:spec_obs_D1g}
\end{figure*}
\begin{figure}
\epsscale{1.0}
\plotone{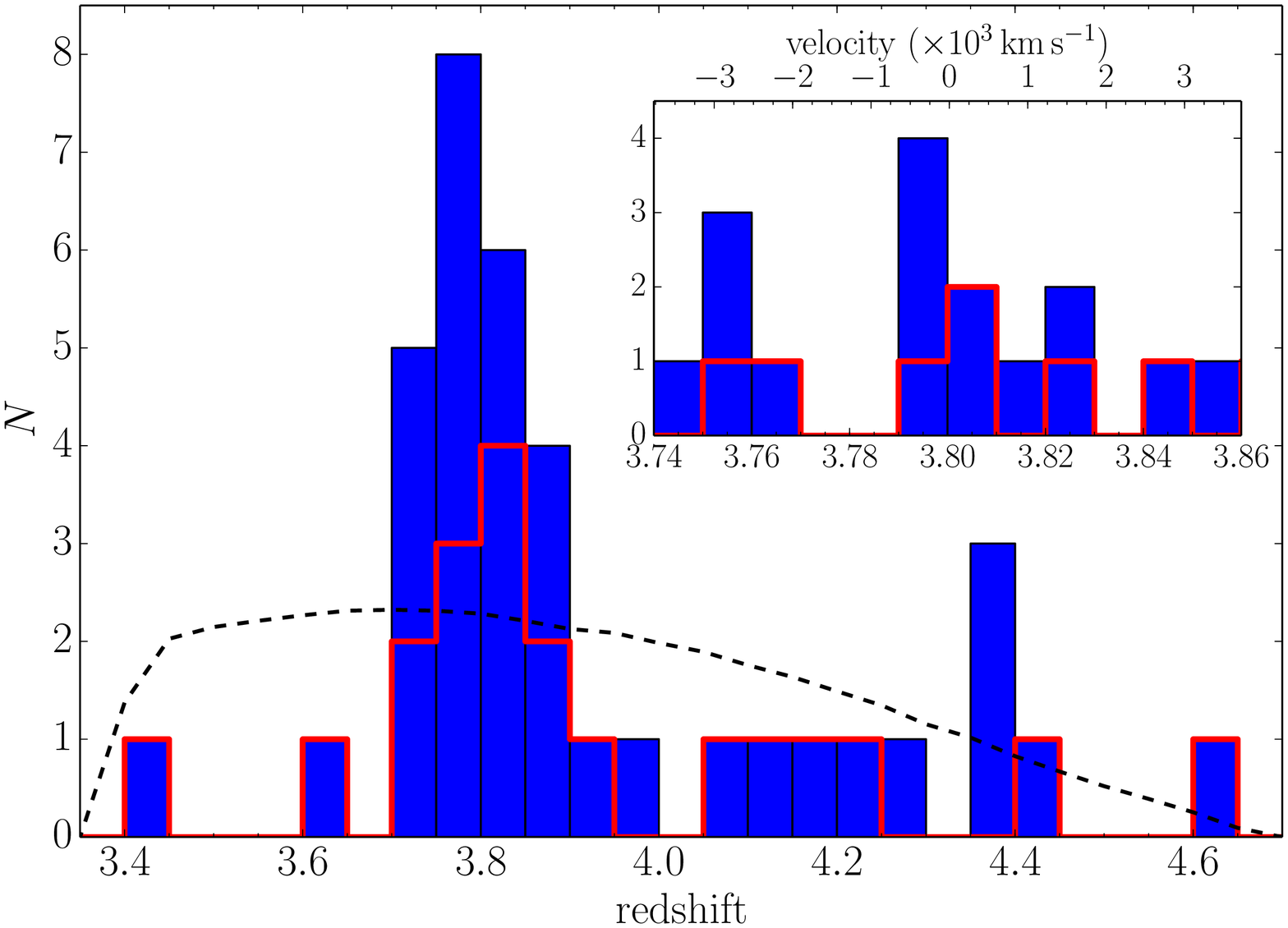}
\caption{Redshift distribution of 36 $g$-dropout with the bin size of $\Delta z=0.05$ in/around the D1GD01 region.
    Blue histogram shows all 36 galaxies, and red line shows galaxies only in the D1GD01 region.
    The inset is a close-up of the protocluster redshift range, with a bin size of $\Delta z =0.01$.}
\label{fig:redshift_D1g}
\end{figure}
Combining the DEIMOS and FOCAS follow-up observations, 123 $g$-dropout galaxies were observed, and the redshifts
of 36 galaxies were determined by detecting Ly$\alpha$ emission lines.
The sky distribution of the observed galaxies is shown in Figure \ref{fig:spec_obs_D1g}.
Figure \ref{fig:redshift_D1g} shows the redshift distribution of confirmed galaxies.
Although galaxies seem to be clustering at $z\sim3.8$, these galaxies are spread over a large projected area,
as shown in Figure \ref{fig:spec_obs_D1g}.
Since the DEIMOS has a wide FoV ($\sim16.7\times5.0\,\mathrm{arcmin^2}$), which is larger than the area of the
D1GD01 region, some $g$-dropout galaxies outside the overdense region were also observed.
When we focus only on galaxies in the overdense region, shown by the red-line histogram in Figure
\ref{fig:redshift_D1g}, the peak around $z=3.8$ becomes lower and only three galaxies are clustering within the
expected redshift range of the protocluster candidate ($\Delta z<0.01$).
Given the total number of confirmed galaxies in the overdense region, the group of three galaxies can be
reproduced even from a random homogeneous distribution with  a probability of 21\%.
 Based on this probability, it cannot be dismissed that the observed redshift distribution is drawn from a
uniform random distribution.
Hence, we cannot currently conclude that there is a protocluster in the D1GD01 region without further observations.
Hereafter, we regard the D1GD01 region as not being a protocluster.
The high surface overdensity observed in this field could be attributed to a coincidental alignment of large-scale
structure on a scale of $\Delta z\sim0.1\mathrm{-}0.2$, which is too large to grow into a single
halo by $z=0$.

\subsubsection{The $g$-dropout protocluster candidate in the D4 field}
\begin{figure}
\epsscale{1.0}
\plotone{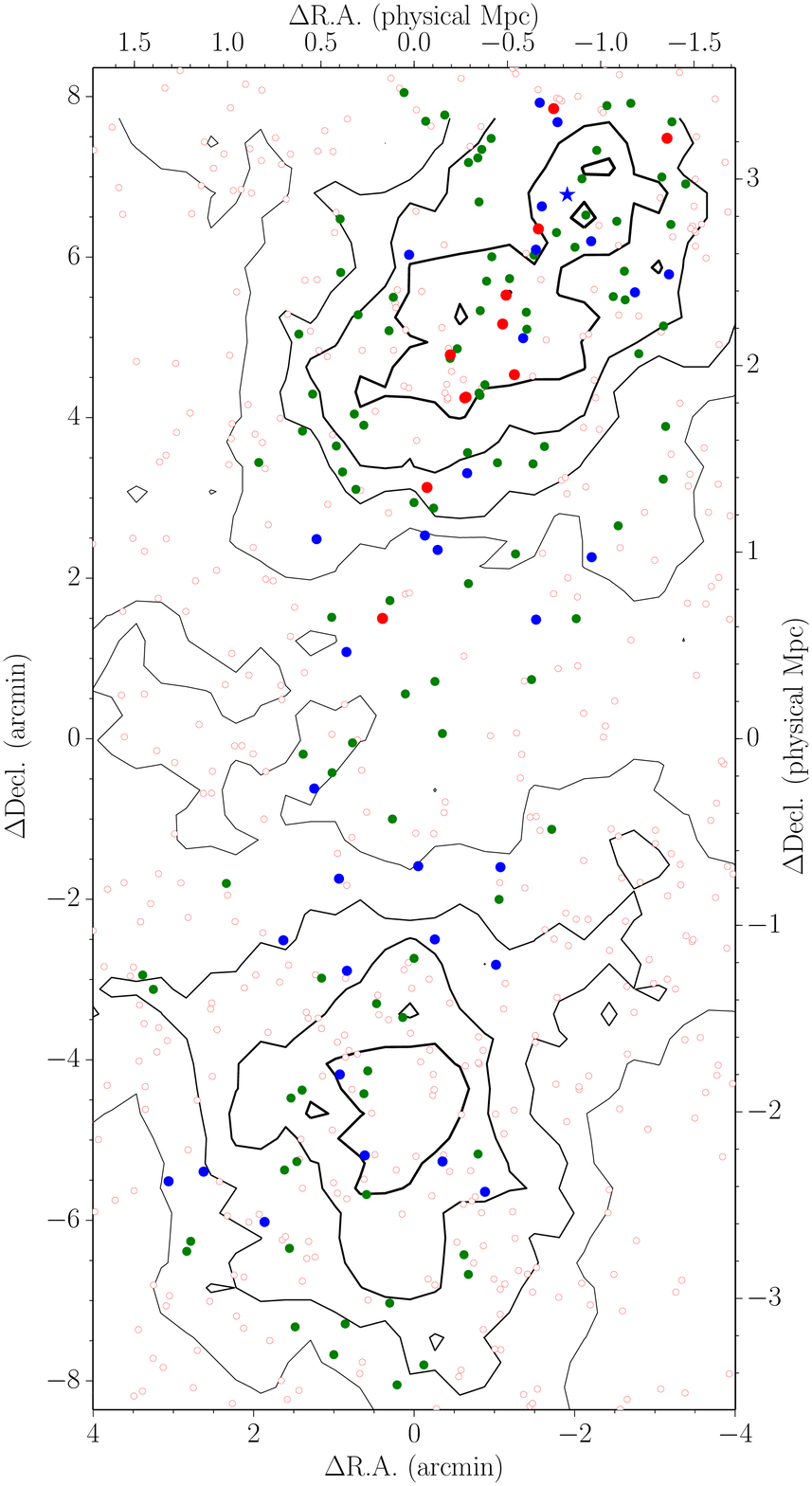}
\caption{Sky distribution of $g$-dropout galaxies and number density contours in/around the D4GD01 region.
    Spectroscopically observed galaxies are marked by filled circles (red: protocluster members, blue: non-members,
    green: Ly$\alpha$ undetected galaxies), and spectroscopically unobserved galaxies are shown by open circles.
    The blue star indicates the position of the AGN.
    The origin (0,0) is $(\mathrm{R.A.}, \mathrm{Decl.})=(22:16:56.3,-17:23:21.9)$, which is defined as the
    center of the figure.
    The lines show the number density contours of $i$-dropout galaxies from $4\sigma$ to $0\sigma$ with a step
    of $1\sigma$.}
\label{fig:spec_obs_D4g}
\end{figure}

\begin{figure}
\epsscale{1.0}
\plotone{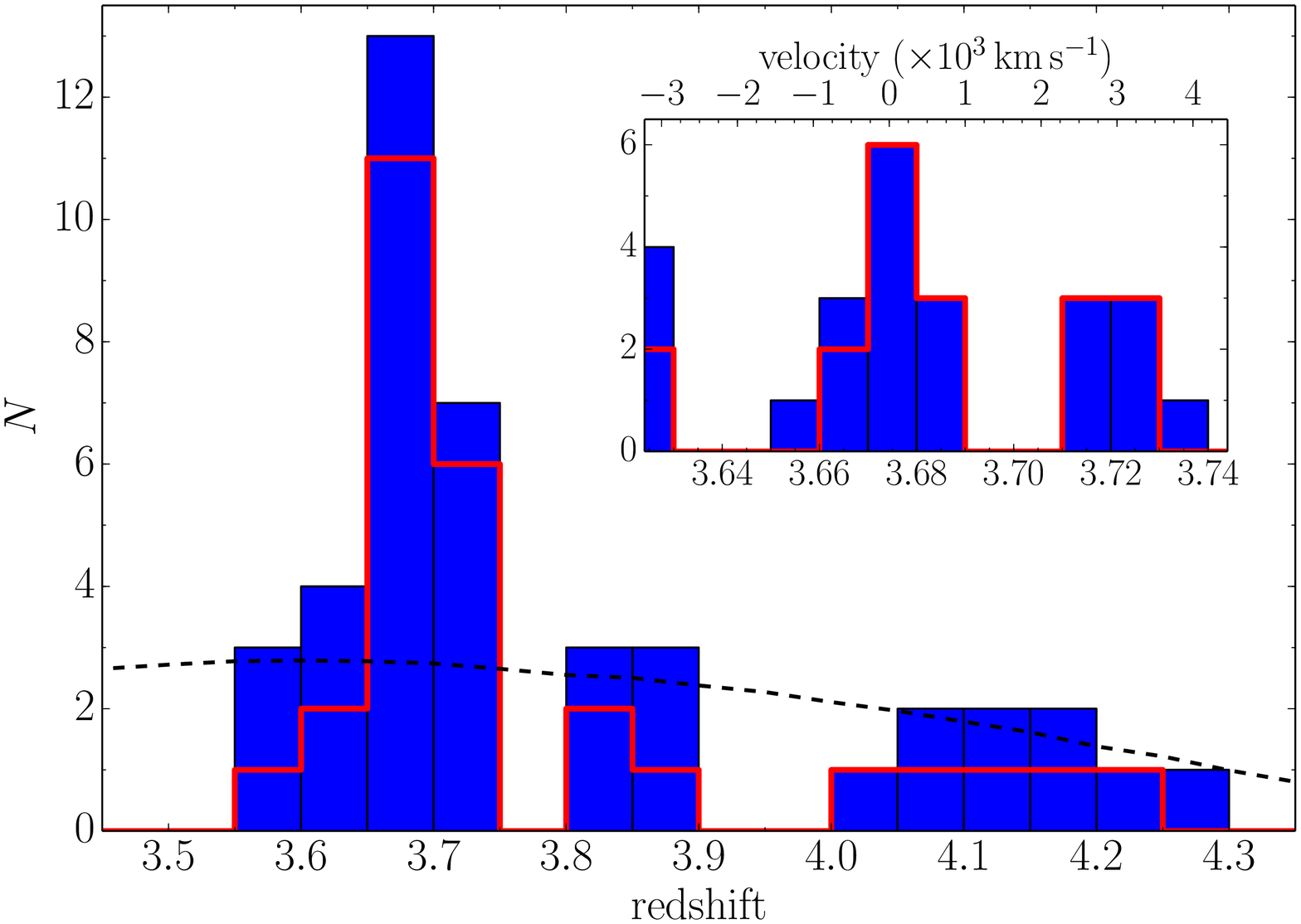}
\caption{Redshift distribution of 42 $g$-dropout galaxies with the bin size of $\Delta z=0.05$ in/around the
    D4GD01 region.
    Blue histogram shows all 42 galaxies, and red line shows galaxies only in the D4GD01 region.
    The inset is a close-up of the protocluster redshift range, with a bin size of $\Delta z =0.01$.}
\label{fig:redshift_D4g}
\end{figure}

Combining the DEIMOS and FOCAS follow-up observations, 144 $g$-dropout galaxies were spectroscopically observed
in the D4GD01 protocluster region, and the redshifts of 42 galaxies were determined by detecting Ly$\alpha$
emission lines.
The sky distribution of the observed galaxies is shown in Figure \ref{fig:spec_obs_D4g}.
The redshift distribution is shown in Figure \ref{fig:redshift_D4g}.
There is a clear excess at $z=3.67$, and, as contrasted with the D1GD01 region, eleven galaxies are clustered
in a narrow redshift range of $\Delta z=0.016$, corresponding to $2.6\,\mathrm{Mpc}$ in physical scale.
Since it is almost impossible ($<0.01\%$) to explain this clustering with a random homogeneous distribution, we
concluded that there is a protocluster at $z=3.67$, which includes eleven member galaxies (ID=10-20).

\begin{figure}
\epsscale{1.0}
\plotone{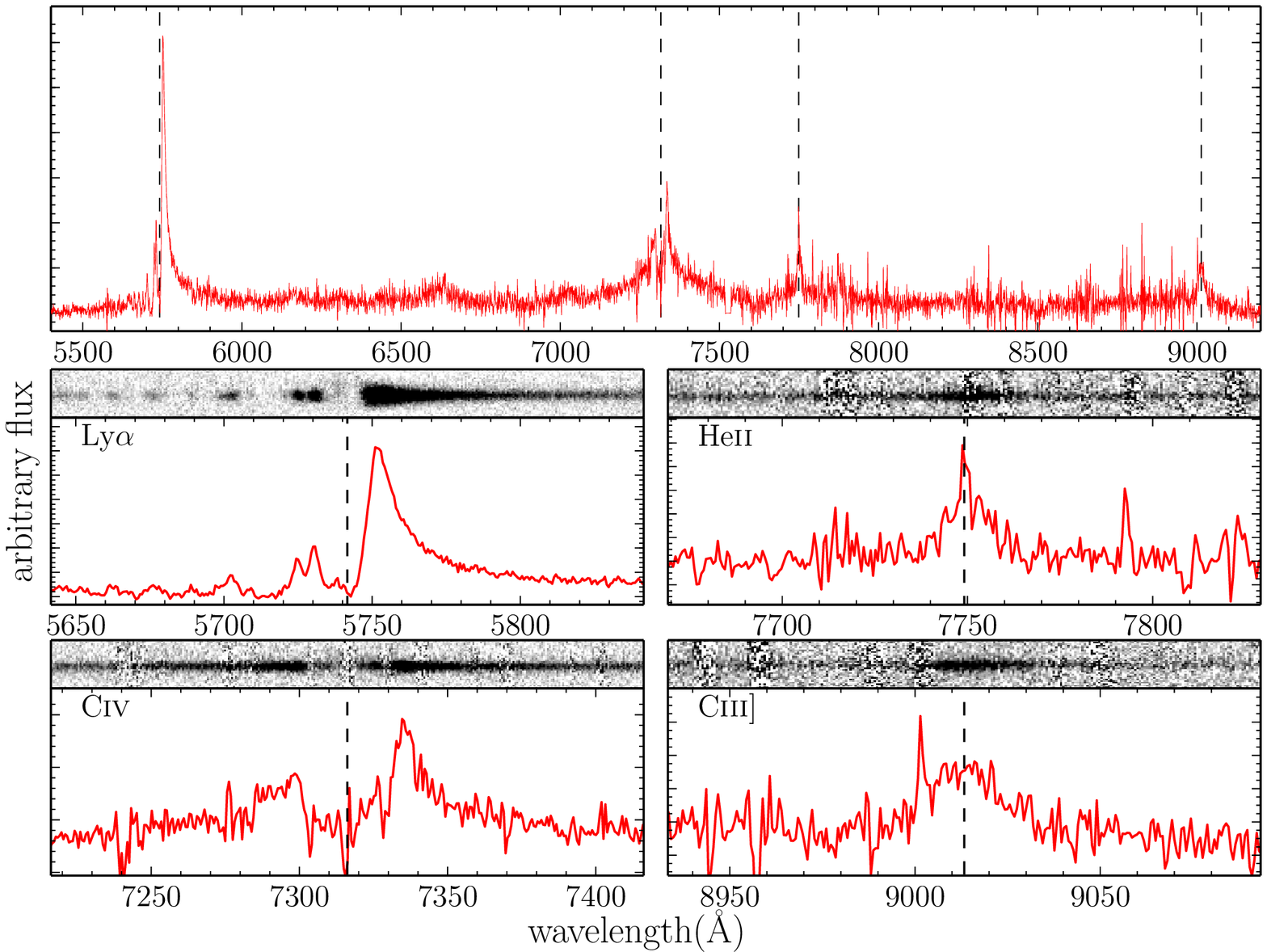}
\caption{Spectra of the AGN in the D4GD01.
    Upper panels shows the full wavelength coverage of the AGN, and lower four panels show emission lines which
    were clearly detected.
    The redshift was estimated by the peak wavelength of He{\sc ii} and C{\sc iii}], and the vertical dashed
    lines in the He{\sc ii} and C{\sc iii}] panels indicate the peak of the emission line.
    On the other hand, the vertical dashed lines in Ly$\alpha$ and C{\sc iv} panels indicate the expected
    wavelength according to the redshift.}
\label{fig:spec_D4g_AGN}
\end{figure}

It should also be noted that an AGN was found in this region at
$(\Delta\mathrm{R.A.},\,\Delta\mathrm{Decl.})=(-1.9,6.8)\,\mathrm{arcmin}$ by our spectroscopy as shown in
Figure \ref{fig:spec_obs_D4g}.
The redshift was derived to be $z=3.72$ based on its He{\sc ii} and C{\sc iii}] emission lines (Figure
\ref{fig:spec_D4g_AGN}).
According to this estimate, the redshift separation between the AGN and the center of the protocluster is
$\Delta z=0.05$, which corresponds to the radial distance of $\sim8\,\mathrm{physical\>Mpc}$.
As the redshift separation is too large, it is unlikely that this AGN is a part of the protocluster
members that will merge into a single halo by $z=0$.

\subsubsection{The $u$-dropout protocluster candidate in the D1 field}
\begin{figure}
\epsscale{1.0}
\plotone{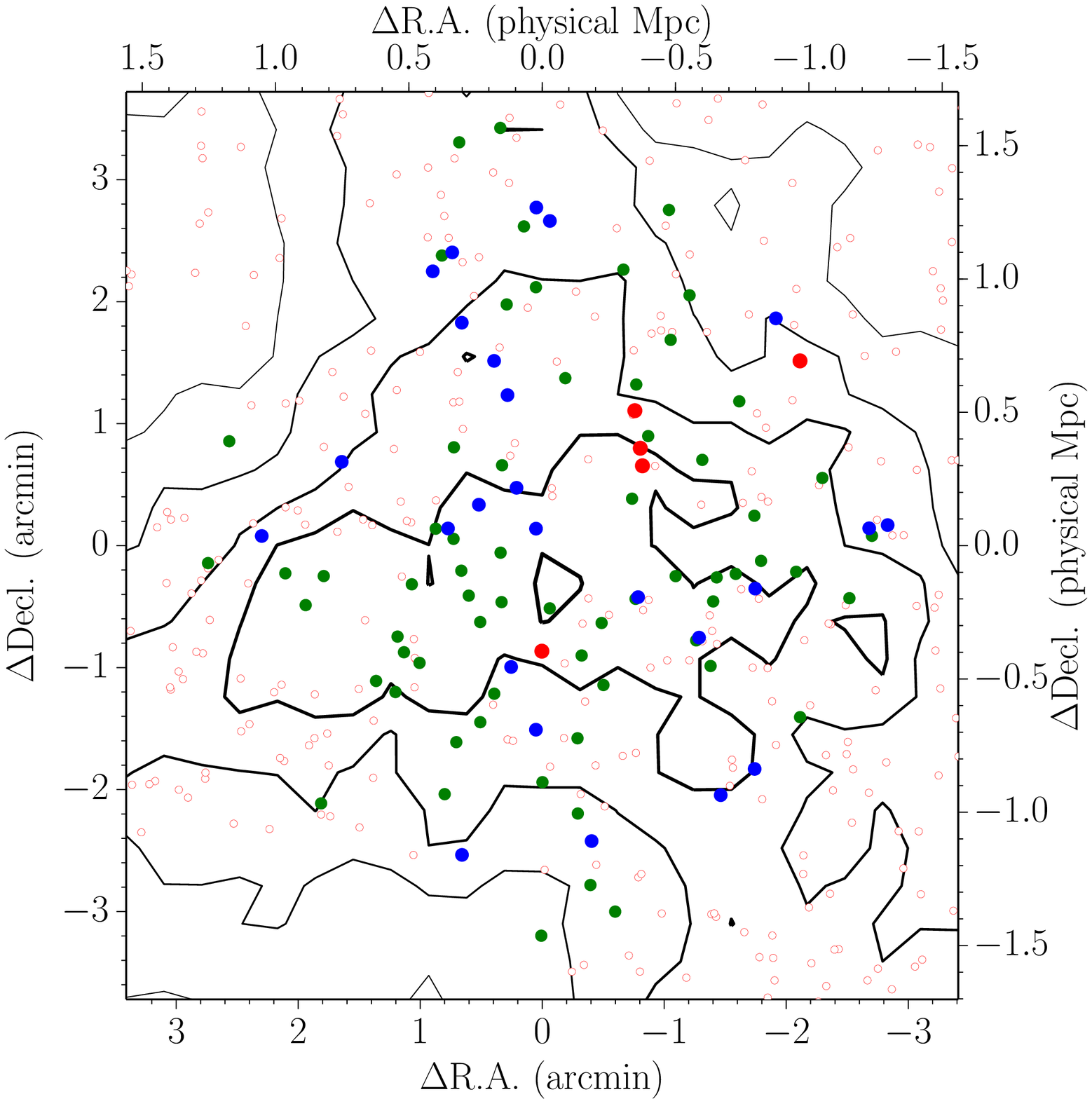}
\caption{Sky distribution of $u$-dropout galaxies and number density contours in the D1UD01 region.
    Spectroscopically observed galaxies are marked by filled circles (red: protocluster members, blue: non-members,
    green: Ly$\alpha$ undetected galaxies), and spectroscopically unobserved galaxies are shown by open circles.
    The origin (0,0) is $(\mathrm{R.A.}, \mathrm{Decl.})=(02:24:35.4,-04:19:40.3)$, which is defined as the
    center of the figure.
    The lines show the number density contours of $i$-dropout galaxies from $4\sigma$ to $0\sigma$ with a step
    of $1\sigma$.}
\label{fig:spec_obs_D1u}
\end{figure}

\begin{figure}
\epsscale{1.0}
\plotone{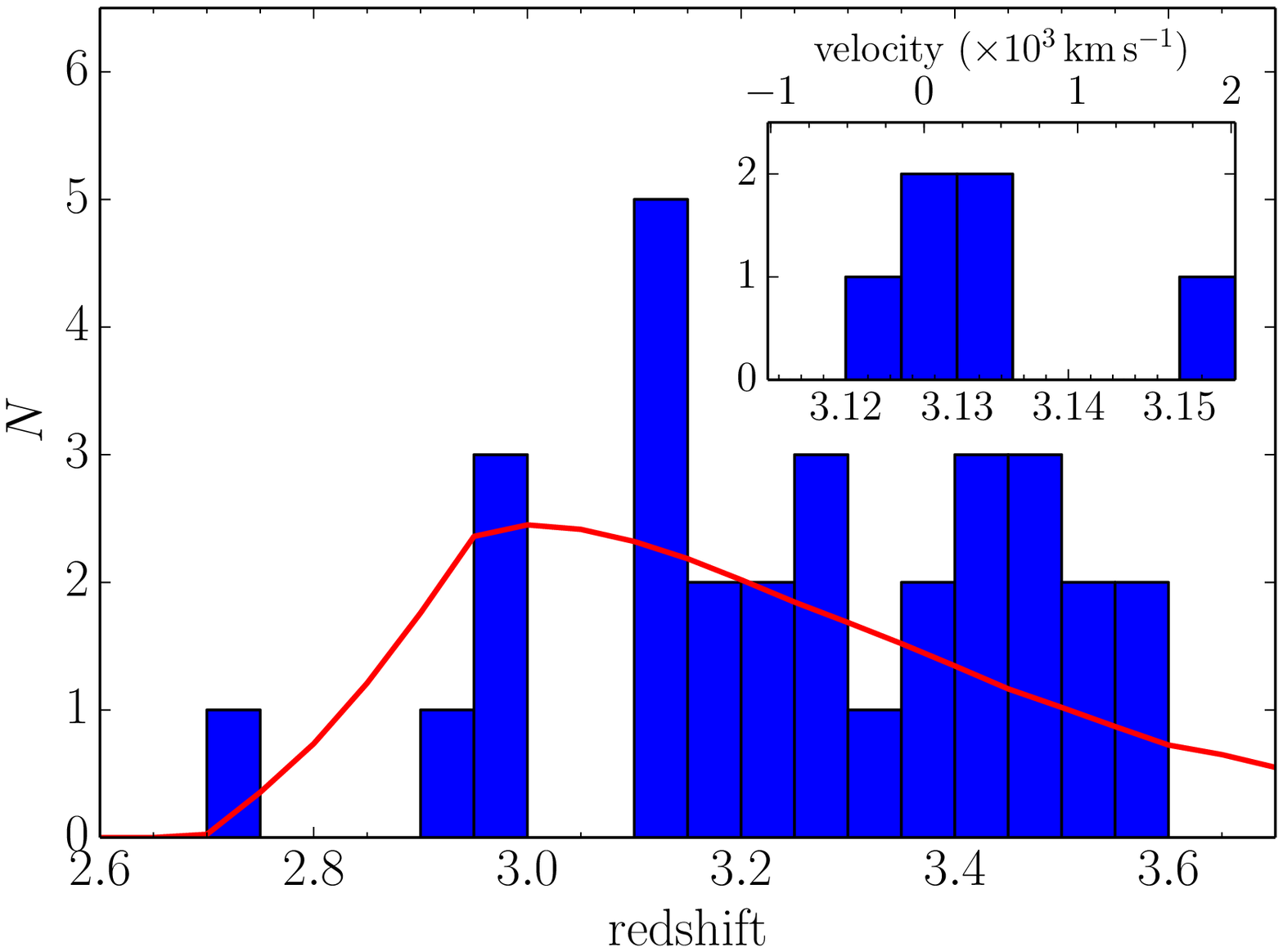}
\caption{Redshift distribution of 30 $u$-dropout galaxies with the bin size of $\Delta z=0.05$ in the D1UD01
    region.
    The inset is a close-up of the protocluster redshift range, with a bin size of $\Delta z =0.005$.}
\label{fig:redshift_D1u}
\end{figure}

We have spectroscopically observed 95 $u$-dropout galaxies in the D1UD01 region, and 30 galaxies have single
emission lines.
The sky distribution of the observed galaxies is shown in Figure \ref{fig:spec_obs_D1u}.
The redshift distribution is shown in Figure \ref{fig:redshift_D1u}.
There is a excess at $z=3.13$, including five galaxies within $\Delta z=0.008$.
 The probability to reproduce this excess by drawing from a uniform random distribution of 30 galaxies is only
0.9\%; thus, the five galaxies were found to be significantly clustered though the absolute excess is only five. 
The spatial and redshift separations among these five galaxies are small enough to merge into a single halo by
$z=0$ compared with the model prediction; therefore we confirmed a protocluster at $z=3.13$, which includes
five member galaxies (ID=6-10).

\subsubsection{The $u$-dropout protocluster candidate in the D4 field}
\begin{figure}
\epsscale{1.0}
\plotone{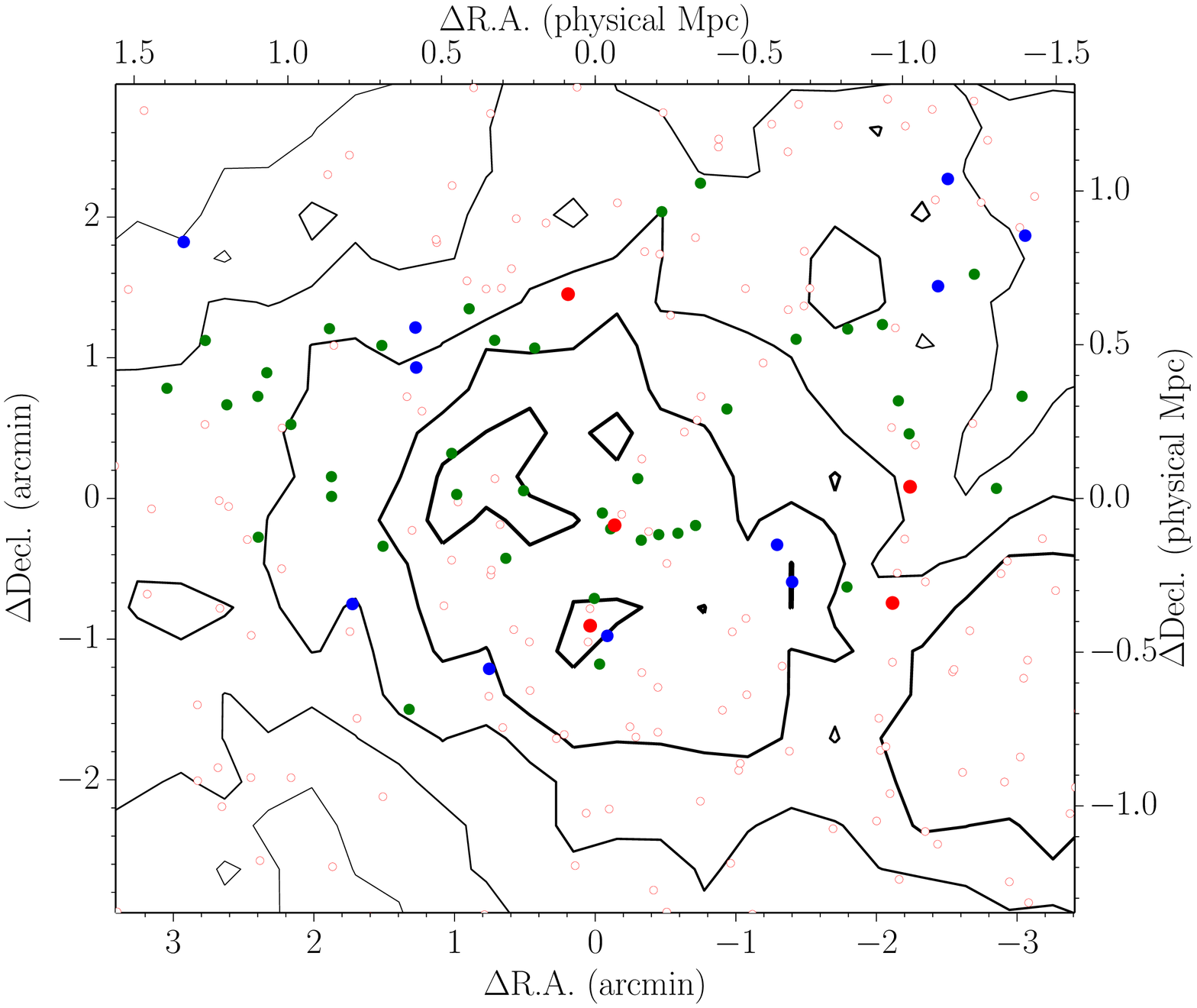}
\caption{Sky distribution of $u$-dropout galaxies and number density contours in the D4UD01 region.
    Spectroscopically observed galaxies are marked by filled circles (red: protocluster members, blue: non-members,
    green: Ly$\alpha$ undetected galaxies), and spectroscopically unobserved galaxies are shown by open circles.
    The origin (0,0) is $(\mathrm{R.A.}, \mathrm{Decl.})=(22:14:04.0,-17:59:11.3)$, which is defined as the
    center of the figure.
    The lines show the number density contours of $i$-dropout galaxies from $4\sigma$ to $0\sigma$ with a step
    of $1\sigma$.}
\label{fig:spec_obs_D4u}
\end{figure}

\begin{figure}
\epsscale{1.0}
\plotone{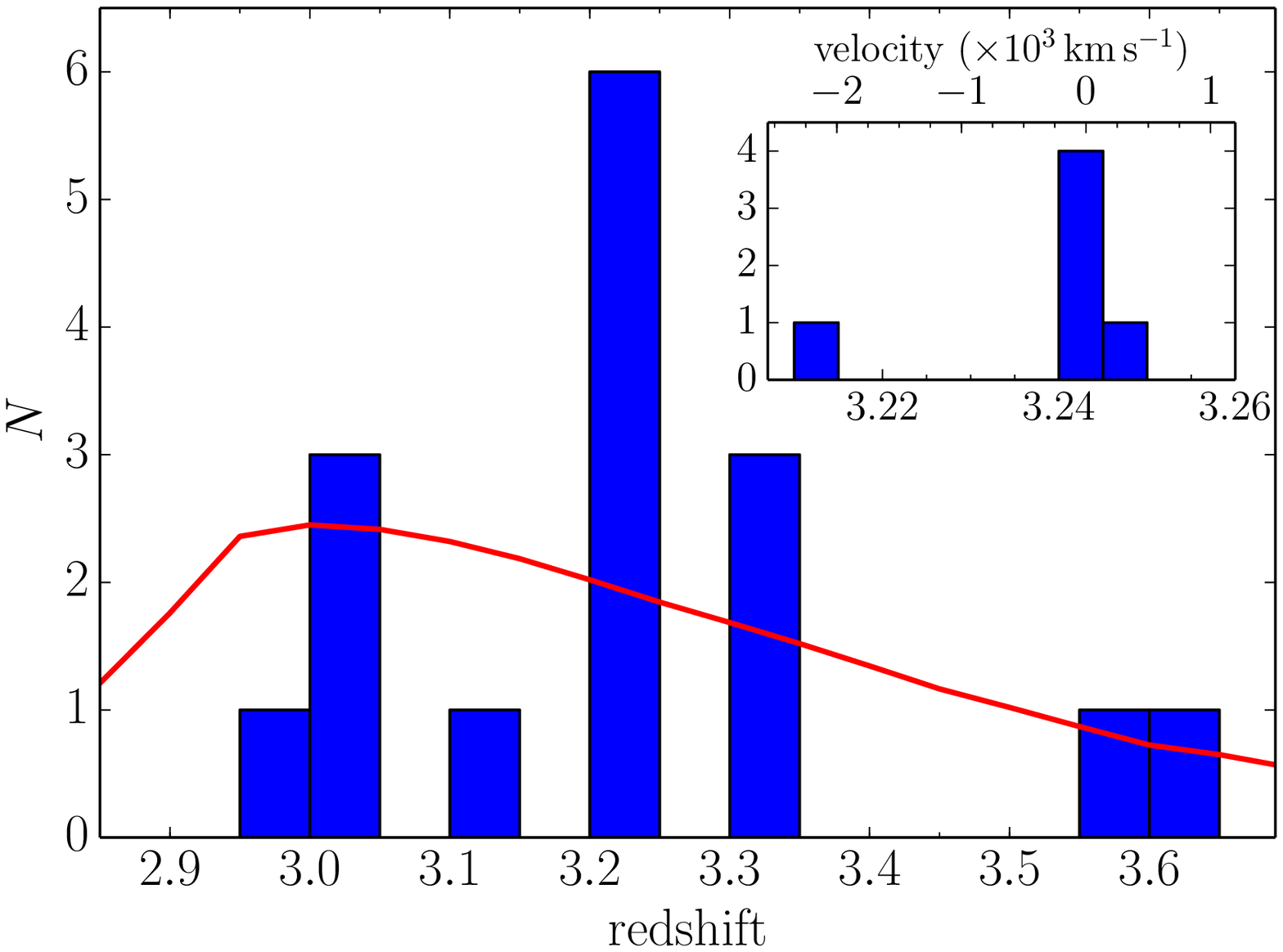}
\caption{Redshift distribution of 16 $u$-dropout galaxies with the bin size of $\Delta z=0.05$ in the D4UD01
    region.
    The inset is a close-up of the protocluster redshift range, with a bin size of $\Delta z =0.005$.}
\label{fig:redshift_D4u}
\end{figure}

We have spectroscopically observed 57 $u$-dropout galaxies in the D4UD01 region, and 16 galaxies have single
emission lines.
The sky distribution of the observed galaxies is shown in Figure \ref{fig:spec_obs_D4u}.
The redshift distribution is shown in Figure \ref{fig:redshift_D4u}.
There is a peak at $z=3.24$, consisting of five galaxies within $\Delta z=0.008$.
The probability to reproduce this excess by an uniform random distribution of 16 galaxies was found to be
less than 0.1\%. 
These five galaxies are expected to merge into a single halo by $z=0$ compared with the model prediction.
Therefore, we confirmed a protocluster at $z=3.24$, which includes five member galaxies (ID=7-11).

\subsubsection{Summary of protocluster confirmation}
Based on these follow-up spectroscopic observations, we were able to confirm three protoclusters in the D4GD01,
D1UD01, and D4UD01 regions.
We could not confirm that the overdensity observed in the D1GD01 region is indeed a protocluster.
Thus,  at least, the success rate of our protocluster search is found to be 3/4 at $z\sim3\mathrm{-}4$, which
is consistent with that of the model prediction ($\gtrsim76\%$ of $4\sigma$ overdense regions are expected to be
real protoclusters).
As for $r$- and $i$-dropout protocluster candidates, it is unclear whether they are real
protoclusters or not because of the small number of spectroscopically confirmed galaxies, though the D3ID01
and D1RD01 regions include close galaxy pairs, which could indicate the existence of protoclusters.
These results suggest that most of the other protocluster candidates will turn out to be genuine protoclusters
once sufficient follow-up spectroscopic observations are performed.
The summary of our protocluster confirmation is described in Table \ref{tab:cluster}.
 These findings do not only increase the number of known protocluster at high redshift, but provide us with
samples that are complementary to previous works in which protoclusters were mainly discovered in QSO or RG
regions.
Although we discovered one AGN in an overdense region by our spectroscopy, it was not associated with any
protocluster.
The precise relation between protoclusters and AGNs merits future investigation.

The radial velocity dispersions of the confirmed protoclusters were calculated by the redshifts of protocluster
members assuming that the redshifts probe line-of-sight velocity.
To calculate dispersion, we used the biweight variance \citep{beers90}, which is an effective method even with
a small sample. 
Since our follow-up spectroscopy is not complete, the effect of small number statistics should be taken into
account in the uncertainty of the radial velocity dispersion.
The uncertainty of radial velocity dispersion was measured by the combination of the velocity error of our
spectroscopic observations and the standard deviation of the measurements by bootstrap sampling the protocluster
members.
In addition to the optical imaging and our spectroscopy, rich multi-wavelength data is available in part of
the CFHTLS Deep Fields.
Although this enables us to make further analysis, such as SED fitting, to derive galaxy properties in more
detail, these studies will be addressed in a future paper.

\section{DISCUSSION} \label{sec:discuss}
The D4GD01 protocluster, of which we confirmed eleven protocluster members, is the most extensively mapped
protoclusters among the three confirmed systems described in Section \ref{sec:spec}.
Thus, we focus on this protocluster in the following discussion of the internal structure of the protocluster
and the properties of its galaxies.

\subsection{Protocluster internal structure} \label{sec:struc}
\begin{figure}
\epsscale{1.0}
\plotone{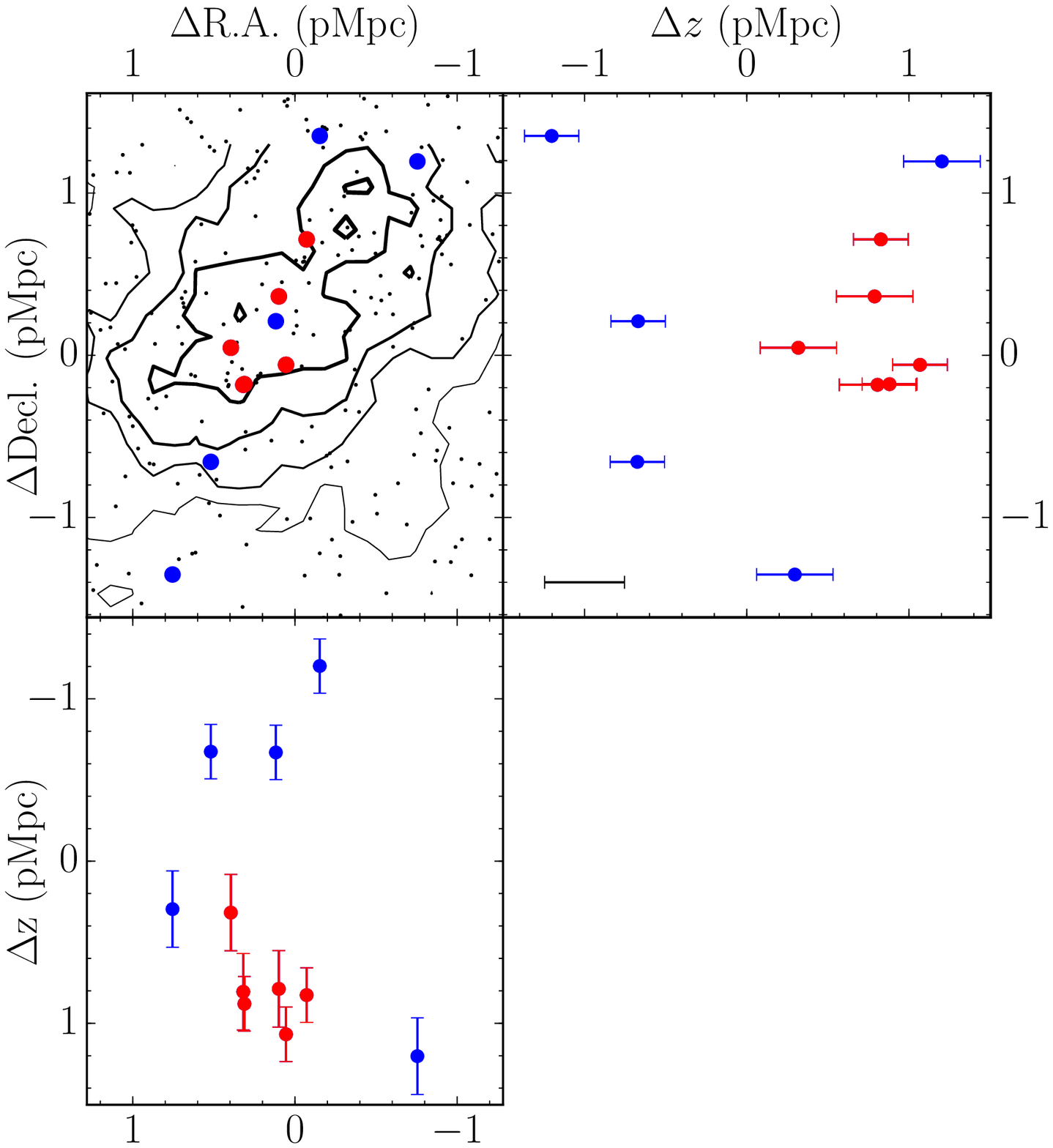}
\caption{Three-dimensional distribution of the protocluster galaxies in the D4GD01 region.
    The filled circles represent the eleven protocluster galaxies (six red ones are galaxies residing in core
    region, and five blue in outskirt region), and the dots are $g$-dropout galaxies.
    Note that the origin (0,0) of this figure is defined as
    $(\mathrm{R.A.}, \mathrm{Decl.})=(22:16:50.4,-17:18:41.6)$.
     The black scale shown in the lower left corner of the top-right panel represents the typical difference
    expected between the apparent (i.e. including the effect of peculiar velocities) and geometrical redshifts.}
\label{fig:3D}
\end{figure}
\begin{figure}
\epsscale{1.0}
\plotone{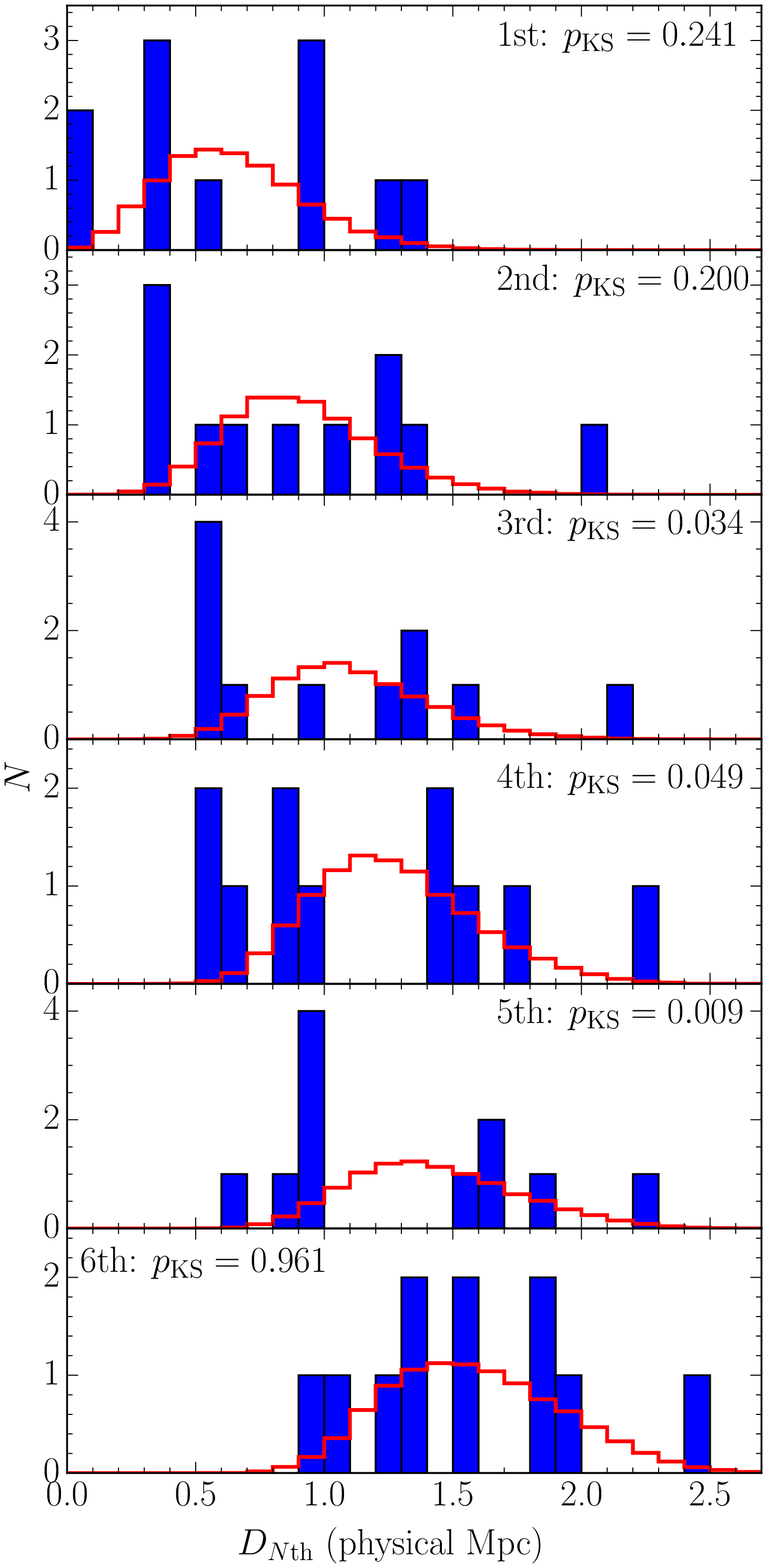}
\caption{Distribution of the separation from the first (top) to sixth (bottom) nearest galaxy in the D4GD01
    protocluster (blue histogram).
    The red line shows an expected distribution assuming that eleven galaxies are randomly distributed in the same
    volume of the D4GD01 protocluster.}
\label{fig:DNth}
\end{figure}
We investigated the three-dimensional distribution of protocluster galaxies in the D4GD01 region, as shown in
Figure \ref{fig:3D}.
In this analysis, the distances are simply estimated from the measured redshift, including any possible peculiar
velocity component.
However, we have checked that this does not significantly affect our estimates given the typical size of
protoclusters
\footnote{In principle, redshift does not completely correspond to radial distance due to the effect of radial
velocity.
However, in the high-redshift protoclusters, nearly all of the member galaxies, which merge into a single dark
matter halo with mass of $>10^{14}\,\mathrm{M\sun}$ by $z=0$, are still embedded in individual host halos at
$z>3$, and their clustering is not probably strong enough to provoke a large peculiar velocity field.
Therefore, we expect that their real three-dimensional distribution at some level can be determined by using their
redshifts as proxies for their relative line of sight distances.
According to the theoretical model of \citet{henriques12}, the deviation of the difference between apparent and
geometrical redshift is $\Delta z\sim0.002$, shown by black scale in Figure \ref{fig:3D}.}.
This protocluster seems to have a region where galaxies are strongly concentrated, reminiscent of a cluster core.
To discuss the internal structure quantitatively, we calculated the spatial separation of galaxies with respect
to the $N$th nearest neighbor.
In Figure \ref{fig:DNth}, the distributions of the separation from the 1st to 6th nearest galaxies of individual
protocluster galaxies are shown, and the red-line histograms indicate the expected distribution if eleven
galaxies are randomly located inside the protocluster. 
The distribution of protocluster galaxies from the 3rd to 5th nearest are significantly different from the random
distribution, whose significances are $p<0.05$ based on the Kolmogorov-Smirnov (KS) test; especially, the
significance of 5th nearest is less than 0.01.
In contrast, there are no significant differences in the distribution of 6th or higher nearest galaxies.
These results suggest that the protocluster has a subgroup consisting of six galaxies; therefore, the subgroup
cannot be seen in the distribution of the 6th or higher nearest galaxies, which are consistent with the random
distribution.
The six galaxies constituting the subgroup are defined by having shorter separation than
$1.0\,\mathrm{physical\>Mpc}$ from the 5th nearest galaxies (ID=11-16).
The distributions of 1st and 2nd nearest galaxies could also be divided into two groups of shorter and longer
separations though they are not so significant ($p\sim0.2$) due to focusing on too small scale.
These six galaxies are indicated by red points in Figure \ref{fig:3D}, and are located near the center of the
protocluster.
There are several galaxies in the region surrounding the core, which could assemble into the core to form a
rich cluster.
This is in clear contrast that we saw in the protocluster at $z\sim6$, which was found to have several small
subgroups, like galaxy pairs \citep{toshikawa14}.
Although we have only one protocluster at each redshift, if they are the progenitor and descendant of each
other, the transformation of protocluster internal structure from $z\sim6$ to $z\sim4$ may be indicative of the
virializing process over cosmic time, whereby protoclusters dynamically evolve into a more and more concentrated
structure.
At $z\sim2\mathrm{-}3$, the virializing process would not be completed yet though there are richer protoclusters
where some massive and passive galaxies have already appeared.
The internal structures of some protoclusters have been more closely investigated using extensive spectroscopy,
and some protoclusters were found to have significant substructure \citep{kuiper11,kuiper12}.
Therefore, even at the same redshift, protoclusters could have a large variety of internal structure.
In this study based on a small number of protoclusters, the difference of internal structure found in the $z\sim6$
and $z\sim4$ protoclusters are assumed to be resulted from the evolutionary phase of the representative
protoclusters.
However, larger samples at each redshift will be required to statistically match progenitors and descendants,
allowing us to study cluster formation over cosmic time \citep[e.g.,][]{chiang13}.

\subsection{Rest-frame UV Properties of the Protocluster Members}
\begin{figure}
\epsscale{1.2}
\plotone{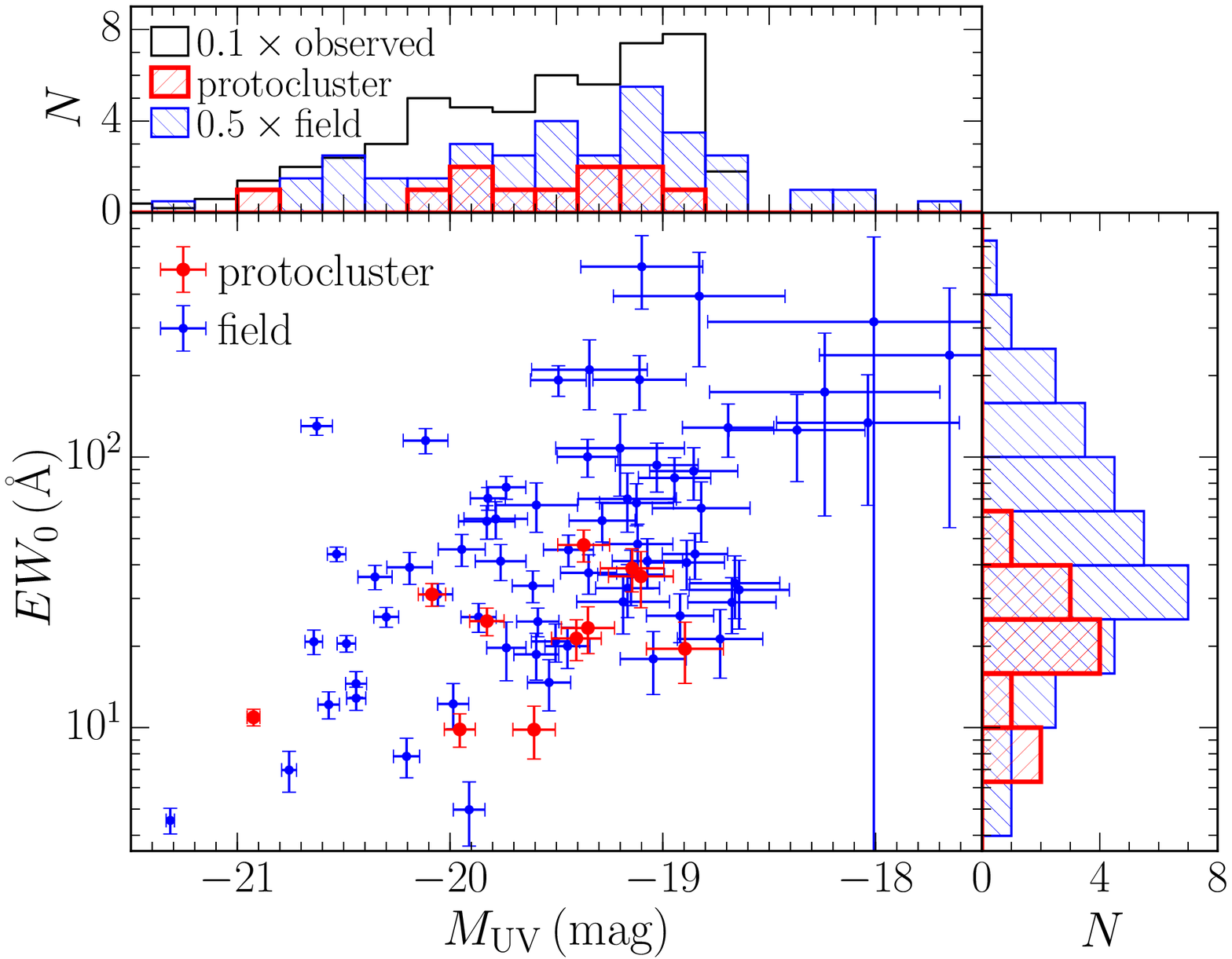}
\caption{$EW_0$ versus $M_\mathrm{UV}$ of spectroscopically confirmed $g$-dropout galaxies in the
    CFHTLS D1 and D4.
    The histograms in the top and right panels show the $EW_0$ and $M_\mathrm{UV}$ distributions of
    the protocluster and field galaxies.
    Red and blue color represent the protocluster and field galaxies  in all three panels, respectively.
    In the histogram of $M_\mathrm{UV}$ (top), spectroscopically observed $g$-dropout galaxies are also
    shown by the black line, which is useful for evaluating the magnitude distribution of the targets and the
    Ly$\alpha$ detection rate in our follow-up spectroscopy.
    For clarity, the histograms of observed and field galaxies were multiplied by a factor of 0.1 and 0.5,
    respectively.
    $M_\mathrm{UV}$ of observed galaxies (black), including Ly$\alpha$ undetected galaxies, are estimated from
    the $i'$-band magnitude, which is free from IGM absorption and Ly$\alpha$ emission at $z\sim3.8$, by assuming
    a flat UV slope; on the other hand, the $M_\mathrm{UV}$ of Ly$\alpha$ detected galaxies (red and blue) was
    calculated based on a combination of the broad-band photometry and the Ly$\alpha$ flux, as described in
    Section \ref{sec:spec_result}.}
\label{fig:M-EW}
\end{figure}

We compared several galaxy properties between protocluster members and coeval field galaxies to investigate
whether there are any differences due to their environment.
The D4GD01 protocluster consists of eleven members, while the number of field $g$-dropout galaxies available
is 67 by combining the galaxies in the D1 and D4 fields.
The protocluster and field galaxies were identified by the same imaging and spectroscopic observations;
thus, some sample bias, if any, would affect both samples in the same way.
This should allow us to make a fair comparison between protocluster and field galaxies.
The average of $L_\mathrm{Ly\alpha}$, $M_\mathrm{UV}$, and $EW_0$ are described in Table
\ref{tab:ave_D4g}.
Since the D4GD01 protocluster has a core-like substructure as mentioned in Section \ref{sec:struc}, the
eleven members were divided into two groups of six galaxies in the core and five galaxies in the outskirt.
Table \ref{tab:ave_D4g} also shows the average properties of these two subgroups.
In Figure \ref{fig:M-EW}, the properties of individual protocluster and field galaxies are plotted on the
$M_\mathrm{UV}$ and $EW_0$ diagram.
We found that protocluster galaxies have significantly smaller $EW_0$ or $L_\mathrm{Ly\alpha}$ luminosity than
field galaxies (KS p-value is $<0.05$); especially, fainter protocluster members in $M_\mathrm{UV}$ are
strongly suppressed in their Ly$\alpha$ emissions compared with field galaxies.
No significant difference between the core and the outskirt in the $EW_0$ and $M_\mathrm{UV}$ distribution
was found.
\begin{deluxetable}{cccc}
\tablecaption{Average of observed properties of $g$-dropout galaxies in the CFHTLS Deep Fields.
    \label{tab:ave_D4g}}
\tablewidth{0pt}
\tablehead{
    \colhead{} & \colhead{$L_\mathrm{Ly\alpha}$} & \colhead{$M_\mathrm{UV}$} & \colhead{$EW_0$} \\
    \colhead{} & \colhead{($10^{42}\,\mathrm{erg\,s^{-1}}$)} & \colhead{(mag)} & \colhead{(\AA)}
}
\startdata
protocluster & $1.46\pm0.76$ & $-19.47\pm0.51$ & $24.48\pm12.20$ \\
field & $2.34\pm1.67$ & $-19.45\pm0.74$ & $41.68\pm39.00$ \\
p-value\tablenotemark{a} & 0.04 & 0.67 & 0.03 \\
\hline
core & $1.69\pm0.75$ & $-19.84\pm0.60$ & $22.45\pm10.03$ \\
outskirt & $1.20\pm0.68$ & $-19.29\pm0.26$ & $27.13\pm14.60$
\enddata
\tablenotetext{a}{Using the KS test, the distribution of observed properties are compared between protocluster
    and field galaxies.}
\end{deluxetable}

Previous studies have found little evidence for significant differences between the properties of galaxies inside
and outside protoclusters, at least at $z\gtrsim4$ \citep{overzier08,overzier09b,toshikawa14}.
It is therefore interesting that we are finding a difference in the $EW_0$ or $L_\mathrm{Ly\alpha}$ distributions
between field and protocluster in the D4GD01 system at $z=3.67$.
This is perhaps an indication that the situation is changing around $z\sim4$.
 One simple mechanism that could reduce the Ly$\alpha$ $EW_0$ would be dust, which traps Ly$\alpha$
photons.
If dust is a major reason for the small $EW_0$, $M_\mathrm{UV}$ is expected to be systematically smaller in the
protocluster because UV flux can also be easily attenuated by dust as in the case with Ly$\alpha$ emission.
However, we did not find any systematic difference in $M_\mathrm{UV}$ between the protocluster and field galaxies 
(see Table \ref{tab:ave_D4g}).
 Although dust attenuation does not seems to be the reason for this result, it is still possible that the
amount of UV flux attenuated by dust is being compensated by higher SFR or other effects related to UV emission;
in that case, we would not find any difference in the distribution of apparent $M_\mathrm{UV}$ between protocluster
and field galaxies even if there were to be a clear difference of dust attenuation.
Therefore, in order to more directly estimate the dust attenuation, we compared the UV slope between protocluster
and field galaxies.
The UV slope of each $g$-dropout galaxy was determined from the $i-z$ color.
However, since the $i'$-band image was significantly deeper ($\sim1\,\mathrm{mag}$) than the $z'$-band image,
some $g$-dropout galaxies were not detected in the $z'$-band.
For the estimate of the UV slope, we therefore only used $g$-dropout galaxies which were detected in the $z'$-band
image with $>2\sigma$ significance ($<26.70\,\mathrm{mag}$).
Although this discordant depth between the $i'$- and $z'$-band images will lead to a bias in the estimate of UV
slope, the protocluster and field galaxies would be equally biased as they were selected by the same criteria and
from the same dataset.
The numbers of $g$-dropout galaxies used in the estimate was nine for the protocluster and 50 for the field.
Note that there is no difference in the fraction of galaxies detected in $z'$-band between protocluster
(9/11) and field (50/67).
The UV slope was calculated from
\begin{equation}
\beta =
-0.4\times \frac{m_i-m_z}{\log_{10}\lambda_{\mathrm{eff},i}-\log_{10}\lambda_{\mathrm{eff},z}}-2.0,
\end{equation}
where $\lambda_\mathrm{eff}$ is the effective wavelength.
The average $\beta$ of the protocluster galaxies was $\beta=-1.88\pm0.38$, and that of the field galaxies was
$\beta=-1.92\pm0.17$; thus,  it would be hard to explain the difference in $EW_0$ simply by dust attenuation.
It should be noted that, the radiative process of Ly$\alpha$ photon to escape from a galaxy is very complicated
affected by many other quantities, such as dust geometry, gas kinematics, and outflow
\citep[e.g.,][]{verhamme08,duval14}.

Neutral hydrogen gas within a protocluster is another possible reason for the small $EW_0$.
\citet{cucciati14} found a large amount of neutral hydrogen gas ($\sim10^{12}-10^{13}\,\mathrm{M_\sun}$) in the
intracluster space of a protocluster by examining spectra of background galaxies of a protocluster at $z=2.9$
that showed absorption at the same wavelength as the observed Ly$\alpha$ of the protocluster.
If the same is true in the D4GD01 protocluster, though our follow-up spectroscopy is not deep enough to check
this even by stacking, a small $EW_0$ could be explained as a result of resonant scattering by the
intracluster neutral hydrogen gas.
While the UV photons can penetrate neutral hydrogen gas, Ly$\alpha$ emission is scattered and diffused,
consistent with our results.
Suppose that a nearly mature protocluster, such as the D4GD01 protocluster, had already accumulated
significant cold intracluster gas at $z=3.67$; the intracluster gas would come either from the outside of the
protocluster drawn in by the strong gravitational potential of the protocluster, or could be brought in with the
evolved member galaxies themselves through in- and outflows.

Although it is difficult to identify the cause, we have found that the  average $EW_0$ of $z=3.67$
protocluster  members is significantly smaller than that of field galaxies.
However, it is still under debate how Ly$\alpha$ $EW_0$ depends on environments.
Actually, in contrast with our study, \citet{yamada12} have found larger $EW_0$ in the SSA22 protocluster at
$z=3.09$.
\citet{kuiper12} reported a protocluster being composed of two subgroups at $z=3.13$, where the one subgroup has
larger $EW_0$, but the other has smaller.
Furthermore, there are some protoclusters which have no significant difference of Ly$\alpha$ $EW_0$ from that of
field galaxies \citep{mawatari12}.
Since the number of known protoclusters is limited even around $z\sim3$ so far , it will be required to make a
large sample enough to address a general feature.
This may have consequences for the measurement of the Ly$\alpha$ fraction \citep{ono12,treu13}, which is
one of the ways to probe reionization at high redshifts.
 A detailed study of galaxy properties will be made in the future by combining multi-wavelength data.

\section{CONCLUSIONS} \label{sec:conc}
In this study, we have presented a protocluster survey from $z\sim3$ to $z\sim6$ in the CFHTLS Deep Fields.
This survey was performed using wide-field imaging without using the preselection of common protocluster probes
such as RGs and QSOs.
Protocluster candidates were identified by measuring the surface number density of dropout galaxies, and the
follow-up spectroscopic observations identified three real protoclusters.
The major results and implications of this study are summarized below.

\begin{enumerate}
\item We investigated the sky distribution of $u$-, $g$-, $r$-, and $i$-dropout galaxies in the wide-field imaging
    of the CFHTLS Deep Fields, and quantified the local surface number density by counting galaxies within a fixed
    aperture.
    We selected a total of 21 overdense regions with an overdensity significance greater than $4\sigma$ as
    protocluster candidates.
    The number density of protocluster candidates was approximately one candidate per $1\,\mathrm{deg^2}$ area
    for each redshift sample based on a $4\,\mathrm{deg^2}$ survey.
\item We investigated the relation between the overdensity at high redshifts and the descendant halo mass using
    light-cone models constructed from cosmological simulations.
    We selected galaxy samples with the same redshift distribution as the observations, and the same overdensity
    measuring procedure was applied to this simulated sample of dropout galaxies.
    A strong correlation between the overdensity at high redshifts and the descendant halo mass at $z=0$ was
    found, and $\gtrsim76\%$ of the overdense regions with significance over $4\sigma$ are expected to grow
    to dark matter halos with $M>10^{14}\,\mathrm{M_\sun}$ at $z=0$.
    Despite significant projection effects, the model predicts that protoclusters can be identified with high
    confidence by measuring the surface overdensity significance.
    In addition, the model predicts that protocluster members are, on average, spread within a sphere of
    $2\,\mathrm{physical\>Mpc}$ radius.
\item We carried out follow-up spectroscopic observations of eight protocluster candidates between $z\sim3$ and
    $z\sim6$ to confirm whether these were genuine protoclusters.
    The redshifts of all the protocluster members were determined by detecting their Ly$\alpha$ emission lines,
    and no apparent contamination from low-redshift interlopers was found in our spectroscopic observation.
    Although the completeness of slit allocation to dropout galaxies in a protocluster candidate is about
    $30\mathrm{-}60\%$, three of the eight protocluster candidates were confirmed to be genuine protoclusters
    by ascertaining that their member galaxies were clustering both in spatial and redshift directions
    ($<2\,\mathrm{physical\>Mpc}$) with $\sim3\sigma$ significance with more than five members spectroscopically
    identified.
    Spectroscopy revealed that chance alignment of dropout galaxies mimics an overdensity region for one
    candidate.
    There are some signatures of clustering of several galaxies in the other four protocluster candidates at
    $z\sim5\mathrm{-}6$; however, the numbers of spectroscopically confirmed galaxies are still too small to
    conclude that they are genuine protoclusters.
    Although there are still many protocluster candidates to follow-up spectroscopically, our method to
    search for protoclusters utilizing wide-field imaging is reliable and sufficiently effective to construct
    high-redshift protocluster samples based on the success rate of follow-up observations.
\item We investigated the internal structure of the D4GD01 protocluster at $z=3.67$ based on its eleven
    spectroscopically confirmed members.
    The distribution of member galaxies exhibits a core-like structure: half of the members are concentrated
    in a central small region ($<0.5\,\mathrm{physical\>Mpc}$), and the others in the outskirts
    ($\sim1.0\,\mathrm{physical\>Mpc}$).
    The result implies that this protocluster might be on the way to evolve into a virialized
    structure though further protocluster samples are required to confirm a general trend.
\item The D4GD01 protocluster galaxies have significantly smaller $EW_0$ than coeval field
    galaxies, while there is no difference in $M_\mathrm{UV}$.
    We considered two physical mechanisms that may lead to this difference; the first is dust in protocluster
    galaxies, and the second is intracluster neutral hydrogen gas.
    Although we were not able to draw definite conclusions based on current data, the UV slope was found not to
    favor an interpretation whereby the difference in $EW_0$ is attributed only to dust.
    Our finding of a smaller $EW_0$ implies that the properties of protocluster galaxies might be affected by the
    environment already at $z=3.67$.
\end{enumerate}

Although we were successful in finding at least three new protoclusters using wide-field imaging and spectroscopy
in a blank deep field, the sample size is still too small to elucidate a general picture of the structure
formation and evolution of environmental effects.
However, this study is an important benchmark for finding large numbers of protoclusters and tracing the cluster
formation history in upcoming deep, wide surveys using identical techniques.
Using the new instrument Hyper SuprimeCam (HSC) on the Subaru telescope, we are performing an unprecedented
wide and deep survey over the next four years.
The HSC strategic survey consists of three layers: the WIDE layer covers $1400\,\mathrm{deg^2}$ with the $i$-band
depth of $m_i=26.0$, the Deep layer $28\,\mathrm{deg^2}$ with $m_i=26.8$, and the Ultradeep layer
$3.5\,\mathrm{deg^2}$ with $m_i=27.4$.
From this study, we estimate that the number of protoclusters, that the HSC survey will be able to find, will be
$>20$ at $z\sim5\mathrm{-}6$ and $>1000$ at $z\sim3\mathrm{-}4$.
From 2018 onwards, spectroscopic follow-up of tens of thousands of dropout galaxies selected from the HSC deep
layer will be performed using the large multiplexing capability of the Prime Focus Spectrograph, also on the
Subaru telescope.
This will allow us to understand the cluster formation process all the way from reionization to the present-day.

\acknowledgements
The CFHTLS data used in this study are based on observations obtained with MegaPrime/MegaCam, a joint project of
CFHT and CEA/IRFU, at the Canada-France-Hawaii Telescope (CFHT) which is operated by the National Research
Council (NRC) of Canada, the Institut National des Science de l'Univers of the Centre National de la Recherche
Scientifique (CNRS) of France, and the University of Hawaii.
This study is based in part on data products produced at Terapix available at the Canadian Astronomy Data Centre
as part of the Canada-France-Hawaii Telescope Legacy Survey, a collaborative project of NRC and CNRS.
And, this study based on data collected at the Subaru, the W. M. Keck, and the Gemini North telescopes.
The Subaru telescope is operated by the National Astronomical Observatory of Japan.
The W. M. Keck telescope is operated as a scientific partnership among the California Institute of Technology,
the University of California and the National Aeronautics and Space Administration.
The W. M. Keck Observatory was made possible by the generous financial support of the W.M. Keck Foundation.
The Gemini North telescope is operated by the Association of Universities for Research in Astronomy, Inc., under
a cooperative agreement with the NSF on behalf of the Gemini partnership: the National Science Foundation (United
States), the National Research Council (Canada), CONICYT (Chile), the Australian Research Council (Australia),
Minist\'{e}rio da Ci\^{e}ncia, Tecnologia e Inova\c{c}\~{a}o (Brazil) and Ministerio de Ciencia, Tecnolog\'{i}a
e Innovaci\'{o}n Productiva (Argentina).
We are grateful to the Subaru, the W. M. Keck, and Gemini Observatory staff for their help with the observations,
and wish to recognize and acknowledge the very significant cultural role and reverence that the summit of
Mauna Kea has always had within the indigenous Hawaiian community.
The Millennium Simulation databases used in this study and the web application providing online access to them
were constructed as part of the activities of the German Astrophysical Virtual Observatory (GAVO).
We thank the anonymous referee for valuable comments and suggestions which improved the manuscript.
This research was supported by the Japan Society for the Promotion of Science through Grant-in-Aid for Scientific
Research 12J01607 and 15H03645.

\end{document}